\documentclass[a4paper,11pt]{article}
\usepackage[normalem]{ulem}
\pdfoutput=1 

\usepackage{jheppub} 
\usepackage{url}
\usepackage{caption}
\usepackage{subcaption}

\usepackage[latin9]{inputenc}
\usepackage{float}
\usepackage{amsmath}
\usepackage{amssymb}
\usepackage{graphicx}
\usepackage{esint}
\usepackage{hyperref}
\usepackage{comment}
\usepackage{color}
\usepackage{microtype}
\usepackage{cleveref}
\usepackage{breakurl}
\usepackage{bbm}
\newcommand{\be}{\begin{equation}}
\newcommand{\ee}{\end{equation}}
\newcommand{\ben}{\begin{displaymath}}
\newcommand{\een}{\end{displaymath}}
\newcommand{\bea}{\begin{eqnarray}}
\newcommand{\eea}{\end{eqnarray}}
\def\K{K{\"a}hler}
   \newcommand{\rf}[1]{(\ref{#1})}

\def\E{{$E_{7(7)}$}}

\def\be{\begin{equation}}
\def\ee{\end{equation}}
\def\bea{\begin{eqnarray}}
\def\eea{\end{eqnarray}}
\def\ba{\begin{array}}
\def\ea{\end{array}}
\def\bit{\begin{itemize}}
\def\eit{\end{itemize}}

\newcommand{\cN}{\mathcal{N}}

\def\E{{$E_{7(7)}$}}

 \makeatletter

\allowdisplaybreaks

\makeatother

\makeatletter
\DeclareRobustCommand{\rcite}[1]{%
  \rcite@aux#1,\@nil{#1}%
}
\def\rcite@aux#1,#2\@nil#3{%
  \if\relax#2\relax
    Ref.~\cite{#3}%
  \else
    Refs.~\cite{#3}%
  \fi
}
\makeatother

\hypersetup{
    colorlinks = true,
    citecolor = {blue},
    linkcolor = {blue},
    urlcolor = {blue},
}

 \title{\rm { \Large \bf        M-theory Cosmology,  Octonions,   Error Correcting  Codes }}

\author[a,b]{Murat Gunaydin,}
\author[a]{Renata Kallosh,}
\author[a]{Andrei Linde,}
\author[a,c]{and Yusuke Yamada}

\affiliation[a]{Stanford Institute for Theoretical Physics and Department of Physics,\\ Stanford University, Stanford, CA 94305, USA}
\affiliation[b]{Institute for Gravitation and the Cosmos and Department of Physics, \\ Pennsylvania State University,  University Park, PA 16802, USA
}
\affiliation[c]{Research Center for the Early Universe (RESCEU), Graduate School of Science,\\ The University of Tokyo, Hongo 7-3-1
Bunkyo-ku, Tokyo 113-0033, Japan}
\emailAdd{mgunaydin@psu.edu}
\emailAdd{kallosh@stanford.edu}
\emailAdd{alinde@stanford.edu}
\emailAdd{yamada@resceu.s.u-tokyo.ac.jp}
\notoc
\preprint{RESCEU-14/20}

\abstract{  We study M-theory compactified on   twisted 7-tori with $G_2$-holonomy. The effective 4d supergravity has 7 chiral multiplets, each with a unit logarithmic \K\, potential.  We propose  octonion, Fano plane  based superpotentials, codifying the error correcting Hamming (7,4) code. The corresponding 7-moduli models have  Minkowski vacua with one flat direction.  We also propose superpotentials based on octonions/error correcting  codes for Minkowski vacua  models with two flat directions. We update phenomenological  $\alpha$-attractor models of  inflation   with $3\alpha=7,6,5,4,3,1$, based on inflation along these flat directions.
These inflationary models reproduce the benchmark targets for detecting B-modes,   predicting 7 different values of  $r = 12\alpha/N_{e}^{2}$  in the range  $10^{-2}\gtrsim r \gtrsim  10^{-3}$, to be explored by future cosmological observations.
}

\begin{document}

\maketitle

   \newpage

\tableofcontents{}

  \parskip 8pt

 \newpage
 \section{Introduction}\label{intro}

\begin{figure}[H]
\centering
\includegraphics[scale=0.6]{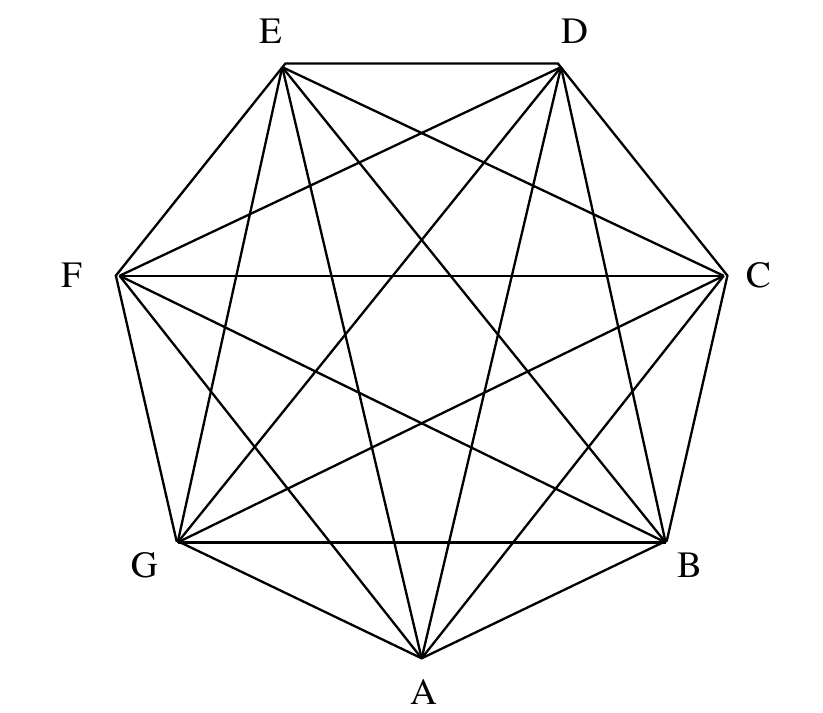}
\caption{\footnotesize  The $E_7$ entanglement diagram from~\cite{Duff:2006ue}. Each of the 7 vertices A,B,C,D,E,F,G represents a qubit and each of the 7 triangles ABD, BCE, CDF, DEG, EFA, FGB, GAC describes a tripartite entanglement.}
\label{fig:heptagon}
\end{figure}

 It was observed in studies of supersymmetric black holes in maximal 4d supergravity that in the context of \E\, symmetry and
 tripartite entanglement of 7 qubits, there is a relation between black holes,  octonions, and  Fano plane~\cite{Duff:2006ue,Levay:2006pt,Levay:2010hna,Borsten:2012fx}.
The entangled 7-qubit system corresponds to 7 parties: Alice, Bob, Charlie, Daisy, Emma, Fred and George. It has 7 3-qubit states and 7 complimentary 4-qubit states. One can see the 7-qubit entanglement in a heptagon with  7 vertices A,B,C,D,E,F,G and  7 triangles  and 7 complimentary quadrangles in Fig. \ref{fig:heptagon}.
The
seven triangles in the heptagon correspond to a multiplication table of the octonions.

It was also observed in~\cite{Levay:2006pt,Levay:2010hna,Borsten:2012fx} that these black hole structures are related to a (7,4)
   error correcting Hamming code \cite{Hamming:1950:EDE}  which is capable of correcting up to 1 and detecting up to 3 errors.
Here we  will update the original 7-moduli cosmological models developed in~\cite{Ferrara:2016fwe,Kallosh:2017ced,Kallosh:2017wnt,Kallosh:2019hzo}
which predict 7  specific targets for detecting primordial gravitational waves from inflation.  We will use the superpotentials  derived from  octonions and $G_2$ symmetry. The updated cosmological models will be  shown to be related to
(7,4)  Hamming error correcting codes, octonions and  Fano planes.

We will continue  an investigation of the 7-moduli model effective 4d supergravity following \cite{DallAgata:2005zlf,Duff:2010vy,Derendinger:2014wwa,Cribiori:2019hrb} and apply them to observational cosmology as in  \cite{Ferrara:2016fwe,Kallosh:2017ced,Kallosh:2017wnt,Kallosh:2019hzo}. Some earlier insights into the corresponding M-theory models have  been obtained in~\cite{Bilal:2001an,Beasley:2002db,House:2004pm}.
 Our cosmological models are based on 11d M-theory/supergravity  compactified on  a twisted 7-tori   with holonomy  group  $G_2$.
A related 7-moduli model  originating from  a 4d  gauged $\cN=8$  supergravity was studied in~\cite{Bobev:2019dik}. It was emphasized there that $\mathbb{Z}_2^3$ invariant 7 Poincar\'e disk  scalar manifold  $\Big[{SU(1,1)\over U(1)}\Big]^7$ is related  to a remarkable superpotential whose structure matches the single bit error correcting (7,4) Hamming code.

In M-theory compactified on a 7-manifold with $G_2$ holonomy the coset space describing our 7 moduli is
$\Big[{SL(2,\mathbb{R})\over SO(2)}\Big]^7$. It corresponds to half-plane variables related to Poincar\'e disk variables by a Cayley transform. In such a 7-moduli model we will establish a relation between octonions\footnote{Octonions were discovered in 1843 by J. T. Graves. He  mentioned them in a letter to W.R. Hamilton dated 16 December 1843.
Hamilton described the early history of Graves' discovery in~\cite{Hamilton:1848}.  A. Cayley  discovered them independently in 1845, in an Appendix to his paper~\cite{Cayley:1845}.  These octonions are related to a non-cyclic Hamming (7.4) error correcting code. Later  Cartan and Schouten \cite{Cartan:1926} proposed a class of octonion notations  in the context of Riemannian geometries admitting an absolute parallelism. These were studied and developed by  Coxeter  \cite{Coxeter:1946}.
},
 Hamming error correcting code, Fano plane and  superpotentials.

The order in which we will describe these relations is defined by the mathematical fact that the smallest of the exceptional Lie groups, $G_2$, {\it  is the automorphism group of the octonions}, as shown by Cartan in 1914~\cite{Cartan:1914}, and  studied in detail by Gunaydin-Gursey (GG)  in~\cite{Gunaydin:1973rs}.
Thus we start with octonions,  defined by their multiplication table.  Under $G_2$ symmetry  the multiplication table of octonions is left invariant.

  There are 480 possible notations   for the multiplication table of octonions assuming that each imaginary unit squares to $-1$, as shown by Coxeter  in 1946 in~\cite{Coxeter:1946} and detailed in \cite{Schray:1994ur}.  They can be divided into two sets of 240  notations related via the Cayley-Dickson construction. Cayley-Dickson construction represents octonions in the form
${\cal O} =q_1 + j q_2$  where $q_1$ and $q_2$ are quaternions and $j $ is an additional imaginary unit that squares to $-1$  and anticommutes with the imaginary units of quaternions  \cite{Cayley:1845,Dickson:1919}.  This  leads to  240 possible notations for the multiplication table  of octonions represented by the same oriented Fano plane as explained in Sec. \ref{sec2}.  In~\cite{Gunaydin:1973rs} $j$ was labelled as $e_7$ and the quaternion imaginary units as $e_i \, (i=1,2,3)$ and the additional imaginary units as $e_{(i+3)} = e_7 e_i$. Modulo the permutation of indices (3,4) and (5,6) of imaginary octonion units  the notations used  in~\cite{Coxeter:1946} and  in~\cite{Gunaydin:1973rs} are equivalent and belong to this set of 240 notations.
One can equally well take ${\cal O} =q_1- j q_2$ to represent octonions since $(-j)^2=-1$ , which will give another 240 possible notations for the multiplication table. In particular, in the conventions  of ~\cite{Gunaydin:1973rs}  this will lead to defining $e_{(i+3)} =- e_7 e_i$. Again one finds 240 possible conventions that can be represented on an oriented Fano plane that differs from the Fano plane of the first set of 240 conventions in that the direction of arrows along the   3 lines of the Fano plane are reversed. For example in Fig. \ref{fig:Fano1} the relevant 3 lines are the ones inside the triangle which cross $e_7$.  Cayley-Graves octonions~\cite{Hamilton:1848,Cayley:1845} belong to this set of 240 conventions.

Thus,   480 notations split naturally into two sets of 240 related by the map $j \rightarrow  -j$ which reverses the directions of the 3  arrows involving $e_7$ in the Fano plane in~\cite{Gunaydin:1973rs}.
Octonion  multiplication table in turn, leads to an oriented  Fano plane and error correcting codes. All these are known mathematical facts, see for example,~\cite{Hamilton:1848,Cayley:1845,Cartan:1914,Dickson:1919,Cartan:1926,Coxeter:1946,Gunaydin:1973rs,Dundarer:1983fe,Dundarer:1991poa,Gursey:1996mj,Schray:1994ur,Baez:2001dm,Planat,Anastasiou:2013cya}.

Here we also note that  Cartan-Schouten-Coxeter \cite{Cartan:1926,Coxeter:1946}  octonion convention  is associated with the so-called cyclic\footnote{The codewords are called cyclic if  the circular shifts of each codeword give another word that belongs to the code. The cyclic error correcting codes, the so-called BCH codes, were invented by    Hocquenghem in 1959~\cite{Hocquenghem:1959} and by  Bose and   Chaudhuri in 1960~\cite{BoseChaudhuri:1960}.} Hamming (7,4) error correcting code.
  The Cayley-Graves  octonions naturally lead to the original non-cyclic Hamming (7,4) error correcting code~\cite{Hamming:1950:EDE}.

  Octonions have made their appearance within the framework of 11d  supergravity and its compactifications  \cite{Englert:1982vs,Gunaydin:1983mi,Awada:1982pk,Englert:1983qe} soon after it was constructed \cite{Cremmer:1978km}. Furthermore  the U-duality  symmetries of maximal supergravity in five, four and three dimensions are described by the exceptional Jordan algebra over split octonions and the associated Freudenthal triple systems \cite{Ferrara:1997uz,Gunaydin:2000xr,Gunaydin:2009pk}. More recently octonions were shown to describe the non-associative algebra
of non-geometric R-flux
background in string theory and their uplifts to M-theory \cite{Gunaydin:2016axc}.

  Of particular interest for our purposes is the fact that {\it  the maximal supersymmetry of M-theory is spontaneously broken by compactification to minimal
   $\cN=1$ supersymmetry in} 4d \cite{Awada:1982pk,Englert:1983qe}.  A  spontaneously induced torsion breaks all supersymmetries but one, and  renders the compact space Ricci-flat.
The supersymmetry breaking torsion was computed explicitly in \cite{Englert:1983qe} and  it was observed that the flattening torsion components are constant and given by the structure constants of octonions.

We start  with the 7-moduli model, following \cite{DallAgata:2005zlf,Duff:2010vy,Derendinger:2014wwa}, i.e.  we look at the model of a   compactification of  M-theory on a $G_2$-structure manifold  with the following \K\, potential and superpotential
\be
K_{7{\rm mod}}= - \sum_{i=1}^7 \log\left(  T^i + \overline{T}^i\right)\, , \, \qquad  W={1\over 2}  M_{ij} T^i T^j\, , \qquad   M_{ii} =0,  \quad \forall \,  i.
\label{ourKW}\ee
Here $M_{ij}$ is a symmetric matrix with 21 independent elements defined by the twisting of the 7-tori, which in general breaks $G_2$ holonomy down to $ \mathbb{Z}_2^3$. We propose to take an octonion based superpotential of the form
\be
\mathbb{WO}= \sum_{\{ijkl \}} (T^i -T^j) (T^k -T^l),
\label{ourW} \ee
\noindent where we  take a sum over  7 different 4-qubit states defining the choice of ${\{ijkl \} }$ in $\mathbb{WO}$.   The sense in which these superpotentials are octonion based will be explained in great detail later. One important property of the superpotentials  $\mathbb{WO}$ is the fact that
the defining matrices $M_{ij}$  have only  $\pm 1$ entries and \be
\sum_j M_{ij} =0\, ,  \qquad  \forall i\, ,
\ee
so that  only 14 terms of the type $T^i T^j$ are present. As a consequence  for these superpotentials  $G_2$ holonomy of the compactification manifold is preserved and  is not broken to $G_2$ structure manifolds with $\mathbb{Z}_2^3$ holonomy. We will also find that the mass eigenvalues of the superpotential around the vacuum are independent of the convention chosen for octonion multiplication.
We will also  find in these models Minkowski vacua with 1 and 2 flat directions  and  apply them for cosmology.

We will present simple examples of eq. \rf{ourW}  based on a cyclic symmetry of octonion multiplication in  clockwise or counterclockwise directions when the imaginary units are represented on the corners of a  heptagon.
 These octonions, Fano planes and superpotentials  have a very simple relation to cyclic Hamming (7,4) code. {\it A single   example of} $\mathbb{WO}$ in eq. \rf{cw} or in the form  \rf{cwgen1} {\it is sufficient for all cosmological applications in this paper}.

 In addition to the simple cyclic choices we propose the general form of  $\mathbb{WO}$, in terms of the structure constants of octonions and generalized cyclic permutation operator, valid for any choice of the 480 octonion conventions.  This general formula is presented in eq. \rf{superpotentialG} and details of the construction with examples are given in Appendix \ref{secA}.

In Sec.~\ref{sec2} we present some basic facts about octonions, Fano planes, Hamming (7,4) codes, $G_2$ symmetry, together with some  useful references. The goal is to provide the information for  understanding how  the mathematical structure of M-theory compactified on a manifold with $G_2$ holonomy, is codified in  our cosmological models using octonions.

  In Sec.~\ref{sec3} we describe, following \cite{DallAgata:2005zlf}, a special case of models with compactification on manifolds with $G_2$ structure which can have Minkowski vacua under the special condition when the holonomy group is extended from $\mathbb{Z}_2^3$  to $G_2$.

In Sec.~\ref{sec4}
we  present a simple derivation of two superpotentials \rf{ourW}. The first one in eq.  \rf{cw} is  based on heptagons with clockwise orientation  using the  Cartan-Schouten-Coxeter notation for octonions  \cite{Cartan:1926,Coxeter:1946}. These models  are shown to be related to a cyclic Hamming (7,4) error correcting code.
 The second one  in eq.  \rf{ccw} is based on heptagons with counterclockwise orientation.  This one is  related
to Reverse Cartan-Schouten-Coxeter notation for  octonions, which we introduce there. These models are related to the original non-cyclic Hamming code. Since our superpotential is quadratic in moduli, the fermion mass matrix in supersymmetric Minkowski vacua is proportional to the second derivative of the superpotential, we study it there.

In Sec.~\ref{sec5} we discuss  octonionic superpotentials for various octonion conventions using the general formula \rf{superpotentialG}.
 We also explain the alternative derivation of new superpotentials via the change of variables, starting with superpotentials in  eq.  \rf{two}   based on heptagons with clockwise   or  counterclockwise orientation. In Appendices \ref{secA} and \ref{secB} we give
  examples  of
relations between most commonly used octonion notations using Fano planes. These include   Cayley-Graves \cite{Hamilton:1848,Cayley:1845},  Cartan-Schouten-Coxeter  \cite{Cartan:1926,Coxeter:1946},  Gunaydin-Gursey  ~\cite{Gunaydin:1973rs},  Okubo notation \cite{Okubo:1990nv} for octonions  and  the ones we have introduced here in Sec.~\ref{sec4}   and called Reverse  Cartan-Schouten-Coxeter notation for octonions.

 For each choice of octonion multiplication convention we have found 2 independent choices of  superpotentials satisfying our physical requirements.  For other octonion conventions we can get the relevant 2 superpotentials either using the general formula, or making the field redefinitions in the superpotentials the same as the ones which lead to a change of octonion conventions  without the sign flip. This limitation is due to the fact that all 7 moduli with \K\, potential in \rf{ourKW} have a positive real part, therefore we do not flip signs of moduli. Meanwhile the general type of mapping from one convention to another  do involve sign flips in general. Starting with Cartan-Schouten-Coxeter conventions we can get models in 240 different conventions, including the one we started with, without sign flip of moduli.  Similarly starting with Reverse Cartan-Schouten-Coxeter convention we can get models with another 240 conventions, including the one we started with, without sign flip of moduli.  It is convenient to use these two starting points in the form given in eq. \rf{two}. For simplicity  we shall refer to these superpotentials as `octonionic superpotentials'.

In Sec.~\ref{sec6} we  study Minkowski vacua in 7-moduli models with octonionic superpotentials \rf{ourW}.
We show   that these  models  have a Minkowski
 minimum at
\be\label{sol}
T^1=T^2=T^3=T^4=T^5=T^6=T^7 \equiv T
\ee
with one flat direction.  All models studied in Sec. \ref{sec4} and in Appendix \ref{secA} are the same cosmologically:  they  have
 a Minkowski minimum with one flat direction as in eq. \rf{sol}.
  The eigenvalues of the superpotential matrix $M_{ij}$ in eq. \rf{ourKW} as well as the eigenvalues of the
mass matrix  at the vacuum in these models have an $[SO(2)]^3$ symmetry. The resulting effective 1-modulus model in  4d supergravity   has the \K\ potential and superpotential
\be
K = - 7 \log\left(  T + \overline{T}\right)\, , \, \qquad  \mathbb{WO}=0.
\label{start}\ee
This is a starting point for building the inflationary cosmological $\alpha$-attractor models~\cite{Kallosh:2013yoa,Kallosh:2015lwa} leading to a  top benchmark for detecting B-modes~\cite{Ferrara:2016fwe,Kallosh:2017ced,Kallosh:2017wnt}. In the past, it was derived in~\cite{Ferrara:2016fwe} by postulating eq.~\rf{sol} and in~\cite{Kallosh:2017ced} by using a phenomenological superpotential $W=\sum_{1\leq i\leq j \leq 7} (T_i-T_j)^2$. Note that such a superpotential does not fit the   $G_2$-structure  compactification M-theory model where $M_{ii}=0$ for each $i$ in~\rf{ourKW}.

The choice in~\rf{ourW}, associated with octonions and the  (7,4) Hamming code, is fundamental, being associated with {\it maximal supersymmetry of M-theory in 11d, spontaneously broken by compactification to minimal $\cN=1$ supersymmetry in} 4d \cite{Awada:1982pk}, \cite{Englert:1983qe}.
It naturally leads to a desirable starting point~\rf{ourKW}, \rf{sol}, \rf{start}  for building a $3\alpha=7$ cosmological model from M-theory compactified on a $G_2$ holonomy manifold.

In Sec. \ref{sec6} we also present the octonion based superpotentials for the models with Minkowski vacua with 2 flat directions. In these vacua some of the $m$  moduli are equal to each other, $T^1=\dots =T^m\equiv  T_{(1)}$, whereas some other $n$ moduli are equal to each other, $T^{m+1}=\dots =T^7\equiv T_{(2)} $ and  $m+n=7$. The superpotentials in these  models are obtained from the 1-flat-modulus models by removing certain terms in \rf{ourW} corresponding to removing some specific codewords from the cyclic Hamming (7,4) code.
In this way we find  2-moduli  effective 4d supergravity with the following \K\, potential
\be
K= - m \log\left(  T_{(1)} + \overline{T}_{(1)}\right) -   n\log\left(  T_{(2)} + \overline{T}_{(2)}\right) \, , \, \qquad  \mathbb{WO}=0.
\label{mn}\ee
We find models with $6+1$, $5+2$, $4+3$ split. The remaining codewords define the remaining terms in the superpotential for these  split models.

On the basis of the M-theory setup, we  construct the updated cosmological models in 4d $\mathcal N=1$ supergravity in Sec.~\ref{sec7}. These models of $\alpha$-attractors  realize inflation  and lead to 7 benchmark points, which are the B-mode detection targets suggested earlier in \cite{Ferrara:2016fwe,Kallosh:2017ced,Kallosh:2017wnt}. Here we show that these updated models originate from M-theory and octonions and error correcting codes.
 It was shown by Planck satellite measurements  \cite{Planck:2018jri} that $\alpha$-attractor models are in good agreement with data available now. These include special cases with the discrete set of values for $3\alpha = 7,6,5,4,3,2,1$ motivated by maximal supersymmetry. Here we updated these models with account of  their relations to M-theory via octonions. Our 1-flat direction models lead to $3\alpha = 7$ case, a top benchmark point, whereas our 2-flat direction models split models lead to remaining cases.

Future cosmological observations like
 BICEP2/Keck \cite{Hui:2018cvg,Ade:2018gkx}, CMB-S4 \cite{Abazajian:2016yjj,Shandera:2019ufi}, SO \cite{Ade:2018sbj},
  \href{https://ui.adsabs.harvard.edu/abs/2019BAAS...51g.286L/abstract}{LiteBIRD }   \cite{Hazumi:2019lys} and PICO \cite{Hanany:2019lle},   
  might potentially  detect the tensor to scalar ratio at a level $r= 5\times10^{-4} (5\sigma) $ and improve constraints on the value of $n_s$, the spectral tilt of the CMB power spectrum. 
  They
  might support or invalidate the M-theory cosmological models and their 7 benchmark points.
  We show these benchmark points in Figs. \ref{7disk2}, \ref{fig:LB} and explain their relation to octonions and to cosmological observables in Sec.   \ref{sec8}.  A short summary of the main results of the paper is given in Sec.   \ref{sec9}.

In Appendix \ref{secA} we present the general formula for octonion superpotentials in \rf{ourW} for any octonion conventions, with examples. In Appendix \ref{secB} we derive the relations between most commonly used octonion conventions. In Appendix \ref{secC} we
give  more details about multiplication tables and Fano planes.
 In Appendix \ref{secD} we present a specific  transformation from  Cayley-Graves  to Cartan-Schouten-Coxeter octonion notations which also rotates a cyclic  Hamming (7,4) code to a non-cyclic one. In Appendix \ref{secE} we show that octonion superpotentials may be used for generating metastable de Sitter vacua in 4d.

\section{Basics on  octonions, Fano planes,  error correcting codes and $G_2$}\label{sec2}
{\it Octonions}

\noindent There are  four normed division algebras: the real numbers ($\mathbb{R}$), complex numbers ($\mathbb{C}$), quaternions ($\mathbb{H}$), and octonions ($\mathbb{O}$).
\begin{figure}[H]
\centering
\includegraphics[scale=0.8]{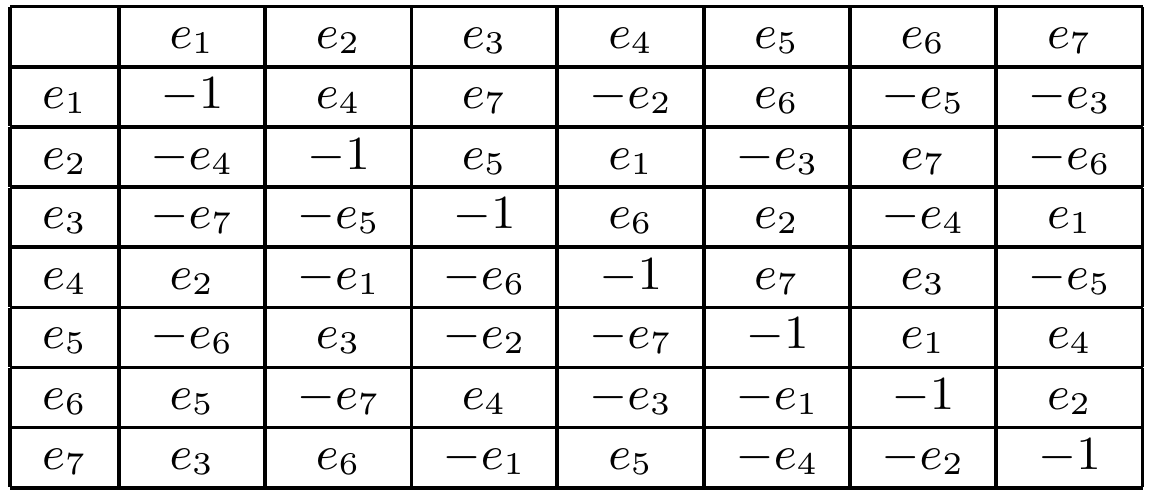}
\caption{\footnotesize  Cartan-Schouten-Coxeter Octonion Multiplication Table }
\label{Table:Mult}
\end{figure}

 \noindent   We will be working with real octonions which are an 8-dimensional division  algebra   spanned by seven imaginary units together with identity  $\{1, e_1, e_2, e_3, e_4, e_5, e_6, e_7\}$
\be
e_i e_j = -\delta_{ij} 1 + f_{ijk} e_k.
\ee
 The structure constants  $f_{ijk} $ are completely antisymmetric. The multiplication  Table in Cartan-Schouten-Coxeter  notation \cite{Cartan:1926,Coxeter:1946}   is given in Fig. \ref{Table:Mult}

\noindent  The non-vanishing structure constants in CSC convention are
\be
 f_{ijk} =+1 \qquad {\rm for}  \qquad \{ijk\} = \{ (124), (235), (346), (457), (561), (672), (713)   \}.
\label{cycl} \ee
 In \cite{Coxeter:1946} the set of 7 triples $ \{ijk\}$  in eq. \rf{cycl} is referred to as 7 associative triads\footnote{The  28 remaining  triads are anti-associative.}.
 The associative triads $ \{ijk\}$ are given by the seven  quaternion subalgebras of octonions  generated by $e_i, e_j, e_k$.
There are 480 possible conventions for  octonion multiplication tables \cite{Coxeter:1946}, i. e. 480 different choices  of the 7 associative triads\footnote{See for example
http://tamivox.org/eugene/octonion480/index.html.}.

 Imaginary octonion units  are anti-commuting. The commutator of two imaginary units is simply
\be
[e_i, e_j] \equiv e_i e_j - e_j e_i = 2 f_{ijk} e_k.
\ee
 Octonions are  not associative and
 the associator  defined as
\be
[e_i, e_j, e_k] \equiv (e_i e_j ) e_k - e_i( e_j e_k)
\ee
does not vanish, in general. The associator is an alternating function of its arguments and hence octonions form an alternative algebra.
The Jacobian of three imaginary units does not vanish and is proportional to their  associator
\be
J(e_m,e_n,e_p) \equiv [e_m,[e_n,e_p]] +[e_p,[e_m,e_n]] +[e_n,[e_p,e_m]] = 3 C_{mnpq} e_q
\label{Cmnpq}\ee
where the completely antisymmetric tensor  $C_{mnpq} $ is given by the structure constants
\be
C_{mnpq} = f_{k[mn} f_{p]kq } \label{via_f}
\ee
and is dual to the structure constants
\be
C_{mnpq} = \frac{1}{6} \epsilon_{mnpqrst} f_{rst}.
\label{eps}\ee
Both the  structure constants and the tensor $C_{mnpq}$ are invariant tensors of the automorphism group  $G_2$ of octonions.

\

 {\it Fano plane and octonions}
 
 \

 \noindent  The Fano plane is  the unique projective plane over a  field of characteristic two which can be embedded projectively in 3 dimensions $\mathbb{Z}_2^3$ \cite{Baez:2001dm} corresponding to an Abelian group of order 8 with the seven points represented by the nontrivial  elements.    It can be used  as  a mnemonic representation of the multiplication table of octonions by identifying its points with the imaginary octonion units and introducing an orientation.

\begin{figure}[H]
\centering
\includegraphics[scale=0.6]{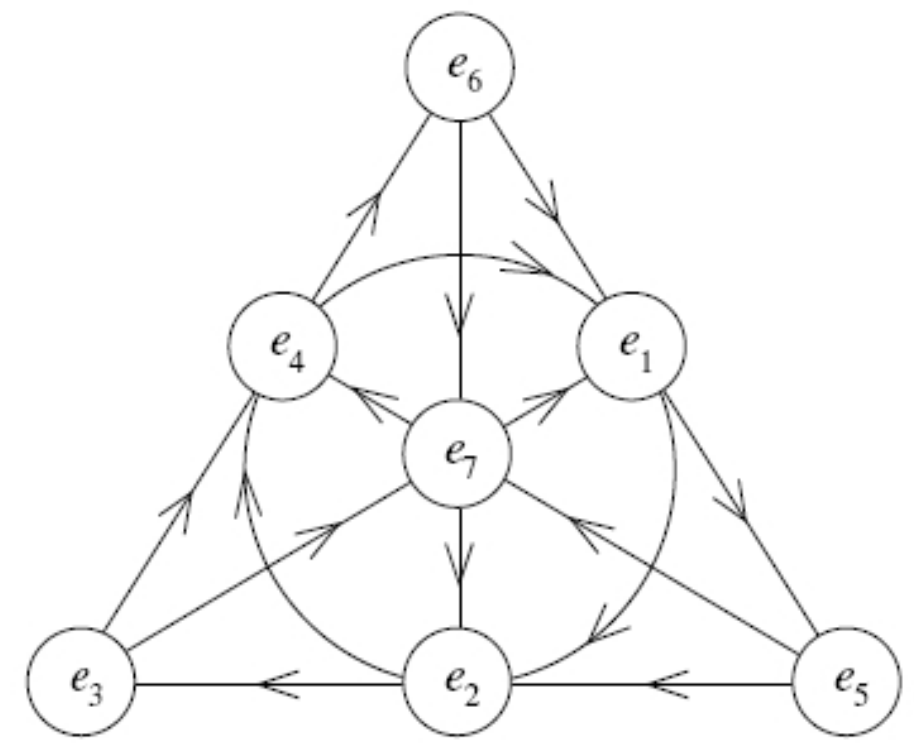}
\caption{\footnotesize  An oriented Fano plane,   Fig. 1 in \cite{Baez:2001dm} for  Cartan-Schouten-Coxeter \cite{Cartan:1926,Coxeter:1946}  octonion conventions. On each of the 7 lines (including the circle) there are 3 points e.g. 1,2, and 4 on a circle. The octonion multiplication rule \rf{cycl} is represented in  the oriented  Fano plane with the arrows indicating the positive directions for multiplication. For example, one can see from the oriented circle that $e_1\cdot e_2= e_4$, which
is also shown in the first term in \rf{cycl} as $f_{124}= +1$.}
\label{fig:Fano1}
\end{figure}

\noindent  Given a multiplication table of octonions, the corresponding Fano plane can be given different orientations depending on the identification of its points with the imaginary units. The three points in each line are identified with the imaginary units $( e_i, e_j , e_k ) $ of a quaternion subalgebra with the positive orientation given  by the cyclic permutation of ($i,j,k $).   There are different ways to build the Fano plane for the same type of triads.  For  example, the original  oriented  Fano plane in  \cite{Coxeter:1946}  for the same set of triples is different from the one in  \cite{Baez:2001dm} shown in Fig.  \ref{fig:Fano1}.

  In the Fano plane every line has 3 points and every point is the intersection point of 3 lines.  Since each line contains 3  points which correspond to the imaginary units of a quaternion subalgebra every imaginary unit belongs to three different quaternion subalgebras. Hence  given an imaginary unit $e_k$ there exist three sets of imaginary units  $(e_k,e_m,e_n) , (e_k, e_p, e_q) , (e_k, e_r, e_s)$ of quaternion subalgebras such that
\be
f_{kmn}=f_{kpq}=f_{krs} =+1
\ee
For example in Fig. \ref{fig:Fano1} given the unit $e_1$ we have $f_{124}=f_{156}=f_{137}=1 $.

\

{\it Error correcting codes, Fano planes, octonions}

\

  \noindent Error correction is a central concept in classical  information theory. When combined with quantum mechanics they lead to quantum error correction especially important in quantum computers. For our purpose only a classical Hamming [7,4,3] code  \cite{Hamming:1950:EDE} is relevant.  It is an example of a linear binary vector space specified by its generator matrix $G$  which allows to produce the 16 codewords. The matrix $H$ is known as a parity matrix, it has the property that $HG^T=0$
\begin{figure}[H]
\centering
\includegraphics[scale=0.65]{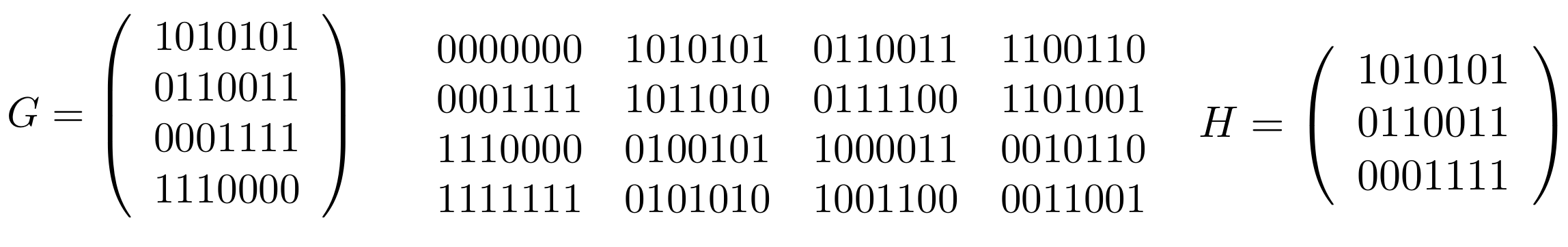}
\caption{\footnotesize  Original Hamming code. The generator matrix $G$ at the left part is used here to produce the 16 codewords of the Hamming code shown in the middle of the figure. They are produced  with account of addition rules $0+0=0, 0+1=1,1+0=1, 1+1=0 $. 1st in the 16 codewords is the zero vector, then the 1st row in $G$. Next we add the 2nd row in $G$ to the first two vectors, which give us the 3rd and the 4th codewords. Next we add the 3rd row in $G$ to the 4 vectors previously obtained, and so on. At the right there is a parity matrix $H$.
}
\label{fig:HammG_16_H}
\end{figure}
In Fig. \ref{fig:HammG_16_H} we have shown in addition to generator matrix $G$ and a parity matrix $H$ all 16 codewords of the Hamming  [7,4,3] code.
The mechanism of error detection and correction using this code is nicely explained in the lecture by Jack Keil Wolf  \href{http://acsweb.ucsd.edu/~afazelic/ece154c/ErrorCorrection-JackWolf.pdf}  {An Introduction to Error Correcting Codes}.

 For our purpose it is useful to observe that in 16 codewords of the Hamming code shown in the middle of the Fig. \ref{fig:HammG_16_H} one of the  codewords is all 0's, one is all 1's. The remaining 14 codewords are split into two groups of 7: one group has   3 0's and 4 1's, the other has  3 1's and   four 0's. They are complimentary to each other when 0 is replaced by 1.
  For example, in the context of the Graves-Cayley octonions one can use the following 7 codewords

 \be
1110000, \, 1001100, \, 0101010,  0010110, \, 0100101, \, 0011001, \, 1000011.
\label{NC} \ee

 All these set of codewords which we discussed so far are known as non-cycling Hamming codes. Namely, one can see in $G$ in Fig. \ref{fig:HammG_16_H} that the 4 codewords are not obtained by recycling any of them.  Same feature is present in all 16 codewords in Fig. \ref{fig:HammG_16_H}.

 Even though in the literature one set of Hamming  code is referred to as cyclic, one can make all of them cyclic with respect to  the action of a cyclic permutation  operator $P$ to be defined later in section 5.1 that enters in equation \ref{superpotentialG}.  For the Cayley-Graves octonions ths permutation operator is $P_{CG} =(1245736)$. Under its action the codewords above get mapped into each other in a unique way:
\bea
P_{CG} (1110000)=(0101010) \cr
P_{CG}(0101010)=(1001100) \cr
P_{CG}(1001100)=(0100101) \cr
P_{CG}(0100101) = (0011001) \cr
P_{CG} (0011001) = (0010110) \cr
P_{CG}(0010110)= (1000011) \cr
P_{CG}(1000011)= (1110000)
\eea

\begin{figure}[H]
\centering
\includegraphics[scale=0.65]{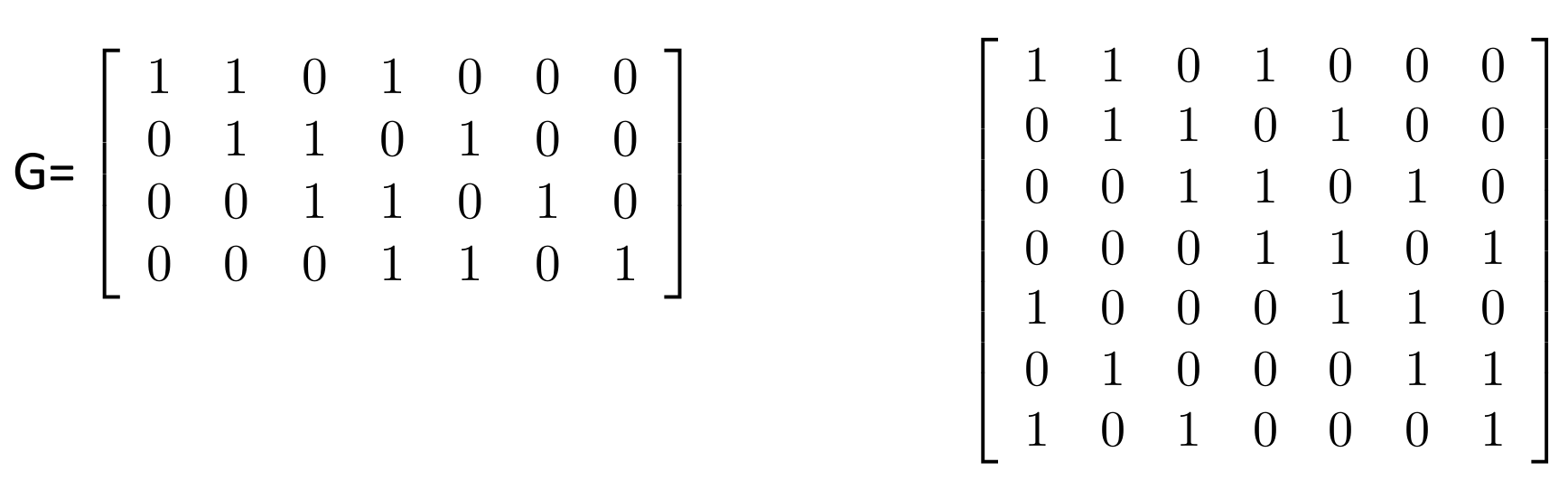}
\caption{\footnotesize  Cyclic  Hamming code. The generating matrix $G$ at the left part has 4 rows with 3 1's each,  where every row is a cyclic permutation of the previous one.  More cyclic permutations generate directly the 7 codewords at the right of the figure, all with 3 1's, 4 0's.
}
\label{fig:Planat}
\end{figure}

Now we describe shortly a cyclic  Hamming (7,4) code following \cite{Planat}, where cyclic codes were discussed in the context of the projective spaces. The generating matrix is shown in Fig. \ref{fig:Planat} at the left.

All codewords can be obtained from any particular one by a cyclic permutation.  In this case the cyclic permutation operator is simply $P_{CSC}=(1234567)$.   We should note that  this $7\times7$ matrix is  also  the incidence matrix of the underlying Fano plane. It is easily related to octonions in  CSC convention \cite{Cartan:1914,Coxeter:1946}
shown in eq. \rf{cycl}.
Each triad is shown in the 7 codewords in Fig. \ref{fig:Planat}: the position of the 1's in the first row is 124, in the 2nd row is 235 and so on, where 1 after 6 is 1 mod 7, and so on.

 One finds the mechanism of error detection and correction using the cyclic (7,4) Hamming error correcting code,  for example, in
 the book by R. Blahut `Algebraic codes for data transmission' \cite{Blahut:2012} to which we refer for details and further references.

The $7\times7$ matrix at the right of the Fig. \ref{fig:Planat} is the set of 7 codewords  which will be often used in our construction of cosmological models. Sometimes we will  change  the order of rows, so that we can easier explain our new octonionic superpotentials, leading to Minkowski vacua with one flat direction.
Sometimes we will remove some rows in this $7\times7$ matrix, to build  different superpotentials, leading to Minkowski vacua with two flat directions.

 In Appendix  \ref{secD} we show that one can bring the cyclic Hamming (7,4)  error correcting code to the original Hamming code by a particular permutation. 

\

{\it $G_2(\mathbb{R})$, its finite subgroups and 480 octonion conventions}

\

\noindent The automorphism group $G_2$ of  the division algebra of octonions is a 14-dimensional subgroup of $SO(7)$ \cite{Cartan:1914,Gunaydin:1973rs,deWit:1983gs,Gunaydin:1995ku,Gunaydin:1995as}. Under $G_2$ the adjoint representation of $SO(7)$ decomposes as
${\bf 21} = {\bf 14} + {\bf 7}$. The 14 generators of $G_2$,  which we call $G^{ij}$,  can be represented using the 21 generators of $SO(7)$, denoted here as $J^{ij}$, and a totally antisymmetric tensor $C_{ijkl}$, related to octonion associator which is defined in eqs.
\rf{Cmnpq}, \rf{via_f}, \rf{eps}
\be
G^{ij}= {1\over 2} J^{ij} +{1\over 8} C^{ij}{}_{kl} J^{kl}.
\ee
The tensors $C_{ijkl}$ and $f_{ijk}$ are subject to  various identities, so that
\be
f_{ijk} G^{jk} =0.
\ee
This means that there are 7 constraints on 21 generators of $SO(7)$, which makes the remaining 14 the generators of $G_2$. They also satisfy the identity
\be
G_{ij} = {1\over 2} C_{ijkl}  G^{kl}.
\ee
Under the action of   $G_2$ the imaginary  octonion units transform in its 7 dimensional representation and  the structure constants of  octonions form an invariant  tensor of $G_2$.   The  group  $G_2$ has some important finite subgroups, in particular,  $G_2(2)$ of order 12096 and
$[{\cal Z}_2]^3 \cdot  {\cal{PSL}}_2 (7)$, of order 1344. We refer the reader to relatively recent papers on the finite subgroups of $G_2$ \cite{Karsch:1989gj,He:2002fp,Koca:2005fn,Luhn:2007yr,Evans:2014qra,Koca:2016ybu,Ramond:2020dgm} which include more details on the discrete finite subgroups of $G_2(\mathbb{R})$. In particular in \cite{Evans:2014qra} in Table 1 one can find the list of all the important  finite subgroups  $\Gamma \subset G_2$.

To explain why there are 480 possible notations   for the multiplication table of octonions, as shown   in~\cite{Coxeter:1946} we can look at the total number of permutations of imaginary octonion units, including the flipping of signs. This will give
\be
{\cal T}=[{\cal Z}_2]^7 \cdot {\cal S} (7).
\ee
 At first sight  this might suggest   $2^7 \cdot  7! = 645120 $ possible choices of conventions. However, some of these choices do not change the multiplication table.  It is due to the fact that the automorphisms of the oriented Fano plane, preserving triads, form a discrete finite subgroup of
$G_2$ which is
\be
{\cal H}=[{\cal Z}_2]^3 \cdot  {\cal{PSL}}_2 (7).
\ee
It is
an irreducible imprimitive group of order 1344. Thus,   there is
a redundancy of order
$ 1344$.  Hence the total number of inequivalent  multiplication tables comes as
\be
{{\cal T}\over \cal H} = {[{\cal Z}_2]^7 \cdot {\cal S} (7)\over [{\cal Z}_2]^3 \cdot  {\cal{PSL}}_2 (7)
} \qquad \Rightarrow \qquad
{645120 \over 1344}= 480.
\ee
  The multiplication table of octonions can be represented by an oriented Fano plane or via the heptagon rules. There are $7!=5040 $ ways of assigning labels from 1 to 7 to the  points in the Fano plane or the corners of a heptagon corresponding to the symmetric group $ {\cal S} (7)$.
Again this might naively suggest that there are   $ 7! = 5040 $ possible choices of conventions that do not involve sign changes. However, some of these choices do not change the multiplication table while preserving the associative triads. The symmetry  group of the unoriented Fano plane is the finite group $ {\cal{PSL}}_2 (7)$ of order 168 which take collinear points into collinear points. Hence there are $5040/168=30$ inequivalent labelling of the unoriented Fano plane. When one represents the octonion multiplication by an oriented Fano plane one can associate the points on the Fano plane with the imaginary octonion units $\pm e_i$. Out of $2^7$ possible sign assignment $ 2^3$  assigments corresponding to the group ${\cal Z}_2^3$ of order 8 do not change the multiplication table\footnote{ This group is generated by conjugation with respect to three imaginary units intoduced in Cayley-Dickson process in going from real numbers to octonions.}. Hence we have $2^4$ possible inequivalent sign assignments leading to $ 30 \times 16=480$ possible inequivalent  conventions for octonion multiplication. These 480 possible conventions split naturally into two sets of 240 related by octonion conjugation which changes the sign of all imaginary units and reverses the direction of all associative triads. Given a convention the octonion multiplication table is left invariant under the finite subgroup $[{\cal Z}_2]^3 \cdot  {\cal{PSL}}_2 (7)$ of $G_2$. However only the action of the Frobenius subgroup
${\cal Z}_7\rtimes {\cal Z}_3 $ of order 21
\be
{\cal Z}_7\rtimes {\cal Z}_3 \subset [ {\cal Z}_2]^3 \cdot  {\cal{PSL}}_2 (7)
\ee
does not involve  sign changes of the imaginary units. This absence of sign flips 	is important since in our models a convention is chosen so  that ${\rm Re} \,T^i>0$ and the transformations of octonions with the sign flips is not allowed for the moduli change of variables.

Given an octonion multiplication convention one can write down a superpotential by associating the moduli with the imaginary units  using the general formula \ref{superpotentialG} which is invariant under the cyclic group ${\cal Z}_7$.  By the action of the cyclic group ${\cal Z}_3$ one can then generate three superpotentials.  The corresponding superpotentials in the "conjugate" convention differ by an overall sign and hence  are physically equivalent.  The  superpotentials in CSC convention are given in \rf{two} and for the GG convention they are given in Appendix A.2.

{\it Cayley-Graves multiplication table and octonions in quantum computation}

\begin{figure}[H]
\centering
\includegraphics[scale=0.8]{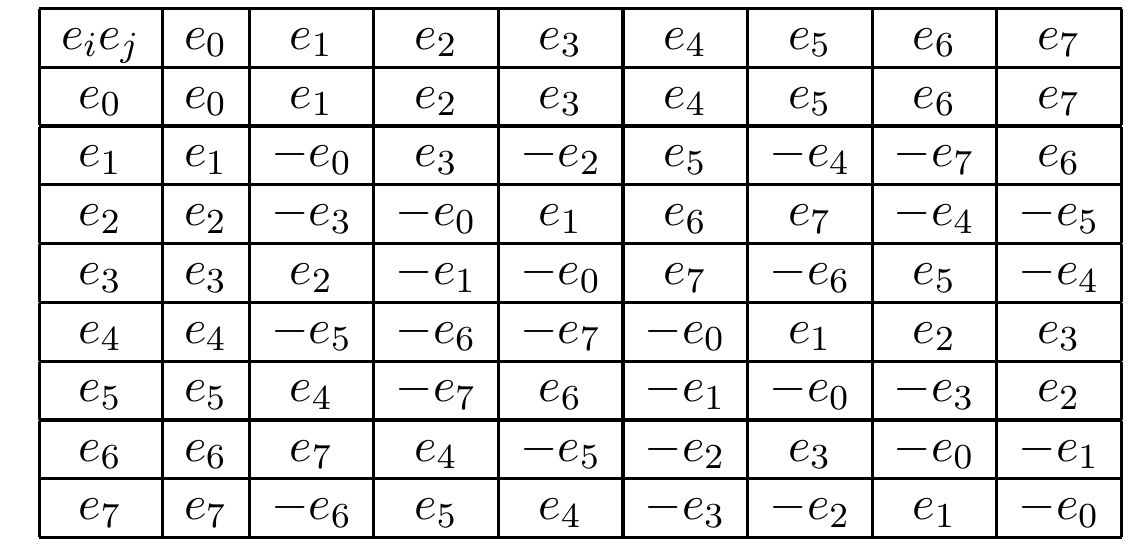}
\caption{\footnotesize  Cayley-Graves Octonion Multiplication Table. It was recently used  in  \cite{Freedman2019} as a starting point for  the proposal on quantum computing with octonions.}
\label{Table:StationQ}
\end{figure}

\noindent To finalize the basics on octonions part of the paper
we would like to add here that there is some interest in using octonions for quantum computation~\cite{Freedman2019}.  The starting point there is Cayley-Graves octonion multiplication table.
The original set of triples defining octonion multiplication table by Graves shown in~\cite{Hamilton:1848} is, using $ijklmno$ order
\be
ijk, \, ilm, \, ion, \, jln, \, jmo, \, klo, \, knm.
\label{Gra}\ee
Translating this into numbers we find
\be
123, \, 145, \, 176, \, 246, \, 257, \, 347, \, 365.
\label{Gra1}\ee

The set of triples by Cayley in
\cite{Cayley:1845} is
\be
123, \, 145, \, 624, \, 653, \, 725, \, 734, \, 176.
\label{Cay}\ee
These two choices agree if we allow a cyclic permutation inside a triple, and change the order of triples.

 In \cite{Freedman2019} the table in Fig. \ref{Table:StationQ}
 was changed   by multiplying by $-1$ all columns except the first one. This in turn allows the authors to work with a specific realization of the Lie multiplication algebra $SO(8)$ of  octonions, which is useful in quantum computing \footnote{ The seminar at Stanford by M. Freedman presenting this work was  stimulating for our interest in the connections between quantum computing and octonions.}.

\section{Twisted 7-tori with  $G_2$ holonomy}\label{sec3}
 We are interested in $\cN=1$ 4d supergravity with chiral multiplets which can be derived  from 11d supergravity \footnote{A derivation of $\cN=1$ 4d supergravity  from 11d supergravity on manifolds with $G_2$ structure, including the mechanism of spontaneous breaking of maximal supersymmetry to the minimal one, was recently performed by A. Van Proeyen, work in progress.}. For this purpose we present a basic information about the twisted 7-tori and the  compact $G_2$ manifolds, which is important for our work. We follow closely the presentation in  \cite{House:2004pm} and   \cite{DallAgata:2005zlf,Derendinger:2014wwa} where the distinction between 7-manifolds with $G_2$ holonomy and $G_2$ structure is explained and many useful references are given.  One starts with  Joyce's
$\mathbb{T}^7$ orbifolds  with $G_2$ holonomy. In general when fluxes in 11d supergravity  and geometrical fluxes describing the twisting of $\mathbb{T}^7$ are added, the deformed backgrounds  no longer have $G_2$ holonomy but rather $G_2$ structure. This still means that $\cN=1$ 4d supersymmetry is preserved when compactification from 11d takes place.
Our vacua  will not require  fluxes in 11d supergravity, however,
the geometric data of the compactified manifold, the twisting of the tori,  will be important.

A manifold with $G_2$ structure is a  7-dimensional manifold which admits a globally defined, nowhere-vanishing spinor $\eta$.  This spinor is covariantly constant with respect to a torsionful connection.
\be
\nabla_m \eta - {1\over 4} \kappa_{mnp} \gamma^{np} \eta=0.
\ee
Here $\nabla_m$ involves a Levi-Civita connection, and the tensor $\kappa_{mnp}$ is the contorsion tensor.
It can be viewed as a  normalized Majorana spinor such that $\bar \eta \eta=1$.

Using this spinor one can construct a globally defined and nowhere vanishing totally antisymmetric tensor
\be
\phi_{mnp} = i  \eta^T \gamma_{mnp}\eta
\label{3form}\ee
where $\gamma_{mnp}$ denotes the antisymmetric product of three gamma matrices with unit norm.
 7 of them are $G_2$ singlets, with different choices of $mnp$ for $m=1,\dots , 7$. They correspond to associate triads for octonions. The remaining 28 correspond to anti-associative triads. This allows to introduce a complexified $G_2$  3-form $T_i\phi^i$ where the 7 complex moduli are contracted with the 7 3-forms in eq. \rf{3form}  labelled by the index $i$.

The 4d superpotential was computed in \cite{House:2004pm} starting with 11d gravitino kinetic term $\Bar \Psi_M \Gamma^{MNP} D_N \Psi_p$. The resulting 4d gravitino mass, which defines the superpotential was given in \cite{DallAgata:2005zlf} in the following form
\be
W= {1\over 2} M_{ij} T^i T^j =-{1\over 8} \int _{X^7} \phi^i \wedge d\phi^j  \, T_i T_j.
\ee
The standard manifolds with $G_2$ holonomy correspond to untwisted tori where $d\phi=0$ and the superpotential is absent. The twist of  the toroidal orbifolds   away from $G_2$-holonomy, describes the  7-manifold with $G_2$-structure and a non-vanishing superpotential, in general with typical AdS vacua.

A special situation takes place when such  $G_2$-structure manifolds have Minkowski vacua, as shown in \cite{DallAgata:2005zlf}. One can introduce the dual 4-forms $\tilde \phi^i$ satisfying
\be
\int _{X^7} \phi^i \wedge \tilde \phi^j = \delta^{ij}.
\ee
One finds in such a case that
\be
d\phi^i = -4M_{ij} \tilde \phi^j
\ee
which leads to an existence of the closed 4-form
$
d  \tilde \phi^i=0
$
and suggest that the manifold has a $G_2$ holonomy unbroken.
For $\Phi=t_i\phi^i$ and for its dual $*\Phi= {V\over t^i} \tilde \phi^i$ at the Minkowski vacum $d\Phi= -4t^i M_{ij} \tilde \phi^j=0$ and $d *\Phi= d[\Big ({V\over t^i} \Big ) \tilde \phi^i] =0$.

The upshot here is that starting with general type $G_2$-structure manifolds one finds Minkowski vacua only in cases that some twisted 7-tori are, in fact,  $G_2$-holonomy manifolds.

The vacua with one flat direction which we will find  have the property that $T^1=T^2=T^3=T^4=T^5=T^6=T^7 \equiv T$ and therefore
\be
 W_{i}= \sum_j M_{ij} T^j|_{T^k=T} = T \sum_j M_{ij}  =0.
\ee
These are exactly the octonion superpotentials we will describe below.
\section{Octonions,  Hamming (7,4) error correcting code, superpotential}\label{sec4}

 \subsection{Cartan-Schouten-Coxeter conventions, clockwise heptagon}\label{sec4.1}
\begin{figure}[H]
\centering
\includegraphics[scale=0.75]{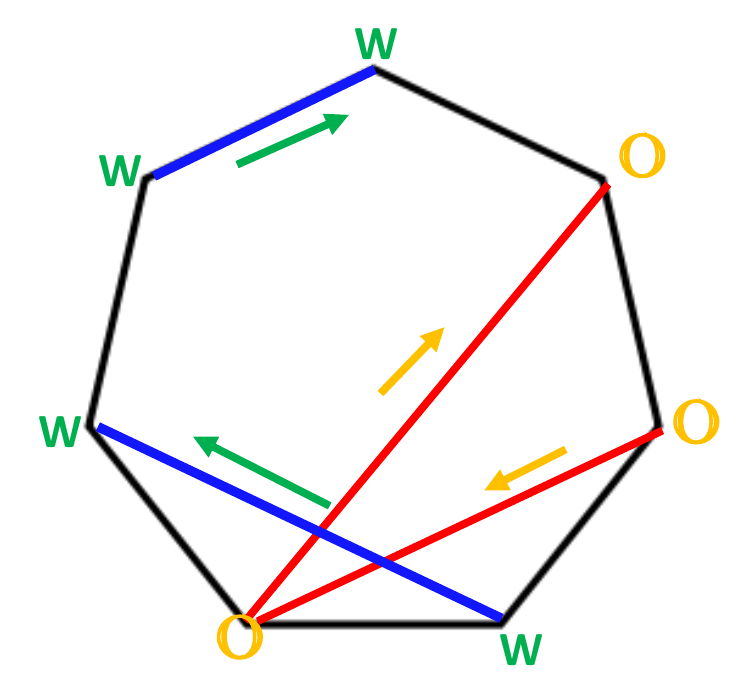}
\caption{\footnotesize   The {\bf clockwise oriented heptagon} for octonion multiplication in Cartan-Schouten-Coxeter  convention \rf{cycl2}.  We start at  the second right octonion $\mathbb{O}$  and call it $e_{r+1}$, we get the 7 associative triads in the form as  $(r+1, r+3, r+7)$. The complimentary quadruples are $(r+2, r+4, r+5, r+6)$. They are shown at the corners with the  oriented clockwise  blue lines. These give $\mathbb{WO}_{\rm cw}$ on the clockwise oriented heptagon.
}
\label{fig:CW_repr}
\end{figure}

  Convention of  octonion multiplication by Cartan-Schouten-Coxeter are
 \be
 e_{r+2} e_{r+4} =  e_{r+3} e_{r+7}= e_{r+5} e_{r+6}=e_{r+1}\, ,  \qquad  \quad  e_r=e_{r+7}.
\label{Cartan} \ee
We make a choice  of associative triads $(ijk)$
   \be
 f_{ijk}^{\rm cw} =+1:  (r+1, r+3, r+7)\, :  \{  (137), (241), (352), (463), (574), (615), (726) \},
\label{cycl2} \ee
\be
i<j<k.
\ee
To build the superpotentials we need the quadruples $(mnpq)$ which are complimentary to associative triads. We take the case
\bea
r+2, r+4, r+5, r+6 : \,
(2456), (3567), (4671), (5712), (6123), (7234), (1345),
\eea
 \be
m<n<p<q.\ee
This leads to the following  superpotential
\be
  \mathbb{WO}_{\rm cw}= \sum_{r=0}^{6}   (T^{r+2}-T^{r+4})(T^{r+5}-T^{r+6}).
\label{cwg} \ee
 This is shown in Fig. ~\ref{fig:CW_repr}. \vskip -0.8 cm
 \noindent Explicitly we have for the  superpotential in the clockwise heptagon picture
    \bea\label{cw}
&&   \mathbb{WO}_{\rm cw} = (T^2 - T^4) (T^5 -T^6)+  (T^3 - T^5) (T^6 - T^7) + (T^4 - T^6) (T^7 -
      T^1)\\
 \cr
      && + (T^5 - T^7) (T^1 - T^2) + (T^6 - T^1) (T^2 - T^3) + (T^7 -
      T^2) (T^3 -T^4) + (T^1 - T^3) (T^4 - T^5).  \nonumber
   \eea
   This superpotential  is easy to compare with the cyclic Hamming error correcting code, a $7\times7$ matrix at the right of the  Fig.~\ref{fig:Planat}.  For our purpose that the codewords represent octonions in eq. \rf{cycl2} we have to move the last row of the $7\times7$ matrix at the right of the  Fig.~\ref{fig:Planat} into the first row.
We have to read the codewords from the top to the bottom and from the left to the right. This will be opposite in the case with counterclockwise heptagon.

 All 3 1's in the codewords in  eq. \rf{Hamming_CW} are in full agreement with the set of triads in eq. \rf{cycl2}. They are also shown in eq. \rf{Hamming_CW} to the left of the codewords.
The 4 zero's in the 7  codewords in  eq. \rf{Hamming_CW},  shown to the right of the codewords, are in perfect agreement with the 7 complimentary  quadruples defining  the superpotential.

 \be
{\rm \bf Triads }\hskip 1.2 cm {\rm \bf Codewords \hskip 0.7 cm     Quadruples  \hskip 0.5 cm \Rightarrow \hskip 0.3 cm \mathbb{WO}_{\rm cw}}  \nonumber \ee
\be
\mathbb{WO}_{\rm cw} = \left(\begin{array}{c|ccccccc|c|c}(137)& \hskip 1 cm1 & 0 & 1 & 0 & 0 & 0 & 1& \hskip 1 cm (2456) & \hskip 1 cm (T^2-T^4)(T^5-T^6) \\(241)&\hskip 1 cm1 & 1 & 0 & 1 & 0 & 0 & 0 & \hskip 1 cm (3567)& \hskip 1 cm(T^3-T^5)(T^6-T^7) \\(352)&\hskip 1 cm 0 & 1 & 1 & 0 & 1 & 0 & 0& \hskip 1 cm(4671)&\hskip 1 cm (T^4-T^6)(T^7-T^1)\\(463)& \hskip 1 cm0 & 0 & 1 & 1 & 0 & 1 & 0 & \hskip 1 cm (5712)& \hskip 1 cm(T^5-T^7)(T^1-T^2)\\(574)& \hskip 1 cm 0 & 0 & 0 & 1 & 1 & 0 & 1& \hskip 1 cm(6123)& \hskip 1 cm(T^6-T^1)(T^2-T^3) \\(615)& \hskip 1 cm1 & 0& 0 & 0 & 1 & 1 & 0& \hskip 1 cm(7234)& \hskip 1 cm(T^7-T^2)(T^3-T^4) \\(726)& \hskip 1 cm0 & 1 & 0 & 0 & 0 & 1 & 1& \hskip 1 cm(1345)&\hskip 1 cm (T^1-T^3)(T^4-T^5) \end{array}\right)
\label{Hamming_CW}\ee

As we have explained in Sec. \ref{sec2}, one can perform the transformations on octonions which preserve the multiplication table. We would like to apply these transformations to 7 moduli $T^i$. Therefore we are only interested in transformations  without flipping the octonion signs, which preserve the multiplication table.
  These form the automorphism of the oriented Fano plane,  without the flipping of signs of octonions, preserving triads.   It is  the  finite Frobenius  subgroup
 $
{\cal Z}_7\rtimes {\cal Z}_3
$ of $G_2$.  It is a subgroup of the collineation group  $PSL(2,7)$ of order 168 of the Fano plane studied in detail
in \cite{Luhn:2007yr}.

 Leaving the full discussion to the general construction given in the next section  let us  show how these symmetries affect our superpotential $\mathbb{WO}_{\rm cw} $ associated with Cartan-Schouten-Coxeter conventions.
 The 7 cyclic permutations in Cartan-Schouten-Coxeter notations are generated by the permutation operator
\be
P_{CSC}=(1234567).
\ee
It describes the transformation
\be
1\rightarrow 2, \quad 2\rightarrow 3,\quad 3\rightarrow 4,\quad 4\rightarrow 5,\quad 5\rightarrow 6,\quad 6\rightarrow 7,\quad 7\rightarrow 1
\label{change1}\ee
which can be represented by the  matrix with the property that $c^7$ is the identity matrix.

\

\be c =  \left(\begin{array}{ccccccc}
0&1&0&0&0&0&0\\
0&0&1&0&0&0&0\\
0&0&0&1&0&0&0\\
0&0&0&0&1&0&0\\
0&0&0&0&0&1&0 \\
0&0&0&0&0&0&1\\
1&0&0&0&0&0&0
\end{array}\right)
\label{matrixPermutation1}\ee
Acting twice with $P_{CSC}$ leads to the mapping
\be
1\rightarrow 3, \quad 2\rightarrow 4,\quad 3\rightarrow 5,\quad 4\rightarrow 6,\quad 5\rightarrow 7,\quad 6\rightarrow 1,\quad 7\rightarrow 2
\ee
etc.  We now take the superpotential \rf{cw} and represent it in the form
\be
 \mathbb{WO}_{\rm cw}=  \sum _i\mathbb{WO}_{\rm cw}^i
\ee
where $\mathbb{WO}_{\rm cw}^1= (T^2 - T^4) (T^5 -T^6)$, $\mathbb{WO}_{\rm cw}^2= (T^3 - T^5) (T^6 - T^7)$ etc.
We now
 act on $ \mathbb{WO}_{\rm cw}$  by the $\mathbb{P}_{CSC}^1=c$ operator in eq. \rf{matrixPermutation1}, which is equivalent to a change of moduli as shown in eq. \rf{change1}. Under this change the first term in  $\mathbb{WO}_{\rm cw}$ becomes the second term there, the second becomes the third etc.
\be
 \mathbb{P}_{CSC}^1[  \mathbb{WO}_{\rm cw}^i] = \mathbb{WO}_{\rm cw}^{i+1}.
\ee
 For all 7 operations
 \be
 \mathbb{P}_{CSC}^k[  \mathbb{WO}_{\rm cw}^i] = \mathbb{WO}_{\rm cw}^{i+k}, \qquad k=1, \dots, 7
\ee
we find that the terms in $\mathbb{WO}_{\rm cw}$ are permuted and as a result, we have the same $\mathbb{WO}_{\rm cw}$ we started with
\be
 \mathbb{P}_{CSC}^k[  \mathbb{WO}_{\rm cw}] =  \mathbb{WO}_{\rm cw}
\ee
for each of the 7 $ \mathbb{P}_{CSC}^k$. This gives an example of the set of permutations of moduli which do not create different set of octonion multiplications tables and do not create different superpotentials.

 The other subgroup of the Frobenius group ${\cal Z}_7\rtimes {\cal Z}_3 $ is the cyclic group ${\cal Z}_3$  of order 3 , also described in \cite{Luhn:2007yr}. It preserves the octonion multiplication table, and does not flip the signs. However, it does not preserve the superpotentials.

\begin{figure}[H]
\centering
\includegraphics[scale=0.8]{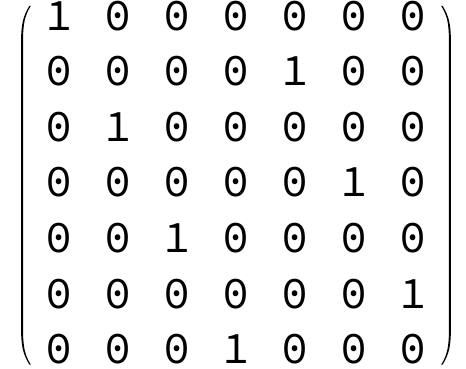}
\caption{\footnotesize   This is a $d$ operator in \cite{Luhn:2007yr}, such that $d^3=1$
}
\label{fig:d}
\end{figure}

In CSC convention it can be given as the following transformation
$
1\rightarrow 1,  2\rightarrow 5, 5\rightarrow 3, 3\rightarrow 2, 4\rightarrow 6, 6\rightarrow 7, 7\rightarrow 4
$.
which corresponds to
\be
(1)(253)(467).
\ee
In matrix notation
\be
d^3=1
\ee
where the matrix $d$ is in Fig.~\ref{fig:d}.

Thus, there are 3 possibilities, the first one is the original $\mathbb{WO}_{\rm cw}=d^3[\mathbb{WO}_{\rm cw}] $, the second one is $d[\mathbb{WO}_{\rm cw}]$, the 3d one is $d^2[\mathbb{WO}_{\rm cw}]$. It stops here since $d^3=1$ and we are back to the original $\mathbb{WO}_{\rm cw}$. We define
\be
d[\mathbb{WO}_{\rm cw}] = \mathbb{WO}_{\rm cw}' \ ,\qquad d^2 [\mathbb{WO}_{\rm cw}] = \mathbb{WO}_{\rm cw}^{''}\ .
\ee
One finds therefore that  there are  3 superpotentials for the Cartan-Schouten-Coxeter conventions.
\be
  \mathbb{WO}_{\rm cw}= d^3[\mathbb{WO}_{\rm cw}]= \sum_{r=1}^{7}  ( T^{r}T^{r+2}-T^{r} T^{r+1})
\label{cwgS} \ee
 \be
\mathbb{WO}'_{\rm cw}=  d[\mathbb{WO}_{\rm cw}] = \sum_{r=1}^{7}   (T^{r}T^{r+1}-T^{r}T^{r+3})  \label{CSC2}
\ee
\be
\mathbb{WO}''_{\rm cw}= d^2 [\mathbb{WO}_{\rm cw}]=  \sum_{n=1}^{7}  (T^{r}T^{r+3}-T^{r}T^{r+2})
 \label{CSC3} \ee
Obviously, these three superpotentials   satisfy
    \be \mathbb{WO}_{\rm cw} +\mathbb{WO}'_{\rm cw} +\mathbb{WO}''_{\rm cw}=0.
\label{iden}    \ee
 Thus only 2 of these superpotentials are independent.

 \subsection{Reverse Cartan-Schouten-Coxeter convention, counterclockwise heptagon}\label{sec4.2}
 Cartan-Coxeter-Schouten belongs to the set of 240 octonion conventions whic can be represented by the same oriented Fano plane modulo the permutations of the labeling of the points on it without sign changes. The other set of 240 conventions can not be reached by permutations alone and require sign flips of some of the imaginary octonion units. That is the reason we introduce the Reverse Cartan-Schouten-Coxeter convention which belongs to the second set of 240 conventions which includes the  Cayley-Graves convention as well. The conventions in the second set of 240 can also be represented by the same oriented Fano plane modulo the permutations of the labelling of the points on it without any  sign flips.

\begin{figure}[H]
\centering
\includegraphics[scale=0.75]{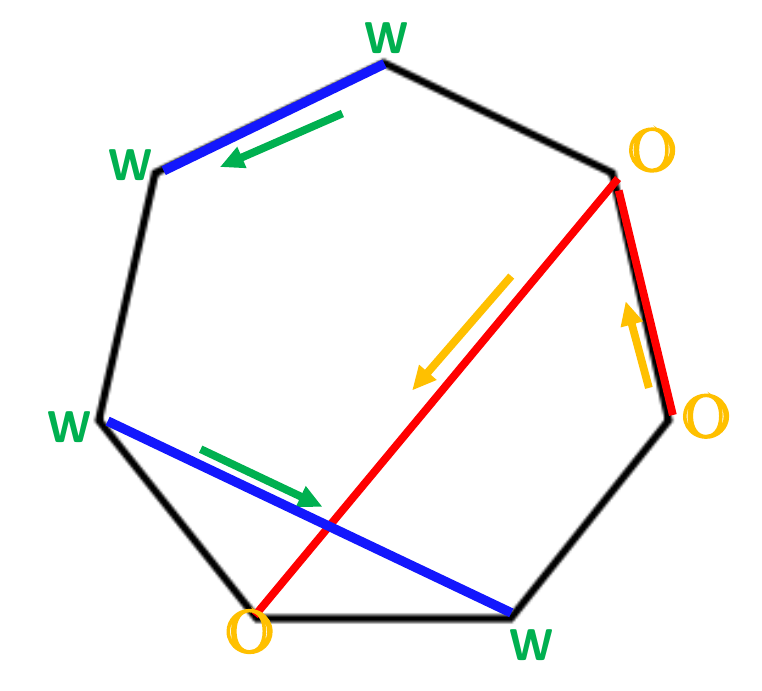}
\caption{\footnotesize  The {\bf counterclockwise oriented heptagon} for octonion notations in eq. \rf{acycl2}.  We start at  the second right octonion  and call it $e_{r+1}$, we get the 7 associative triads in the form as  $(r+1, r+2, r+6)$. The complimentary quadruples are $(r+3, r+4, r+5, r+7)$. They are shown by the corners of the oriented counterclockwise 2 blue lines. These give $\mathbb{WO}_{\rm ccw}$ on the counterclockwise oriented heptagon.
}
\label{fig:CCW_repr}
\end{figure}

Therefore we define the convention for octonion multiplication, following from the counterclockwise heptagon in Fig. \ref{fig:CCW_repr}, as follows:
 \be
 e_{r+3} e_{r+4} =  e_{r+2} e_{r+6}= e_{r+5} e_{r+7}=e_{r+1}\, ,  \qquad  \quad  e_r=e_{r+7}.
 \label{reverse}\ee
We  consider the following set associative triads consistent with the cyclic permutation in the counterclockwise direction   \footnote{A related set of triads was considered in \cite{Cerchiai:2018shs} in the form
 (126), (134), (157), (237), (245), (356), (467) in the context of irreducible representation of $PSL(2,7) $.}
   \be
 f_{ijk}^{\rm ccw} =+1:  (r+1, r+2, r+6)\, :  \{  (126), (237), (341), (452), (563), (674), (715) \},
\label{acycl2} \ee
 \be
i<j<k.
 \ee
To build the superpotentials we need the quadruples $\{mnpq\}$ which are complimentary to these associative triads which we take  to be
 $((r+3),(r+4),(r+5),(r+7))$, so that
  \be
m<n<p<q.
 \ee
 This defines the superpotential as follows:
\be
  \mathbb{WO}_{\rm ccw}= \sum_{r=0}^{6}   (T^{r+3}-T^{r+4})(T^{r+5}-T^{r+7})
  \label{ccwg} \ee

\noindent Explicitly we have for the  superpotential in the counterclockwise heptagon
  \bea
    \label{ccw}
&&   \mathbb{WO}_{\rm ccw} =  (T^3 - T^4) (T^5 - T^7) + (T^4 - T^5) (T^6 -
      T^1)+ (T^5 - T^6) (T^7 - T^2)\\
 \cr
      &&  + (T^6 - T^7) (T^1 - T^3) + (T^7 -
      T^1) (T^2 -T^4) + (T^1 - T^2) (T^3 - T^5) +(T^2 - T^3) (T^4 -T^6) .\nonumber
      \eea
 This one is easy to compare with the cyclic Hamming error correcting code, a $7\times7$ matrix at the right of the  Fig.~\ref{fig:Planat}. As in clockwise case we have to move the last row of the $7\times7$ matrix at the right of the  Fig.~\ref{fig:Planat} into the first row.
However, to see the triads and quadruples in eq. \rf{Hamming_CCW}, now { \it we have to read the codewords from the bottom to the top and from the right to the left.} This is opposite to the clockwise heptagon case in \rf{Hamming_CW} were we  read the codewords from the top to the bottom and from the left to the right.

All 3 1's in the codewords in  eq. \rf{Hamming_CCW} are in full agreement with the set of triads in eq. \rf{acycl2}. They are also shown in eq. \rf{Hamming_CCW} to the left of the codewords.
The 4 zero's in the 7  codewords in  eq. \rf{Hamming_CCW},  shown to the right of the codewords, are in perfect agreement with the 7 complimentary  quadruples defining  the superpotential.

\

    \be
{\rm \bf Triads }\hskip 1.2 cm {\rm \bf Codewords \hskip 0.7 cm     Quadruples  \hskip 0.5 cm \Rightarrow \hskip 0.3 cm \mathbb{WO}_{\rm ccw}}  \nonumber \ee
\be
\mathbb{WO}_{\rm ccw} = \left(\begin{array}{c|ccccccc|c|c}(715)& \hskip 1 cm1 & 0 & 1 & 0 & 0 & 0 & 1& \hskip 1 cm (2346) & \hskip 1 cm (T^2-T^3)(T^4-T^6) \\(674)&\hskip 1 cm1 & 1 & 0 & 1 & 0 & 0 & 0 & \hskip 1 cm (1235)& \hskip 1 cm(T^1-T^2)(T^3-T^5) \\(563)&\hskip 1 cm 0 & 1 & 1 & 0 & 1 & 0 & 0& \hskip 1 cm(7124)&\hskip 1 cm (T^7-T^1)(T^2-T^4)\\(452)& \hskip 1 cm 0 & 0 & 1 & 1 & 0 & 1 & 0 & \hskip 1 cm (6713)& \hskip 1 cm(T^6-T^7)(T^1-T^3)\\(341)& \hskip 1 cm 0 & 0 & 1 & 1 & 0 & 0 & 1& \hskip 1 cm(5672)& \hskip 1 cm(T^5-T^6)(T^7-T^2) \\(237)& \hskip 1 cm1 & 0& 0 & 0 & 1 & 1 & 0& \hskip 1 cm(4561)& \hskip 1 cm(T^4-T^5)(T^6-T^1) \\(126)& \hskip 1 cm0 & 1 & 0 & 0 & 0 & 1 & 1& \hskip 1 cm(3457)&\hskip 1 cm (T^3-T^4)(T^5-T^7) \end{array}\right)
\label{Hamming_CCW}\ee

In the counterclockwise case there are also 3 possibilities, the first one is the original $\mathbb{WO}_{\rm ccw}=d^3[\mathbb{WO}_{\rm ccw}] $, the second one is $d[\mathbb{WO}_{\rm ccw}]$, the 3d one is $d^2[\mathbb{WO}_{\rm ccw}]$.  We define
\be
d[\mathbb{WO}_{\rm ccw}] = \mathbb{WO}_{\rm ccw}' \ ,\qquad d^2 [\mathbb{WO}_{\rm ccw}] = \mathbb{WO}_{\rm ccw}^{''}\ .
\ee
One finds therefore that  there are  3 superpotentials for the Reverse-Cartan-Schouten-Coxeter conventions. In fact, they coincide with the ones we have found in the clockwise case.
\be
  \mathbb{WO}_{\rm ccw}= d^3[\mathbb{WO}_{\rm ccw}]= \sum_{r=1}^{7}  ( T^{r}T^{r+2}-T^{r} T^{r+1})=  \mathbb{WO}_{\rm cw}
\label{cwgen1} \ee
 \be
\mathbb{WO}'_{\rm ccw}=  d[\mathbb{WO}_{\rm ccw}] = \sum_{r=1}^{7}   (T^{r}T^{r+1}-T^{r}T^{r+3}) =\mathbb{WO}'_{\rm cw} \label{cwgen2}
\ee
\be
\mathbb{WO}''_{\rm ccw}= d^2 [\mathbb{WO}_{\rm ccw}]=  \sum_{n=1}^{7}  (T^{r}T^{r+3}-T^{r}T^{r+2})=\mathbb{WO}''_{\rm cw}
 \label{cwgen3} \ee
Obviously, these three superpotentials   satisfy
    \be \mathbb{WO}_{\rm ccw} +\mathbb{WO}'_{\rm ccw} +\mathbb{WO}''_{\rm ccw}=0.
\label{iden1}    \ee
 Thus only 2 of these superpotentials are independent.

  To summarize we can construct 2 independent superpotentials using the general formula \ref{superpotentialG} within each of the 480 octonion multiplication conventions. However the superpotentials obtained in the first set of 240 CW conventions  are not independent of the superpotentials obtained in the second set of 240 CCW conventions related to the first set by octonion conjugation.  Indeed under octonion conjugation the superpotential given by the formula \rf{superpotentialG} simply picks up an overall minus sign and the directions of all the arrows in the oriented Fano plane get reversed.
We shall, however, take  as independent one superpotential associated with Cartan-Schouten-Coxeter convention , $\mathbb{WO}_{\rm cw}$, belonging to the first set of 240 conventions,  and a different one, $\mathbb{WO}'_{\rm ccw}$,  associated with  Reverse-Cartan-Schouten-Coxeter convention belonging to the second set of 240. These are
\be
\mathbb{WO}_{\rm cw}= \sum_{r=1}^{7}  ( T^{r}T^{r+2}-T^{r} T^{r+1}),  \qquad \mathbb{WO}'_{\rm ccw}=   \sum_{r=1}^{7}   (T^{r}T^{r+1}-T^{r}T^{r+3})\ .
\label{two}\ee
From these two superpotentials we can reach any other superpotential by a change of moduli variables, which can be also obtained directly for a total 480 possible conventions, using the general formula in \rf{superpotentialG}.

 \subsection{Octonion  fermion mass matrix   and mass eigenstates }

 The superpotential for models with $G_2$ structure has 21 terms, it is described a symmetric 7x7 matrix with all diagonal terms vanishing \cite{House:2004pm,DallAgata:2005zlf}.
  It is given in eq.
 \rf{ourKW}
 \be
W={1\over 2}  M_{ij} T^i T^j
 \ee
 and $ M_{ij}  =-{1\over 4} \int _{X^7} \phi^i \wedge d\phi^j  $.  The masses of fermions in supergravity in Minkowski vacua are defined  by the second derivative of the superpotential
\be
m_{ij} ={1\over 2} \, e^{K\over 2} M_{ij}
\ee
and the mass term of the chiral fermions in the supergravity at the vacuum with $W=W_{i}=0$ is
\be
{\cal L}^f_{m}= {1\over 2} \, e^{K\over 2} \chi^i M_{ij} \chi^j +{\rm h.c.}
\ee
 Our octonion superpotentials for models with $G_2$ holonomy have explicitly 14 terms given in eqs. \rf{cwgen1}- \rf{cwgen3}  and can be written in the following form
\be
 \mathbb{WO}={1\over 2}  \mathbb{M}_{ij} T^i T^j.
\label{octW} \ee
The matrix  $ \mathbb{M}_{ij}$  for the case $\mathbb{WO}_{\rm cw}$ for CSC octonion notations is
\be \mathbb{M}_{ \mathbb{WO}_{\rm cw}} =  \left(\begin{array}{ccccccc} 0&-1&1&0&0&1&-1\\
-1&0&-1&1&0&0&1\\
1&-1&0&-1&1&0&0\\
0&1&-1&0&-1&1&0\\
0&0&1&-1&0&-1&1 \\
1&0&0&1&-1&0&-1\\
-1&1&0&0&1&-1&0
\end{array}\right).
\label{matrixWCSC}\ee
 One can see that in each row of this matrix the sum of entries vanishes. This is a condition for Minkowski vacua so that eq. $\sum_j M_{ij} =0\, ,    \forall i$ is satisfied.

 The matrix  $ \mathbb{M}_{ij}$  for the case $\mathbb{WO}'_{\rm ccw}$ for RCSC octonion notations is
 \be \mathbb{M}_{ \mathbb{WO}'_{\rm ccw}} =  \left(\begin{array}{ccccccc} 0&1&0&-1&-1&0&1\\
1&0&1&0&-1&-1&0\\
0&1&0&1&0&-1&-1\\
-1&0&1&0&1&0&-1\\
-1&-1&0&1&0&1&0 \\
0&-1&-1&0&1&0&1\\
1&0&-1&-1&0&1&0
\end{array}\right).
\label{matrixWRCSC}\ee
Here again  a condition for Minkowski vacua,  $\sum_j M_{ij} =0\, ,    \forall i$ is satisfied.

 The non-vanishing eigenvalues of the $ \mathbb{M}$  matrices in both cases above  solve a double set of cubic equations
\bea
&&x^3-7x-7=0,\cr
\cr
&&y^3-7y-7=0.
\eea
Numerically this gives for a set of $x_1, y_1;\, \, x_2, y_2;\, \,  x_3, y_3$ and a massless one, the following values
\be
3.04892, \, 3.04892; \, \, -1.69202, \, -1.69202; \, \, -1.3569, \, -1.3569; \, \, 0.
\label{fermion}\ee
They show an $[SO(2)]^3$ symmetry which is a symmetry of the massive fermion eigenstates.
Moreover, the mass eigenstates of fermions are the same for any of the 480 choices of octonion notations!

The scalar mass eigenvalues in $\cN=1$  supergravity models with octonion superpotentials and one flat direction are
\begin{equation}
(m_{\rm sc}^{\rm can})^2 :   \   \  t^{-3}\{1.16, \, 1.16, \, \, \, 0.35, \, 0.35, \, \, \, 0.23, \, 0.23, \, \, \, 0 \}
\label{eigenSc}\end{equation}
 where $t={\rm Re}(T)$.
Here we have taken into account the correct kinetic term normalization factors. They have a simple relation to fermion mass eigenvalues
in agreement with $\cN=1$ supersymmetry. To see this we have to take into account that canonical fermion masses, with account of a difference in kinetic term normalization are
\be
m_{\rm f}^{\rm can} :   \     t^{-3/2}\{1.07796, \, 1.07796, \, \, \, -0.59822, \, -0.59822, \, \, \, -0.479735, \, -0.479735, \, \, \, 0 \}.\ee
One can check that
\be
(m_{\rm f}^{\rm can})^2 = (m_{\rm sc}^{\rm can})^2
\ee
as it should according to $\cN=1$ supersymmetry. In all models with octonion superpotential $  \mathbb{WO}$  the mass eigenvalues are the same, i. e. they are preserved when new octonion convention is used.

One way to see it is to  perform the change of fermion variables reproducing the transformation on octonions, permutations and sign flips, ${\cal T}=[{\cal Z}_2]^7 \cdot {\cal S} (7)$. It means that the effective matrix $M_{ij}$ will take a different value whenever our transformation is not part of the subgroup ${\cal H}=[{\cal Z}_2]^3 \cdot  {\cal{PSL}}_2 (7)$. In this way
 we can get a total of $ {[{\cal Z}_2]^7 \cdot {\cal S} (7)\over [{\cal Z}_2]^3 \cdot  {\cal{PSL}}_2 (7)} \rightarrow  480 $ different matrices, starting from any one in eqs.  \rf{matrixWCSC} or \rf{matrixWRCSC}. This number of different $\mathbb{M}_{ij}$  matrices is in agreement with the fact that there are 480 different multiplication tables of octonions. The upshot here is: our octonion superpotentials have an eigenstate  mass spectrum invariant under a change of octonion conventions.

\section{ Octonions and General Construction of Superpotentials  }\label{sec5}

 The Fano plane has 7 lines and each line contains three points such that each point is the intersection point of three lines. When we use oriented Fano planes to describe octonion multiplication three points on each line go over to the imaginary units of a quaternion subalgebra.

Given an associative triad $(ijk)$ the corresponding structure constants  $f_{ijk}$ are those  of a quaternion subalgebra generated by  $e_i, e_j$ and $e_k$. The remaining associative triads
 can be obtained by the ``cyclic'' permutation operator $P$ given by the labelling of the imaginary units on a heptagon. For example in Gunaydin-Gursey (GG) \cite{Gunaydin:1973rs} labelling of the real octonions the operator $P$ is
$
P_{GG} = (1243657)
$.
For the Cartan-Schouten-Coxeter  (CSC) labelling  \cite{Cartan:1926,Coxeter:1946} the operator $P$ is simply
$
P_{CSC} = (1234567)
$.
  For the Cayley-Graves labelling of octonions the cyclic permutation operator is
$
P_{CG}=(1245736)
$.

 An imaginary unit $e_k$ is contained in three  different quaternion subalgebras. Let us denote the structure constants of these three quaternion subalgebras as $f_{kij}, f_{kmn} $ and $f_{kpq}$ and assume they are in cyclic order such that
\be f_{kij}=f_{kmn}=f_{kpq} =1 \label{triadcondition}
\ee
where the indices $i,j,m,n,p,q$ are all different.

Given a quaternion subalgebra with imaginary units $e_k, e_i, e_j$ we associate a term in the superpotential of the  form
\be
f_{kij} \Longrightarrow    f_{kij} f_{kmn} f_{kpq} (T^m-T^n)(T^p-T^q)   \label{term}
\ee
where all the 7 indices appearing above are different and satisfy   \eqref{triadcondition} with  no sums over indices. The full superpotential is obtained by acting on a given term of this form by the cyclic permutation operator $P$ repeatedly  and summing over:
\be
W(kij) = \sum_{r=0}^6  (P)^r \{  f_{kij} (T^m-T^n)(T^p-T^q)  \} \label{superpotentialG}
\ee
where $P$ acts on all the seven indices inside the sum.  Note that the superpotential is invariant under the cyclic group ${\cal  Z}_7$ generated by the permutation operator $P$ by construction.
Note that we labelled the superpotential by the the associative triad $(kij)$ since it is determined by it uniquely. One finds that
\be
W(kij) +W(jki) +W (ijk) =0. \label{Constraint}
\ee
Furthermore one has
\be
W(kij) =- W(kji).
\ee
Therefore of the six superpotentials defined by the triad $(kij)$ and its permutations one finds only two independent ones.

The cyclic permutation operator $P$ can be used to make any Hamming code associated with a given convention of octonions cyclic. The codewords associated with the different associated triads get mapped into each other under the action of $P$.

 The three 3 superpotentials we have discussed above  for Cartan-Schouten-Coxeter conventions are particular cases of   the general formula \ref{superpotentialG}.   We combine them in a different order so that
 \bea
\mathbb{WO}_{\rm cw}
     =  \sum_{n=0}^{6}  f_{n+1,n+3,n+7} f_{n+1,n+2,n +4} f_{n+1,n+5,n+6} (T^{n+2}-T^{n+4})(T^{n+5}-T^{n+6}) \label{cwM}
    \eea
 \be
\mathbb{WO}'_{\rm cw}=  \sum_{n=0}^{6}  f_{n+1,n+2,n+4} f_{n+1,n+3,n +7} f_{n+1,n+5,n+6} (T^{n+3}-T^{n+7})(T^{n+5}-T^{n+6})  \label{cw'M}
\ee
\be
\mathbb{WO}''_{\rm cw}=  \sum_{n=0}^{6}  f_{n+1,n+5,n+6} f_{n+1,n+2,n +4} f_{n+1,n+3,n+7} (T^{n+2}-T^{n+4})(T^{n+3}-T^{n+7}).
 \label{cw''M} \ee
 These three superpotentials   satisfy
    \be \mathbb{WO}_{\rm cw} +\mathbb{WO}'_{\rm cw} +\mathbb{WO}''_{\rm cw}=0.
\label{iden2}    \ee
This follows from the fact that
\be
C_{mnpq} (T^{m}-T^{n})(T^{p}-T^{q})=0
\ee
where the completely antisymmetric tensor  $C_{mnpq} = f_{k[mn} f_{p]kq }$ is defined in eq. \rf{via_f}. Thus only 2 of these superpotentials are independent. It is a property satisfied for any of the different 480 conventions: there are 3 superpotentials for the same set of conventions, and they always satisfy  the constraint
\rf{Constraint}.

 However this does not imply that we get $480\times 2= 960$ different superpotentials.  An additional interesting property of the general formula for the superpotentials is that one can change the octonion convention set by odd permutation of indices
\be
f_{ijk} \quad \Rightarrow \quad f_{jik}.
\ee
This can be achieved by the sign flip of 3 octonions which belong to any particular line on the Fano plane, for example in Cartan-Schouten-Coxeter case in Fig. \ref{fig:Fano1}, one can change the signs of the following octonions
\be
e_6 \quad \Rightarrow \quad -e_6 \, , \qquad e_1 \quad \Rightarrow \quad -e_1 \,, \qquad  e_5 \quad \Rightarrow \quad -e_5,
\ee
or any other 3 octonions on one line in the Fano plane. This will change every term in the original multiplication table to the one with the opposite sign (octonion conjugation). As one can see from eqs. \rf{cwM}-\rf{cw''M} each superpotential will change the sign:
\be
\mathbb{WO}_{\rm cw} \Rightarrow - \mathbb{WO}_{\rm cw}\ , \qquad \mathbb{WO}'_{\rm cw} \Rightarrow -\mathbb{WO}'_{\rm cw}\ , \qquad \mathbb{WO}''_{\rm cw} \Rightarrow -\mathbb{WO}''_{\rm cw}\ .
\ee
 This means that we  get a total of 480 different superpotentials when we consider all possible  inequivalent conventions.
And since the potential is quadratic in superpotentials this change of octonion conventions will not affect the potential.

 These 480 different superpotentials corresponding to 480 different conventions can be understood via change  of variables starting from our simple cases in eqs. \rf{cwgen1} and \rf{cwgen2} consistent with the symmetries of the Lagrangian that preserve the underlying octonionic structure. This explains why all these models are equivalent : in particular, one finds that all of them have exactly the same masses of eigenstates at the vacuum.

 \section{  Octonionic  superpotentials and Minkowski vacua}\label{sec6}
  \subsection{Vacua with 1 flat direction}\label{sec5.1}

Supersymmetric Minkowski vacua of the superpotentials in the form shown in eq.  \rf{ourKW} were studied in \cite{DallAgata:2005zlf}. With
\be
T^i=t^i+{\rm i} \, a^i
\ee
 it was shown there that $D_i W= W=0$ means that $M_{ij} t^i =0$  and   $a^i=0$ at the minimum.
Therefore $t^i$ should be a null vector of the matrix $M$: there is at least one flat direction in such Minkowski vacua.

The potential $\mathbb{VO}$ is defined by the 7-moduli  $K_{7{\rm mod}}$ in eq. \rf{ourKW} and $\mathbb{WO}$ in eq. \rf{ourW}.
\be
\mathbb{VO}\equiv e^{K_{7{\rm mod}}}( |D\, \mathbb{WO}|^2- 3 |\mathbb{WO}|^2).
\ee
Specific superpotentials which we studied here are shown in eqs.  \rf{cw}, \rf{ccw}, \rf{cwgen1}, \rf{cwgen2} and in other examples in Appendix \ref{secA}.  We have checked that these potentials have supersymmetric Minkowski vacua when all moduli are equal to each other, as in eq. \rf{sol}
\be
D_i \mathbb{WO}= \mathbb{WO}=0 : \qquad T^i- T^j=0 \quad \forall \,  i,\, j
\ee
with one flat direction $T={1\over 7} \sum_{i} T^i$. This equation is invariant under the permutation group ${\cal S} (7)$ of order $7!=5040$.
 
As  expected, the eigenvector of the massless direction corresponds to
\begin{equation}
t^1+t^2+\cdots+t^7, \qquad \text{and}\qquad a^1+a^2\cdots+a^7.
\end{equation}
This leads to an effective theory of a single modulus supergravity of the form
\be
K = - 7 \log\left(  T + \overline{T}\right)\, , \, \qquad  \mathbb{WO}=0.
\label{startA}\ee
In order to check the stability of the minimum, we consider the Hessian of the scalar potential. We  find the $14\times14$ Hessian matrix of the potential with respect to $\{ t^i,a^i\}$ ($i=1,\cdots,7)$. Since the vacuum is supersymmetric, in each case there is  the same mass eigenvalue for $t^i$ as for $a^i$, so we show here only 7 of them, one flat direction being massless.

On the line in the moduli space $t^i=t$, $a^i=a$ for all $i=1,\cdots,7$, the eigenvalues of the matrix are given by eq. \rf{eigenSc}.
The 6 massive eigenvalues are pairwise equal, i. e. we have an $[SO(2)]^3$ symmetry in the mass matrix at the minimum of the potential.

Of course our $14\times14$ mass matrix for scalars and pseudoscalars has the standard $[SO(2)]^7$ symmetry since each of the scalars has the same mass as the pseudoscalar.  However, this additional $[SO(2)]^3$ symmetry in the mass matrix of the scalars and separately pseudoscalars is a feature of our  vacua. The $14\times14$ scalar/pseudoscalar mass matrix has the following eigenvalues
\begin{equation}
  t^{-7}\{0.58, \, 0.58, 0.58, \, 0.58,\, \, \, 0.18, \, 0.18, 0.18, \, 0.18, \, \, \, 0.11, \, 0.11, 0.11, \, 0.11,\, \, \, 0, 0 \}.
\label{eigenScAx}\end{equation}
They solve the following cubic equation
\be
y^3-14y^2+49y-49=0.
\ee
It is related to the cubic equation for fermion masses $x^3-7x-7=0$ as follows
\be
(x^3-7x)^2=49 \quad \Rightarrow \quad x^6- 14 x^4 + 49x^2=49|_{x^2=y} \quad \Rightarrow \quad y^3-14y^2+49y-49=0.
\ee

As we will see in the following, inflationary potential added to this setup does not spoil the stabilization and we will show that the trajectory discussed here can be realized effectively.

It is interesting that directly in 4d one can suggest many superpotentials for our 7 moduli which also lead to one flat direction Minkowski vacua, but not associated with octonions. The simplest example is the case with
\be
W=(T^n-T^{n+1})(T^{n+2}-T^{n+3}).
\ee
The scalar mass eigenvalues are
\begin{equation}
t^{-7}\{1.40, \, 1.40, \, \, \, 1.21, \, 1.21, \, \, \, 0.07, \, 0.07, \, \, \, 0 \}.
\label{eigenA}\end{equation}
These masses still have an $[SO(2)]^3$ symmetry, however, the eigenvalues are different from the ones which come from models with octonion superpotentials. Therefore, these models are 4d models which are not clearly originating from the 11d supergravity compactified on 7-tori with $G_2$ holonomy. Meanwhile all 480 octonion conventions always lead to models with the same scalar mass eigenstates given in eq. \rf{eigenScAx}.

\subsection{Using cyclic Hamming (7,4)  code to get vacua with 2 flat directions}\label{sec5.2}

{\it  6.1 split}

\begin{figure}[H]
\centering
\includegraphics[scale=0.6]{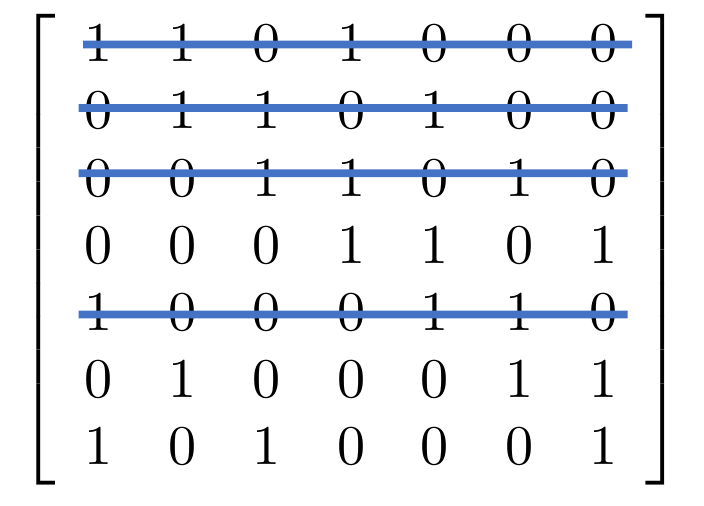}
\caption{\footnotesize We start with a cyclic Hamming code (7,4) in Fig. \ref{fig:Planat}. We take out all codewords which have in a position 7 a 0. This leaves us with the superpotential in eq. \rf{61}}
\label{fig:Cyclic61}
\end{figure}

We can  start with the cyclic  error correcting Hamming (7,4) code  taken,  for example from \cite{Planat} and shown here in Fig. \ref{fig:Planat}. The 7 codewords are
\be
1101000, \, \, 0110100, \, \, 0011010, \, \,   0001101,   \,   \, 1000110,  \,  \, 0100011, 1010001.
\label{Planat}\ee
This corresponds to  triads  in eq. \rf{cycl}  with the formula $(r, r+1, r+3)$ with the first index in triads  taken in the range 1 to 7.
We see the first codeword of the code as  (124), which are positions of 1's. The second codeword is the case (235), which are positions of 1's, etc.  The set of quadruples defined by the positions of zero's in the codewords is
 now:
 \be
 \{pqrs\} = \{  { (3567)},  {(4671)},  {(5712)}, (6123),  {(7234)}, (1345), (2456)   \}.
\label{quadr}\ee
The corresponding superpotential is the same as in eq. \rf{cw}, with the difference that the first term in \rf{cw} becomes the last one.
Now we propose to define a model with 2 flat directions by dropping some codewords in the code in eq. \rf{Planat} in a specific way: we leave only quadruples without 7.  The ones with 7 are underlined below
  \be
 \{pqrs\} = \{ \underline { (3567)}, \underline {(4671)}, \underline {(5712)}, (6123), \underline {(7234)}, (1345), (2456)   \}.
\label{quadUnder}\ee
The remaining quadruples form the superpotential
\begin{eqnarray}
\mathbb{WO}(6,1) =  (T^6 -T^1) (T^2 -T^3)+ (T^1 -T^3) (T^4 -T^5)+ (T^2 -T^4) (T^5 -T^6).
\label{61}\end{eqnarray}
One can see this procedure also using the Hamming code from eq. \rf{Planat} where we take out all lines which have $0$  in the position 7. We show it in Fig. \ref{fig:Cyclic61}.

The potential does not depend on $T^7$ and has a Minkowski minimum with two flat directions. One is $T^7$ and the other is $T^1+T^2+T^3 + T^4+ T^5 +T^6$.
The model has a (6,1)
 split
  \be
K= - 6 \log\left(  T_{(1)} + \overline{T}_{(1)}\right) -   \log\left(  T_{(2)} + \overline{T}_{(2)}\right) \, , \, \qquad  \mathbb{WO}(6,1) =0,
\label{6.1}\ee
where
\be
T^1=T^2=T^3 = T^4= T^5 =T^6\equiv T_{(1)}\qquad T^7 \equiv T_{(2)}.
\ee
Two of the positive mass matrix eigenvalues coincide, the vacuum has an unbroken  $SO(2)$ symmetry.

{\it 5.2 split}

We take the 7 quadruples in eq. \rf{quadr}, $(n+2,  n+4 ,  n+5,  n+6 )$, defining our superpotential and leave only the ones which do not have 6 and 7. These with both  6 and 7 are underlined:
\be
 \{pqrs\} = \{ \underline { (3567)}, \underline {(4671)},  {(5712)}, (6123),  {(7234)}, (1345), (2456)   \}.
\label{quadUnder1}\ee
The remaining quadruples form the superpotential
\begin{eqnarray}
\mathbb{WO} ( 5.2) = && (T^5 -T^7) (T^1 -T^2)+ (T^6 -T^1) (T^2 -T^3)+(T^7 -T^2) (T^3 -T^4)\cr
\cr
&& + (T^1 -T^3) (T^4 -T^5)+ (T^2 -T^4) (T^5 -T^6).
\label{52}\end{eqnarray}
\begin{figure}[H]
\centering
\includegraphics[scale=0.6]{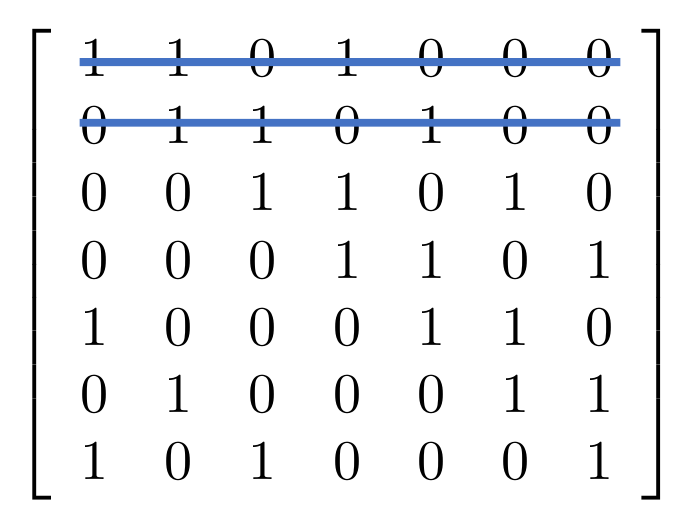}
\caption{\footnotesize We start with a cyclic Hamming code (7,4) in Fig. \ref{fig:Planat}. We take out all codewords which have have $0$  in the position 6 and 7.This leaves us with the superpotential in eq. \rf{52}.}
\label{fig:Cyclic52}
\end{figure}

One can see this procedure also using the Hamming code in eq. \rf{Planat} where we take out all lines which have $0$  in the position 6 and  7.  We show it in Fig. \ref{fig:Cyclic52}.

The model has a Minkowski minimum with two flat directions. One is $T^5+T^6$ and the other is $T^1+T^2+T^3 + T^4+ T^7 $.
The model has a (5,2)
 split
 \be
K= - 5 \log\left(  T_{(1)} + \overline{T}_{(1)}\right) -  2 \log\left(  T_{(2)} + \overline{T}_{(2)}\right) \, , \, \qquad  \mathbb{WO}(5,2) =0,
\label{5.2}\ee
where
\be
T^1=T^2=T^3 = T^4= T^7 \equiv T_{(1)},\qquad T^5=T^6 \equiv T_{(2)}.
\ee
Two of the positive mass matrix eigenvalues coincide, the vacuum has an unbroken  $SO(2)$ symmetry.

{\it 4.3 split}

\begin{figure}[H]
\centering
\includegraphics[scale=0.6]{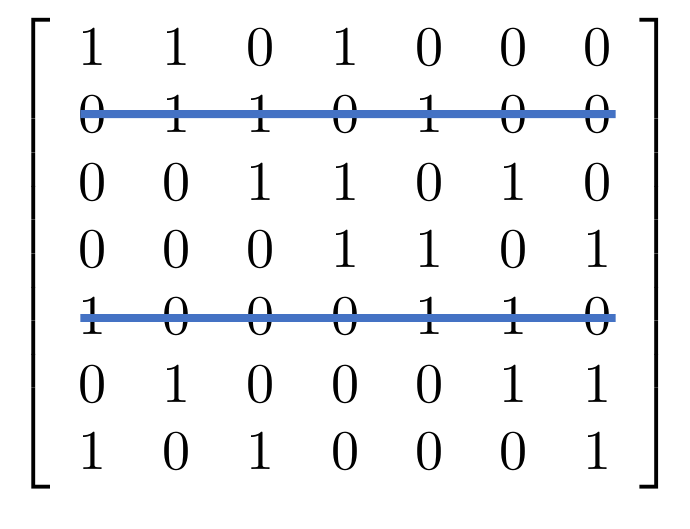}
\caption{\footnotesize We start with a cyclic Hamming code (7,4) in Fig. \ref{fig:Planat}. We take out all codewords which have 0's in a position 4 and 7. This leaves us with the superpotential in eq. \rf{43}}
\label{fig:Cyclic43}
\end{figure}

We take the 7 quadruples in eq. \rf{quadr}, $(n+2,  n+4 ,  n+5,  n+6 )$, defining  $W$  and remove  the ones which  have 4 and 7.  Note that the pattern would require to exclude the lines which do not have the same 3 numbers, or the same 3 zero's in the code. But these do not exist, therefore we find the relevant split (4.3) model not following the underlying procedure which was working in the previous two cases of  split models. So, we exclude the quadruples which have  4  and 7. This we show in Fig. \ref{fig:Cyclic43}. The underlined quadruples in eq. \rf{quadUnder2} are the ones we remove from the superpotential
\be
 \{pqrs\} = \{ (3567), \underline {(4671)},  (5712), (6123),   \underline  {(7234)}, (1345), (2456)   \},
\label{quadUnder2}\ee
The remaining quadruples form the superpotential
\begin{eqnarray}
\mathbb{WO}( 4.3) = && (T^3 -T^5) (T^6 -T^7)+(T^5 -T^7) (T^1 -T^2)+ (T^6 -T^1) (T^2 -T^3)\cr
\cr
&& + (T^1 -T^3) (T^4 -T^5)+ (T^2 -T^4) (T^5 -T^6).
\label{43}\end{eqnarray}
The model has a Minkowski minimum with two flat directions. One is $T^2+T^3+T^6$ and the other is $T^1+T^4+T^5 + T^7 $.
The model has a (4.3)
 split
 \be
K= - 4 \log\left(  T_{(1)} + \overline{T}_{(1)}\right) -  3 \log\left(  T_{(2)} + \overline{T}_{(2)}\right) \, , \, \qquad  \mathbb{WO}(4.3) =0,
\label{4.3}\ee
where
\be
T^1=T^4=T^5 = T^7 \equiv T_{(1)}\qquad T^3=T^3 =T^6\equiv T_{(2)}.
\ee
The mass matrix massive eigenvalues are all different, there is no unbroken  $SO(2)$ symmetry.

\section{Cosmological Models with B-mode Detection Targets }\label{sec7}
\subsection{ Strategy}

So far we have studied Minkowski vacua in 4d $\cN=1$ supergravity derived from compactification of M-theory on twisted 7-tori. Such Minkowski vacua
are known to have only non-negative mass eigenvalues, with some flat directions (zero mass eigenvalues) possible \cite{BlancoPillado:2005fn}. We have presented these eigenvalues in eqs. \rf{eigenSc} for relevant models.

To construct a phenomenological cosmological model, describing the observations, we need to make some changes of these models, which  transfer a flat direction into an almost flat inflationary potential, and replace Minkowski vacuum with de Sitter vacuum at the exit from inflation.

For this purpose we will do the following changes:
\begin{itemize}
\item  Change the \K\, potential frame to the one where the \K\, potential has an inflaton shift symmetry,  only slightly broken by the superpotential, as proposed in~\cite{Carrasco:2015uma}.
  \item Introduce a nilpotent superfield $S$, which in $\cN=1$ supergravity allows to get dS minima  with spontaneously broken supersymmetry~\cite{Bergshoeff:2015tra,Hasegawa:2015bza}. This Volkov-Akulov superfield $S$ brings a new supersymmetry breaking parameter, $F_S$.
We also introduce a parameter $W_0$ to make a gravitino mass at the minimum non-vanishing. This allows to describe the theory with a cosmological constant $\Lambda = F_S^{2}-3W_{0}^{2}$ at the minimum of the potential.
\item Introduce a phenomenological  inflationary potential along the flat direction. We will do it using a generalized version of the geometric approach to inflation developed in   \cite{McDonough:2016der,Kallosh:2017wnt}, where the properties of the inflaton potential are encoded in the choice of the \K\, potential for the nilpotent superfield,  $G_{S\bar{S}}(T, \bar T) S\bar{S}$.
  \end{itemize}

Thus we are looking for $\cN=1$ supergravity cosmological models which are compatible with cosmological observations. The reason why we call these cosmological models phenomenological is the choice of the \K\, potential for the nilpotent superfield,  $G_{S\bar{S}}(T, \bar T) S\bar{S}$:  at present it is not known how to derive it from fundamental theory.

In the absence of the nilpotent superfield  $S$ and at vanishing  $W_0$    our cosmological model   becomes the model with Minkowski vacua in M-theory with flat directions equivalent to the one derived from M-theory in \cite{DallAgata:2005zlf,Derendinger:2014wwa} and specified to the case of octonions in this paper.

We start with  M-theory on $G_2$ holonomy manifolds\cite{DallAgata:2005zlf,Derendinger:2014wwa} with the choice of $K$ and $W$ in \eqref{ourKW}. We specify the choice of  $W={1\over 2} M_{ij} T^i T^j$  to be the one which originate from  octonions, for example any one in eqs.   \eqref{cw},  \eqref{cwgen1} or in \rf{cwgen2},   which we studied above.
\begin{align}
K=- \sum_{i=1}^7 \log\left(  T^i + \overline{T}^i\right),\qquad
W={1\over 2} M_{ij} T^i T^j=  \mathbb{WO}(T^i) .
\label{Guian}\end{align}
For any of our octonionic superpotentials  an equivalent model can be obtained from \rf{Guian} using a \K\, transform, which results in
\be
\mathcal{K} (T^i, \overline T^i)=-{1\over 2} \sum_{i=1}^7\log \left(\frac{(T^{i}+\overline{T}^i)^2}{4T^{i}\overline{T}^{i}}\right),  \qquad   \mathcal{W}^{\rm oct}\equiv {1\over \sqrt{\prod_{i=1}^7 (2 \, T^i)}}  \mathbb{WO}(T^i)\ .
\label{useful}\ee
The potentials in models \rf{Guian} and \rf{useful} are the same, due to \K\, invariance, and lead to a Minkowski vacuum with one flat direction and unbroken supersymmetry. Also the fermion masses are the same under \K\, transform.
In what follows we will  study the models which in absence of the nilpotent superfield and new parameters $F_S$ and $W_0$ and inflationary potentials are the ones in eq. \rf{useful},  which are equivalent to the original models in M-theory in eq. \rf{Guian}.

\subsection{  \boldmath  E-models} \label{emodel}

We construct  our 7-moduli cosmological models where inflation is induced by terms in the \K\ potential where the nilpotent superfield $S$  interacts with the inflaton superfield   \cite{McDonough:2016der,Kallosh:2017wnt}.

Consider the \K\,  and super-potential given by
\begin{align}\label{Kal}
K=&\sum_{i=1}^7-\frac12\log\left(\frac{(T^i+\bar{T}^i)^2}{4T^i\bar{T}^i}\right)+G_{S\bar{S}}(T^i,\bar{T}^i)S\bar{S},\\
W=&  \mathcal{W}^{\rm oct}+W_{0}+F_S\,  S ,
\label{sup}\end{align}
where $S$ is a nilpotent superfield and $ \mathcal{W}^{\rm oct}$ is an arbitrary  holomorphic function of the 7 moduli $T^i$. 
We will take
\be
G_{S\bar{S}}= \frac{F_S^{2}}{F_S^{2}+V_{\rm infl} (T^i,\bar{T}^i)} .
\ee

Then one can show that for  real fields  $T^i=\bar{T}^i = t^{i}$ the scalar potential is given by
\begin{equation}
V=F_{S}^{2}-3(W_{0}+ \mathcal{W}^{\rm oct}(T^i))^{2}+  \sum_{i=1}^7(T^{i}+ \bar T^{i})^{2}  |\partial_i  \mathcal{W}^{\rm oct}|^{2}_{|_{ \bar T^{i}\to T^{i}}} +  V_{\rm infl} (T^{i}, \bar T^{i})_{|_{ \bar T^{i}\to T^{i}}} 
\end{equation}

Along   of the superpotential along the flat direction one has $ \mathcal{W}^{\rm oct} = \partial_i   \mathcal{W}^{\rm oct} = 0$, and therefore in for the real inflaton field(s) $T^{i}= \bar T^{i}= t^{i}$ one has
\begin{equation}
V=F_{S}^{2}-3W_{0}^{2}+  V_{\rm infl} (T^{i}, \bar T^{i})_{|_{ \bar T^{i}\to T^{i}}}  
\label{general}
\end{equation}
Note that $F_{S}^{2}-3W_{0}^{2} = \Lambda$, where $\Lambda \sim 10^{{-120}}$ is  tiny cosmological constant. This  yields
\begin{equation}
V= \Lambda + V_{\rm infl}  (T^{i}, \bar T^{i})_{|_{ \bar T^{i}\to T^{i}}} 
\label{generalinfl}
\end{equation}
During inflation one can safely ignore $\Lambda \sim 10^{{-120}}$, and our equation for the potential along the flat directions reduces to 
\begin{equation}
V=  V_{\rm infl}  (T^{i}, \bar T^{i})_{|_{ \bar T^{i}\to T^{i}}}   \ .
\label{generalinfl0}
\end{equation}
This is a great simplification. Generically, one could expect  the parameters of the M-theory to be large, in Planck mass units, whereas the height of the inflaton potential at the last stages of inflation should be tiny, $V_{\rm infl} \sim 10^{{-10}}$. Therefore it is very important that in the approach developed above, the inflaton potential along the flat directions is almost completely sequestered from the UV dynamics responsible for the M-theory potential.

This simplification was achieved because we started with the superpotential and its derivatives vanishing along some direction, and then we made a \K\ transformation which made the\K\ potential vanish in the same direction.  After this investigation was completed,  an equivalent but more compact procedure  leading to the same results  was developed in \cite{Kallosh:2021fvz,Kallosh:2021vcf}. In that method, one does not transform the original \K\ potential and the  superpotential $W={1\over 2} M_{ij} T^i T^j= \mathbb{WO} (T^i)$ \rf{Guian} to the form \rf{useful}. Instead of that, it is sufficient to slightly modify the term describing interaction of the moduli fields $T^{{i}}$ with the nilpotent field $S$ in the superpotential:
\begin{eqnarray}
K&=&- \sum_{i=1}^7 \log\left(  T^i + \overline{T}^i\right) +G_{S\bar{S}}(T^i,\bar{T}^i)S\bar{S} \ ,   \\
W&=&{1\over 2} M_{ij} T^i T^j +(W_{0}  +F_S\,  S)\, {\cal V}^{1/2}, \qquad  {\cal V} =\prod_{i=1}^{7} (2  T^i) \ .
\label{Guian2}\end{eqnarray}

From the point of view of inflation, the main role of the M-theory (UV dynamics) is to define geometry with the proper kinetic terms, and form a potential with supersymmetric flat directions. Once it is done, the inflaton dynamics (IR) is completely independent of the full M-theory potential, but is determined by the M-theory related kinetic terms, and by the phenomenological inflaton potential $V=F (T, \bar T)$ along the flat directions.

Moreover, an important feature of the $\alpha$-attractor models with inflation along the flat directions in M-theory is that the observational predictions of these models are very stable with respect to the choice of the particular potential $V=F (T, \bar T)$ as long as these potentials are not singular. That is why one can talk about specific predictions of a broad class of such models.

  Now  we will present several examples of inflationary models corresponding to different choices of potentials and flat directions in the octonionic models discussed in this paper.  Note, however, that the results obtained in this subsection are quite general. They may apply to a broad class of theories with any superpotential $Q$ with supersymmetric flat directions, such that $Q = \partial_i  Q = 0$. Therefore our methods of constructing phenomenological inflationary models can be used not only in the octonionic models in the M-theory context, but in a more general class of models with supersymmetric flat directions, which may exist, e.g.,  in type IIB or type IIA string theory.
  
  \

{ \it $3\alpha= 7$ E-model }

\noindent Now we focus on a special choice of $Q$ in our general formula
\be
Q=\mathcal{W}^{\rm oct} \ ,
\ee
where $\mathcal{W}^{\rm oct}$  defined in eq. \rf{useful}.
The potential of the original M-theory model \rf{useful} has a flat direction at $T^i=T^j$, it is
\be
T\equiv T^1=T^2=T^3=T^4=T^5=T^6=T^7 \ ,
\ee
where   $T=t+{\rm i}a$. The full potential  vanishes for any values of $t$ and at $a=0$.

In our general cosmological models the flat direction is now lifted due to dependence on $T$ via $F (T, \bar T)$ in eq.~\rf{general}. At $T^i=T^j= \bar T^i$, which is a minimum of our potential in   \rf{general}, one has
\be
Q=\mathcal{W}^{\rm oct}=0\ ,\qquad  \partial_i Q= \partial_i  \mathcal{W}^{\rm oct}=0 \ .
\ee
Thus, the potential becomes a function of the inflaton only and we will have it in the form of the inflationary potential as well as cosmological constant at the exit
\begin{equation}
V|_{T^i=T^j=\bar{T}^i }=F (t, t)=\Lambda + V_{\rm infl}(t).
\label{generalOct1} \end{equation}

Here  we propose an M-theory cosmological model, supplemented by  a nilpotent superfield~$S$ and a gravitino mass $W_0$ with the following choice of the inflationary potential
\be\label{emodpot}
F (T, \bar T)= \Lambda + m^2 (1-T)(1-\overline T) \ ,
\ee
 where the cosmological constant at the exit from inflation is
 \be
\Lambda= F_S^{2} - 3 W_0^2 >0
 \ee
and we choose it to be positive.
\begin{figure}[H]
\centering
\includegraphics[scale=0.45]{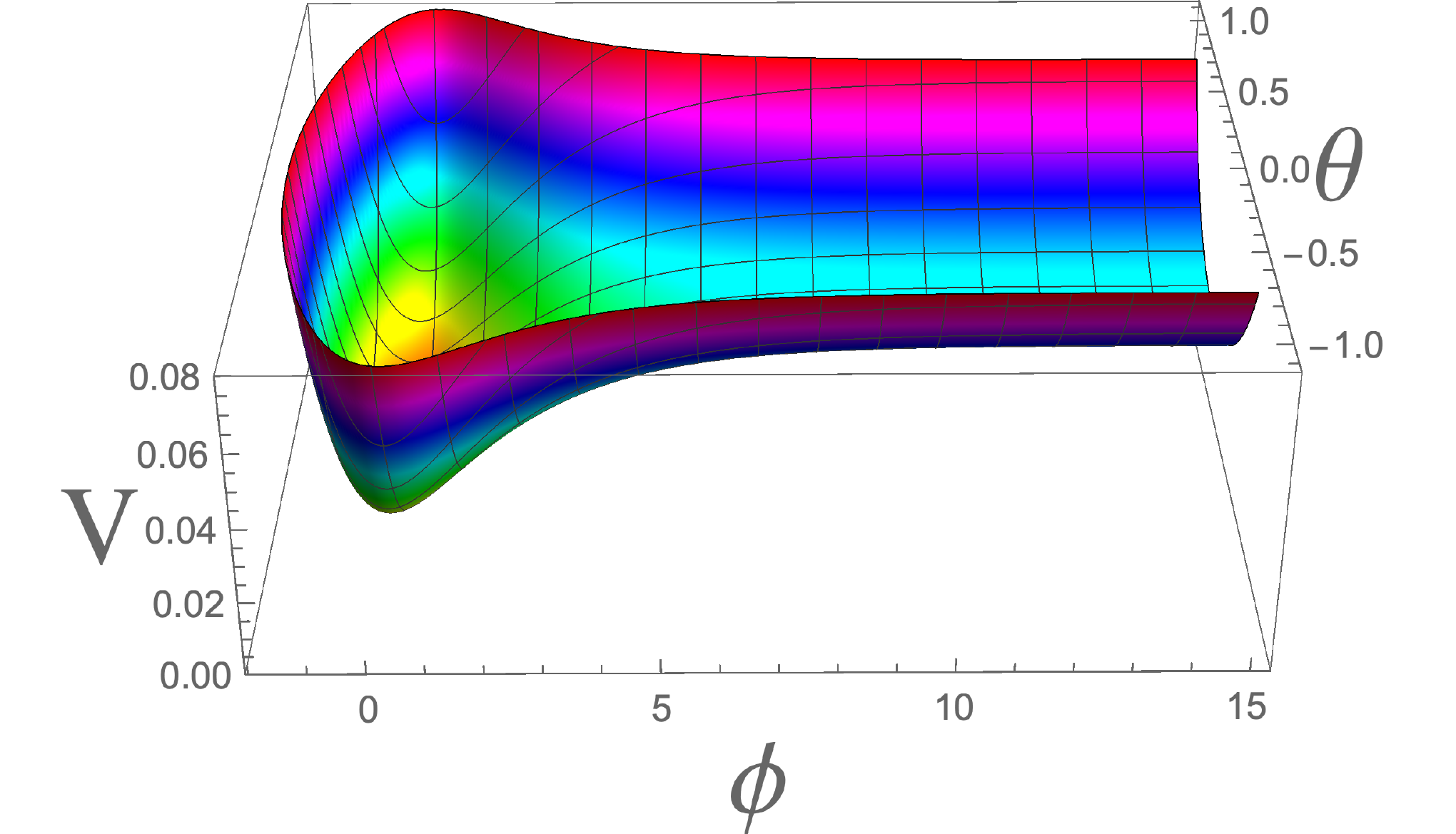}
\caption{\footnotesize  The potential $V_{\rm infl}$ of the E-model with one flat direction for $m = 0.2$, $W_{0} = 0.1$, in Planck mass units. Inflation begins at large $\phi$, and ends when the field reaches the vicinity of the minimum at $\phi = 0$.}\label{f1}
\end{figure}
We take any of the octonionic superpotentials $\mathbb{WO} $  where we have  models with 1 flat direction, in  eqs.     \eqref{cw},  \eqref{cwgen1}, \rf{cwgen2},  or  the octonionic superpotentials in Appendix \ref{secA}. For these cases with one flat direction
 the   effective \K \,   potential is $K= -7 \log (T+\overline T)$, as explained in Sec. \ref{sec5.1}. It is convenient to use the variables $\phi$ and $\theta$:
\be
T =  e^{-\sqrt{\frac{2}{7}}\phi}\,\Big(1  + i \sqrt{\frac{2}{7}}\, \theta\Big) \ .
\ee
Here $\phi$ is a canonical inflaton field, and $\theta$ has a canonical normalization in the vicinity of $\theta = 0$, which corresponds to the minimum of the potential with respect to $\theta$. The full shape of the potential is shown in Fig. \ref{f1}.

 The potential of the canonically normalized inflaton field $\phi$  at $\theta = 0$ in the small $\Lambda$ limit is
\be
V_{\rm infl}= m^2\left(1-e^{-\sqrt{\frac{2}{7}}\phi}\right)^2.
\ee
This is a potential of  {\it the top benchmark target $3\alpha=7$ for B-mode detection} \cite{Ferrara:2016fwe,Kallosh:2017ced,Kallosh:2017wnt} in E-models.

The mass of the axion field $\theta$ in the vicinity of $\theta = 0$ at large $\phi$ approaches the asymptotic value
\be
m^{2}_{\theta} = 2 m^{2} + 4W_{0}^{2} \approx 6 H^{2} + 4W_{0}^{2} \ ,
 \ee
where the Hubble constant at large $\phi$ is given by $V_{\rm total}/3 = m^{2}/3$. Thus  $m^{2}_{\theta} \gg H^{2}$, which stabilizes this field at its minimum $\theta = 0$.

We have checked that the 7-moduli model described above works well. All extra 13 fields like  ${\rm Im} \, T^i$ and combinations of ${\rm Re} \, T^i $ orthogonal to ${\rm Re} \,T$ have positive masses,\footnote{We have found for some  parameters that these masses are positive. In general, following the analysis in~\cite{Kallosh:2017wnt}, one can add terms with bisectional curvatures, which enforce these masses to be positive.} and roll to their minima, where they vanish and decouple from the inflationary evolution.
It is useful to present this  phenomenological model here explicitly, namely
 \begin{align}
K=&\sum_{i=1}^7-\frac12\log\left(\frac{(T^i+\bar{T}^i)^2}{4T^i\bar{T}^i}\right)+\frac{F_S^{2}}{F_S^{2}+m^2 (1-T)(1-\overline T) }S\bar{S} ,\cr
W=& \mathcal{W}^{\rm oct}(T^i)+W_0 + F_{S} \, S.
\label{Emodel}\end{align}
To recover  the original M-theory model  from this phenomenological  cosmological model we need to remove the nilpotent superfield contribution, as well as to remove the supersymmetry breaking parameters $F_S$ and $W_0$, and inflationary parameter $m^2$
\be
S=0\, ,  \qquad F_S=0\, ,  \qquad W_0=0\, ,  \qquad m^2=0.
\ee
This brings our phenomenological model \rf{Emodel} to the M-theory model in \rf{useful} which, in turn, is equivalent to the one in \rf{Guian}, which is directly derived from M-theory via compactification  on a twisted 7-tori,  ${\mathbb{T}^7/ \mathbb{Z}_2^3}$.

We  stress here that the model presented in \rf{Emodel} has a very simple relation to M-theory and is in agreement with the current cosmological data. Moreover, they will be tested by the future B-mode searches.

{ \it $3\alpha=6,5,4,3,2,1$ E-models }

Here we study M-theory models with Minkowski vacua with two flat directions, as described in sec. \ref{sec5.2}.  We start there with the superpotential associated with the cyclic  Hamming (7,4) code in Fig. \ref{fig:Planat}.

 In this sec. \ref{sec5.2} we have models which originate from CSC  octonions and the cyclic Hamming (7,4) code. The
truncated superpotentials in eqs. \rf{61},  \rf{52}, \rf{43}  split the 7 moduli into groups of 6 and 1,  5 and 2, and 4 and 3.  It is convenient now to give the corresponding models the name $m+n$ models where $m+n=7$. Choosing the superpotential
\be
\mathcal{W}^{\rm oct}_{m,n}  (T^i)= {1\over \sqrt{\prod_{i=1}^7 (2 \, T^i)}}  \mathbb{WO} _{m,n}
 \ee
 leads to the effective K\"ahler potential for moduli directions
 \be
 K= - m \log\left(  T_{(1)} + \overline{T}_{(1)}\right) -  n \log\left(  T_{(2)} + \overline{T}_{(2)}\right)
 \ee
with 3 cases
\be\label{comb}
1+ 6\, , \qquad 2+5\, ,  \qquad 3+4 \ .
\ee
From this perspective, the case discussed in the previous section may be called $0+7$. We now use the same form of the \K\, potential  as in eq. \rf{Kal}, however, the function $F$ in $G_{S\bar S}$ depends on 2 flat directions.
\be
G_{S\bar{S}}= {F_S^{2}\over F_S^{2}+  V_{\rm infl}(T_{(1)},\overline T_{(1)},T_{(2)}, \overline T_{(2)})  }.
\label{GSS}\ee
The superpotential $\mathbb{WO}_{m,n} $ is now based on one of the superpotentials  \rf{61},  \rf{52} and in eq. \rf{43}. These superpotentials at the vacuum have the following properties:   $  \mathbb{WO}_{m,n} =0$ and $\partial _i  \mathbb{WO}_{m,n} =0$  at $T^1= \dots =T^m\equiv T_{(1)}$,  $T^{m+1}= \dots =T^n \equiv T_{(2)}$. Our choice of the function $F $ is
\be
F = \Lambda + V_{\rm infl} \ .
 \ee
Here $V_{\rm infl}$ is the phenomenological potential describing two inflaton fields, $T_{(1)}$ and $T_{(2)}$, corresponding to the two flat directions. As before, we will consider the simplest quadratic potential for each of the fields, with a minimum at  $T_{(i)} = c_{i}$, where $c_{i}$ are some real numbers:
\be\label{pott}
V_{\rm infl} = m_{1}^{2}(c_{1}-T_{(1)})(c_{1}-\overline T_{(1)}) + m_{2}^{2}(c_{2}-T_{(2)})(c_{2}-\overline T_{(2)}).
\ee
Importantly, all qualitative results will remain the same for a broad choice of such phenomenological potentials, as long as they are not singular in the limit $T_{(i)} \to 0$, and have a minimum at $T_{(i)} = c_{i}$.

The fields $T_{(i)}$ can be represented in terms   of the canonically normalized fields $\phi_{i}$, such that $T_{(i)}  = e^{-\sqrt{\frac{2}{3\alpha_{i}}}\phi_{i}}$. The corresponding potential becomes
\be
V_{\rm infl}  = \sum_{i} m_{i}^{2}\left(1-e^{-\sqrt{\frac{2}{3\alpha_{i}}}\phi_{i}}\right)^{2}.
\ee
Here without loss of generality we absorbed the constants $c_{i}$ into a redefinition (shift) of the fields $\phi_{i}$. One can take  any of the three combinations of $i$ mentioned in \rf{comb}: 1 and 6,  2 and 5, or 3 and 4.

\begin{figure}[H]
\centering
\includegraphics[scale=0.4]{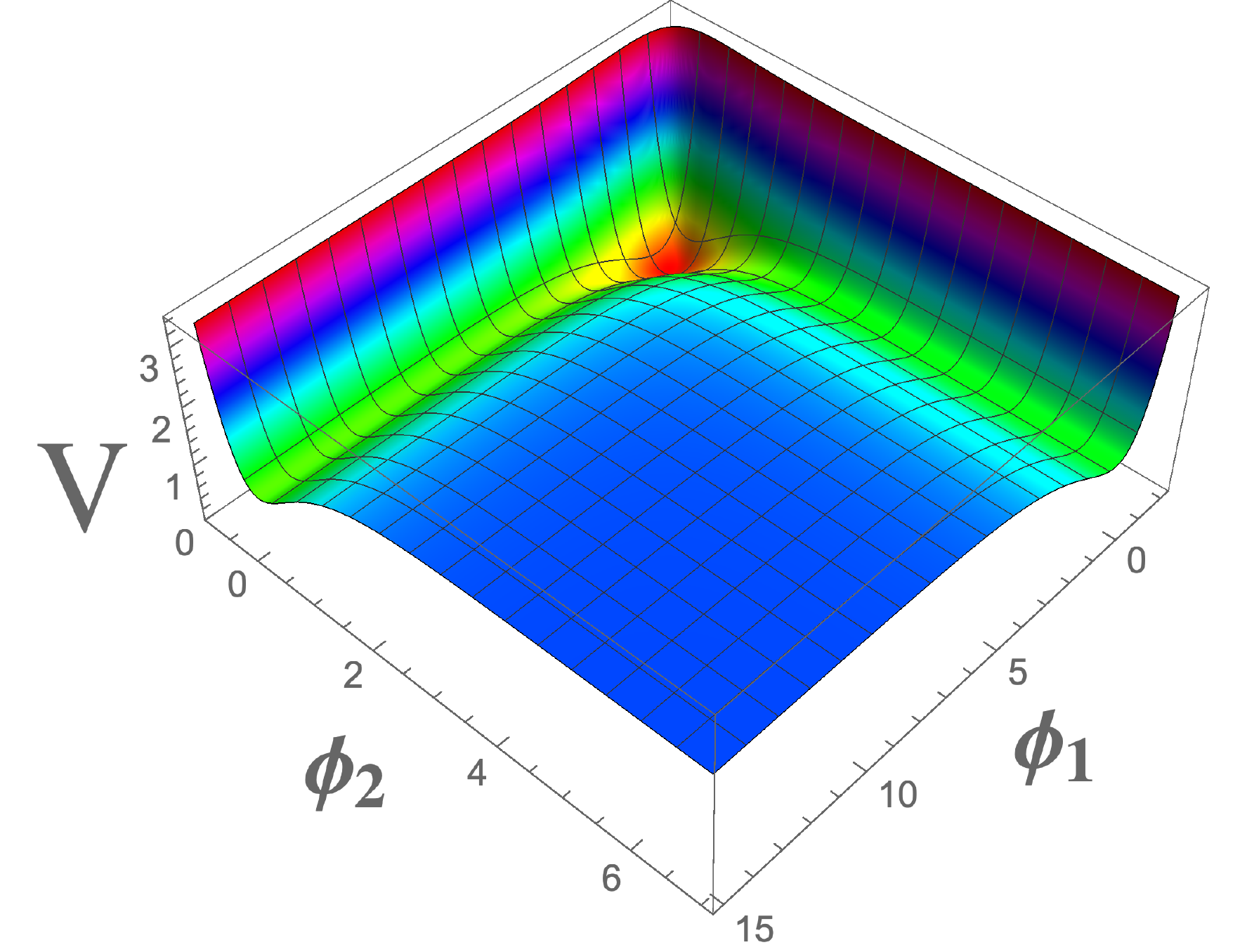}
\caption{\footnotesize  The potential $V_{\rm infl}$ of the E-model with two flat directions, $\phi_{1}$ and $\phi_{2}$, with $3\alpha_{1}= 6$ and $3\alpha_{2}= 1$. }\label{EE}
\end{figure}

The resulting potential has two inflationary valleys, which merge at the minimum of the potential at $\phi_{i} = 0$, see Fig. \ref{EE}. The fields may start their evolution anywhere at the blue inflationary plateau, but then fall to one of the green valleys due to a combination of classical rolling and quantum diffusion. This is a specific version of the process of cascade inflation studied in \cite{Kallosh:2017ced, Kallosh:2017wnt}.

 As a result, the inflationary universe may become divided into many different exponentially large parts, with inflationary perturbations corresponding to one of the two different $\alpha_{i}$.  The particular example shown in Fig.  \ref{EE} corresponds to the split of the 7 moduli into groups of 6 and 1 discussed above, and the universe divided into exponentially large parts   with $3\alpha_{1}= 6$ and $3\alpha_{2}= 1$.

 If we would consider models with flat directions 2 and 5, we would get  $3\alpha_{1}= 5$ and $3\alpha_{2}= 2$. Finally, for the combination 3 and 4 we would get  $3\alpha_{1}= 4$ and $3\alpha_{2}= 3$.
  All of these cases together give us a choice of inflaton potentials
\begin{equation}
V=  \Lambda + m^2\left(1-e^{-\sqrt{\frac{2}{n}}\phi}\right)^2 \ , \qquad n=6,5,4,3,2,1 \ .
\label{Vinf}\end{equation}

If we now add the model 0+7, we get all the benchmarks covered,  using the octonion superpotentials in M-theory
\begin{equation}
V=  \Lambda + m^2\left(1-e^{-\sqrt{\frac{2}{3\alpha}}\phi}\right)^2, \qquad 3\alpha=7,6,5,4,3,2,1 \ .
\label{tot}\end{equation}

\subsection{  T-models } \label{tmodel}
Here we will use the Cayley transform  to switch from the half plane variables $T^i$  to the disk variable $Z^i$ as shown in \cite{Carrasco:2015uma}
\begin{equation}
T^i=\frac{1+Z^i}{1-Z^i}.
\end{equation}
In $3\alpha=7$ case we  define using eq. \rf{useful}
\be
 \mathcal{K}(Z^i, \overline Z^i) =-{1\over 2} \sum_{i=1}^7\log \left(\frac{(T^{i}+\overline{T}^i)^2}{4T^{i}\overline{T}^{i}}\right)= -{1\over 2} \sum_{i=1}^7\log \left(\frac{(1-Z^i \overline Z^i)^2}{(1-Z^i)^2(1-\bar{Z}^i)^2} \right) \ ,
\label{useful0}\ee
and
\be
 \mathcal{W}^{\rm oct} (Z^i) \equiv \mathcal{W}^{\rm oct} \Big (T^i (Z^i)\Big)  \ .
\label{useful1}\ee
The minimum at $T^i=T^j$ becomes the minimum at $Z^i=Z^j$ since $T^i-T^j = 2{Z^i-Z^j\over (1-Z^i) (1-Z^j)}$. The flat direction is at  $Z\equiv {1\over 7} (Z^1+\dots + Z^7)$. We define the T-models as follows
 \begin{align}
K=&-{1\over 2} \sum_{i=1}^7\log \left(\frac{(1-Z^i \overline Z^i)^2}{(1-Z^i)^2(1-\bar{Z}^i)^2} \right)+\frac{F_S^{2}}{F_S^{2}+V_{\rm infl}  }S\bar{S} ,\cr
W=& \mathcal{W}^{\rm oct}(Z^i)+W_0 + F_{S}\, S.
\label{top71}\end{align}
Here we take the following value of the inflationary potential
\be
  V_{\rm infl}(Z, \overline Z)=m^2 Z \overline Z  .
\ee
The total potential for the canonically normalized inflaton now is
\begin{equation}
V_{total}=  \Lambda + m^2\tanh^2 {\frac{\phi}{\sqrt {14}}} \ ,
\label{top}\end{equation}
which is the top benchmark for B-mode detection in the $\alpha$-attractor T-models.
\begin{figure}[H]
\centering
\includegraphics[scale=0.35]{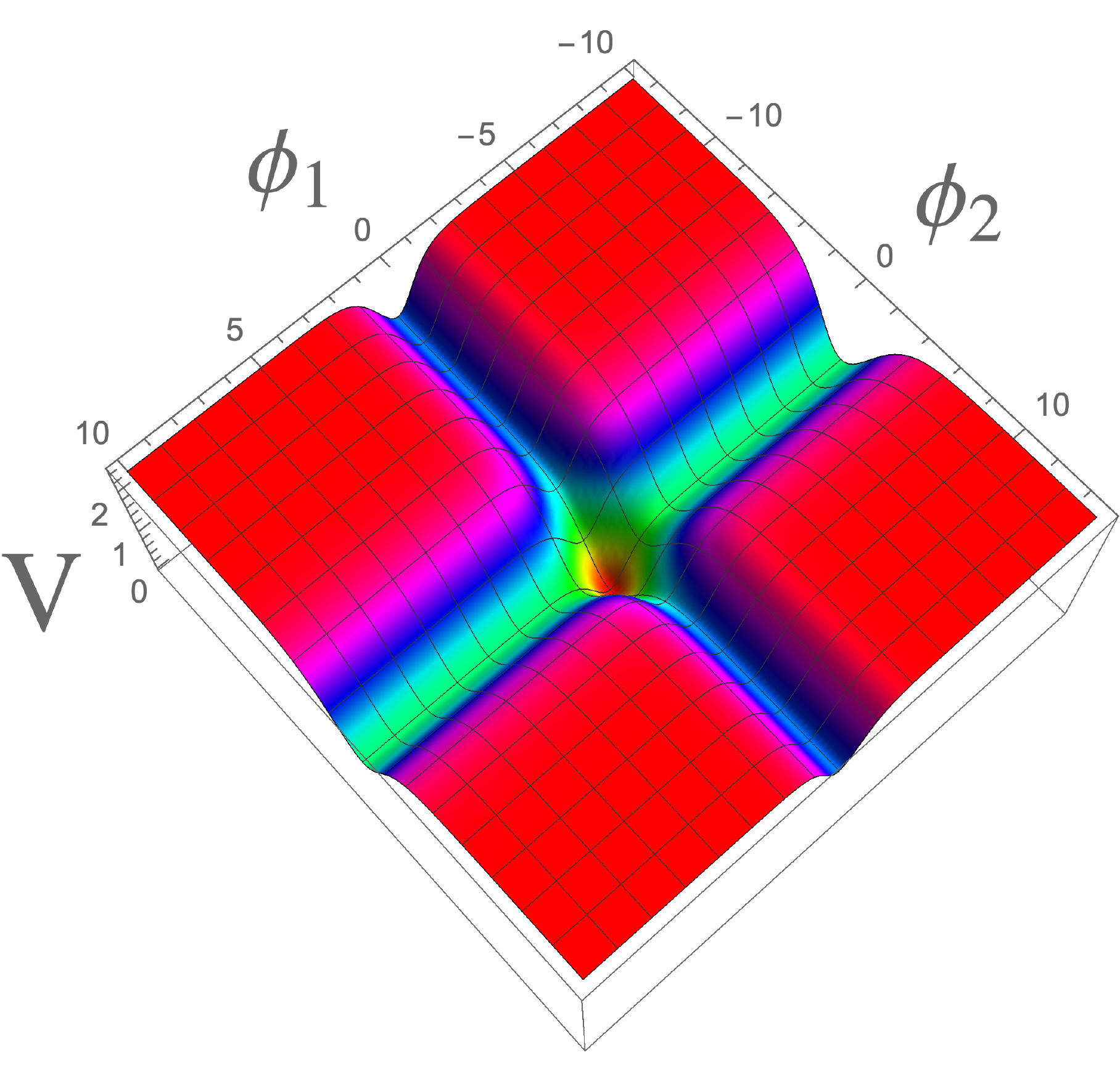}
\caption{\footnotesize  The potential $V_{\rm infl}$ of the T-model with  flat directions  $\phi_{1}$ and $\phi_{2}$, with $3\alpha_{1}= 1$ and $3\alpha_{2}= 6$. }\label{TT}
\end{figure}

The cases with $3\alpha=6,5,4,3,2,1$ can be obtained by analogy with E-models, using   different $W$'s with 2 flat directions $Z_{(1)}$ and $Z_{(2)}$, with
\be
  V_{\rm infl}= m_{1}^2  Z_{(1)} \overline Z_{(1)}+ m_{2}^2  Z_{(2)} \overline Z_{(2)}\ .
\ee
Here one can take any of the two combinations of flat directions mentioned in \rf{comb}: 1 and 6,  2 and 5, or 3 and 4, as we did in \rf{pott}.  In terms of canonical variables $\phi_{1}$ and $\phi_{2}$, this yields the potential
\begin{equation}
V_{\rm total}=  \Lambda +  m^2_{1}\tanh^2 {\sqrt {1 \over 6\alpha_{1}}}\,  \phi_{1}+   m^2_{2}\tanh^2 {\sqrt {1 \over 6\alpha_{2}}}\,  \phi_{2}\, .
\label{top2}\end{equation}
Including the case above $3\alpha=7$  this covers all T-model benchmark targets: $3\alpha= 7,6,5,4,3,2,1$.

Just as for the E-models, we illustrate  in Fig. \ref{TT} for a particular potential with $3\alpha_{1}= 1$ and $3\alpha_{2}= 6$.

One can also consider more complicated models, where the 2 different fields $Z_{({i})}$ corresponding to the 2 flat directions can interact with each other. We assume that these phenomenological interaction terms are much smaller than the typical terms appearing in the original M-theory potential, so they do not affect the structure of the flat directions of the M-theory. However, these terms may force  the two fields $Z_{({i})}$ corresponding to the two flat directions evolve simultaneously \cite{Kallosh:2017ced,Kallosh:2017wnt}.

\begin{figure}[H]
\centering
\includegraphics[scale=0.4]{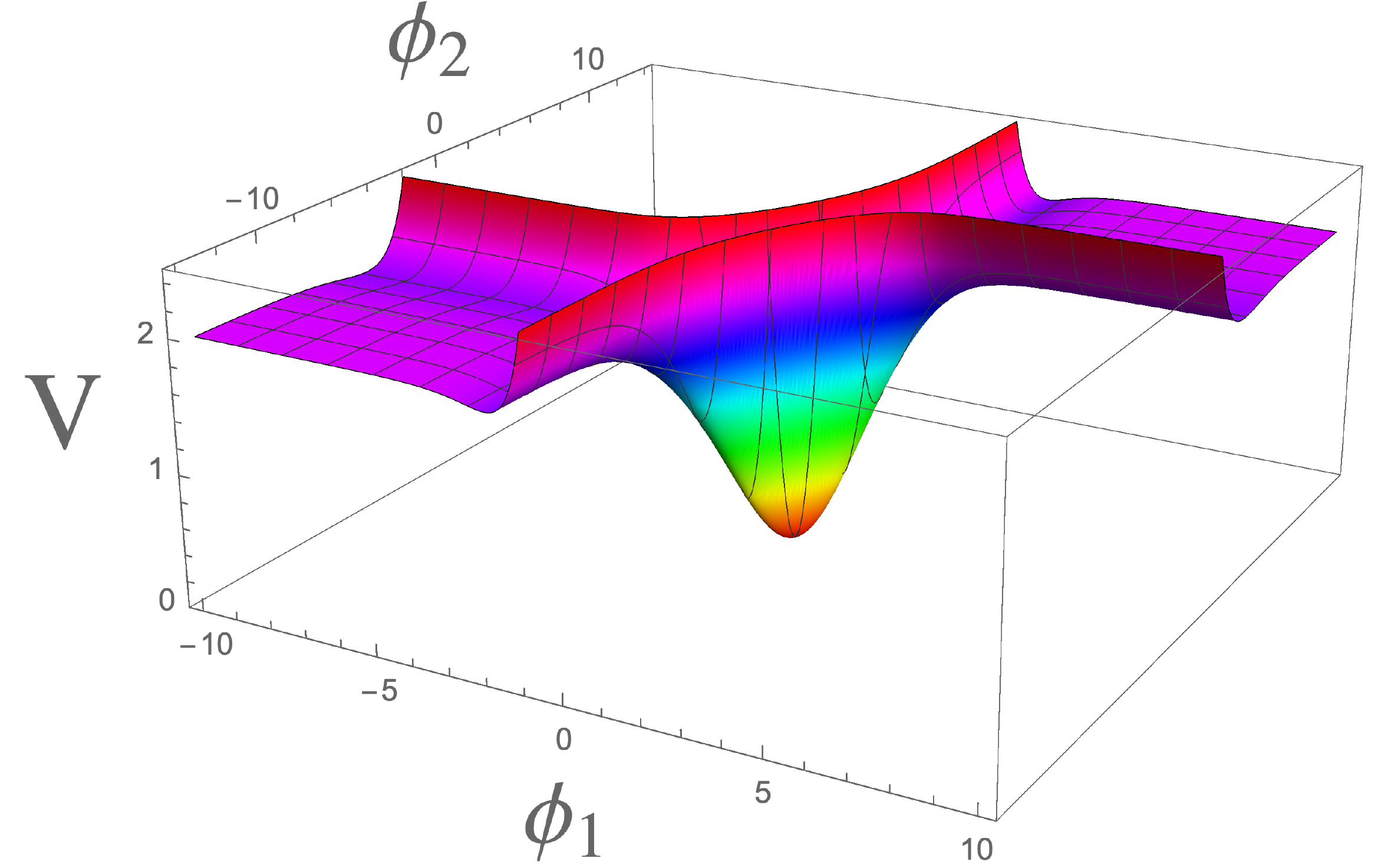}
\caption{\footnotesize  The potential $V_{\rm infl}$ of the T-model \rf{MERGER} with two merging flat directions, with $3\alpha_{1}= 1$ and $3\alpha_{2}= 6$ and $\mu = 3 m$, in terms of the canonical variables $\phi_{1}$ and $\phi_{2}$. }\label{TTM}
\end{figure}

For example, consider a model with the following phenomenological potential along the flat directions $Z_{(1)}$ and $Z_{(6)}$  in terms of the canonical variables $\phi_{1}$ and $\phi_{2}$  with $3\alpha_{1}= 1$ and $3\alpha_{2}= 6$:
\be \label{MERGER}
  V_{\rm infl} =   \Lambda+ m^2  \left(\tanh^2  { \phi_{1}\over \sqrt 2}  +\tanh^2   {\phi_{2}\over \sqrt{12}}\right)+ \mu^{2}  \left(\tanh   { \phi_{1}\over \sqrt 2}  -\tanh    {\phi_{2}\over \sqrt{12}}\right)^{2}.
\ee
This potential, for small $\Lambda$ and $\mu = 3m$, is shown in Fig. \ref{TTM}.

At large $\phi_{i}$ the potential has two flat directions, with $3\alpha_{1}= 1$ and $3\alpha_{2}= 6$, just as in Fig.
\ref{TT}. In our case, they correspond to the dark purple valleys in Fig. \ref{TTM}. At large $\phi_{i}$, these valleys are parallel either to the axis $\phi_{1}$, or to the axis $\phi_{2}$.   However, at smaller values of $\phi_{i}$, these two valleys merge into one, with $3\alpha = 3\alpha_{1}+3\alpha_{2} =7$. The motion in this direction describes the later stages of inflation. The greater the value of $M$, the earlier this merger takes place, see the description of a similar regime in  \cite{Kallosh:2017ced,Kallosh:2017wnt}.

This means that for small $\mu$, observational predictions for these model will correspond either to $3\alpha = 1$ or to $3\alpha = 6$, depending on initial conditions, but for large $\mu$ the last stages of inflation will be described by the single field model with $3\alpha = 7$.

A similar result applies also for merger of directions with $3\alpha_{1}= 2$ and $3\alpha_{2}= 5$, or with $3\alpha_{1}= 3$ and $3\alpha_{2}= 4$. In all of these cases the effective value of $\alpha$ after the merger is given by its maximal number $3\alpha = 7$ corresponding to a single flat direction with $Z^{i} = \bar Z^{i} = Z$ for all $i$.

\section{Observational consequences: Inflation and dark energy}\label{sec8}

\noindent The CMB targets for the future B-mode detectors were discussed recently in~\cite{Kallosh:2019hzo},
in particular   with regard to $\alpha$-attractor inflationary models~\cite{Kallosh:2013yoa}. These models lead to 7 discrete benchmark points~\cite{Ferrara:2016fwe,Kallosh:2017ced,Kallosh:2017wnt} for inflationary observables $n_s$, tilt of the spectrum,   and $r$, a ratio of tensor to scalar fluctuations during inflation.

\begin{figure}[!h]
\vspace{-1mm}
\hspace{-3mm}
\begin{center}
 \includegraphics[scale=0.335]{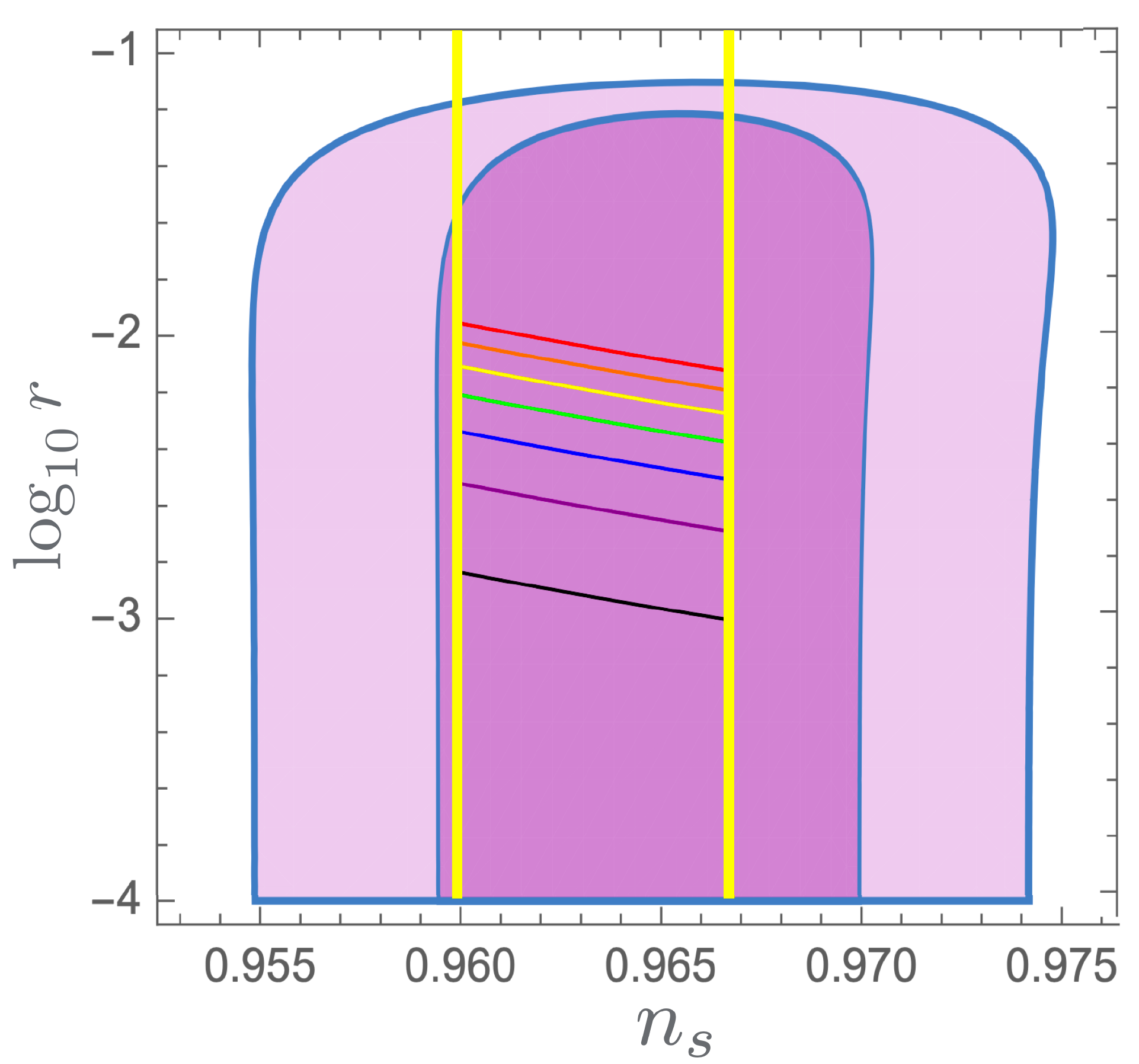}  \hskip 30pt
\includegraphics[scale=0.34]{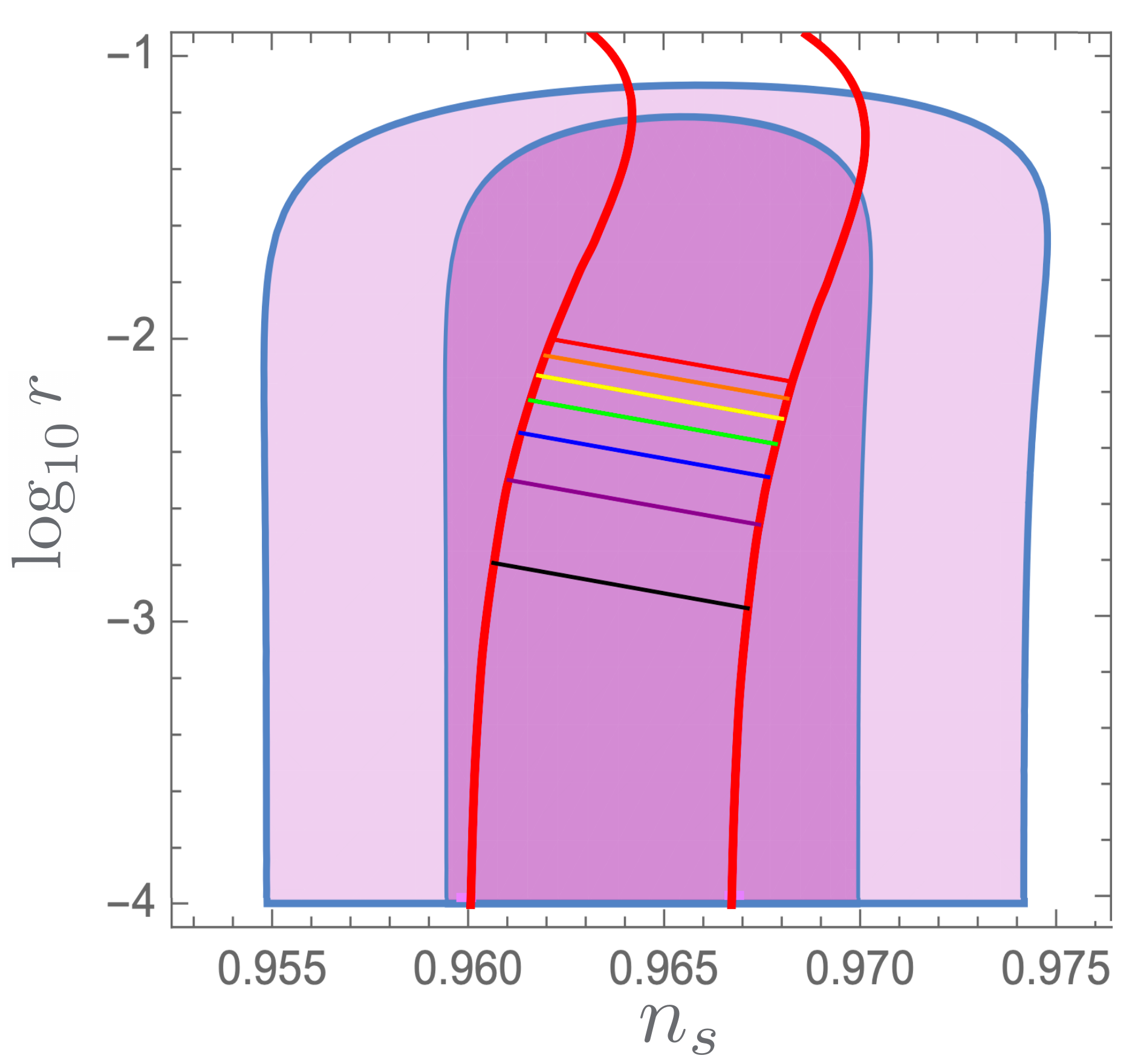}
\end{center}
\vspace{-.12cm}
\caption{\footnotesize
  $\alpha$-attractor inflationary models  benchmarks originating from  M-theory and octonions, as derived here in sec.~\ref{sec4} and  as plotted in~\cite{Kallosh:2019hzo}. The simplest T-models   derived here in sec.~\ref{tmodel} are shown on the left, the simplest E-models , derived  in sec.~\ref{emodel} are on the right panel. The 7-disk model~\cite{Ferrara:2016fwe,Kallosh:2017ced,Kallosh:2017wnt} allows 7
 discrete values:   $3\alpha=7,6,5,4,3,2,1$.  The predictions are shown for the number of e-foldings in the range $50 < N_{e}< 60$. }
\label{7disk2}
\end{figure}

\begin{figure}[H]
\centering
\includegraphics[scale=1]{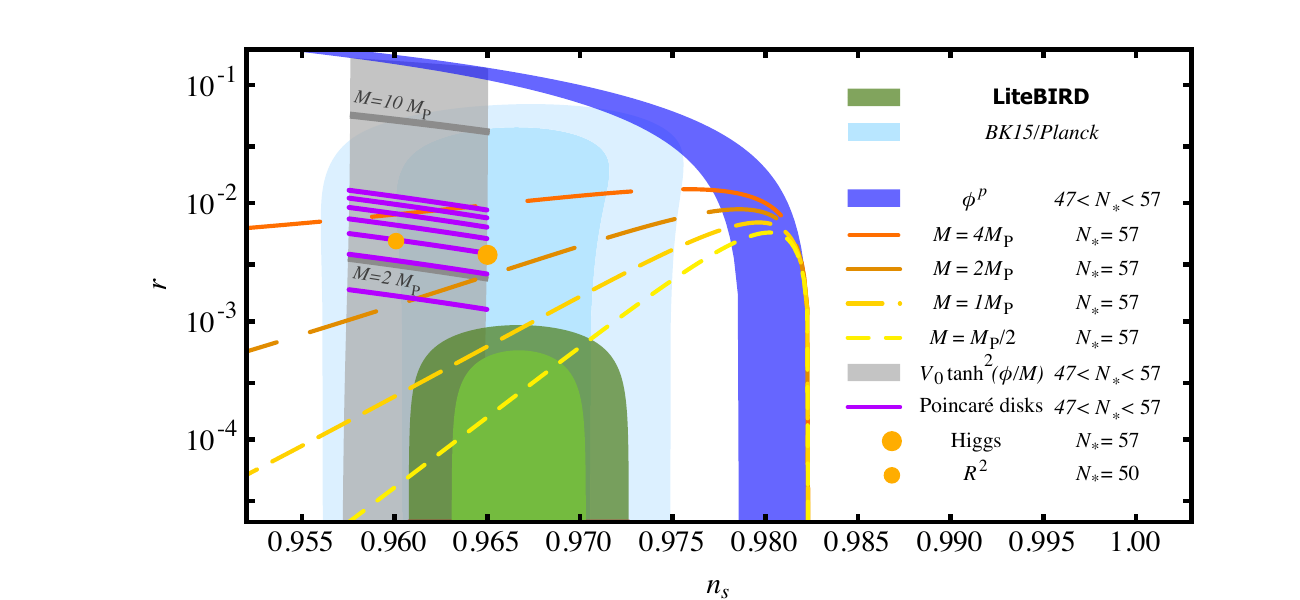}
\caption{\footnotesize  This is a figure A.2 from the Astro2020 APC White Paper~\href{https://ui.adsabs.harvard.edu/abs/2019BAAS...51g.286L/abstract}{LiteBIRD: an all-sky cosmic microwave background probe of inflation} with a forecast of Litebird constraints in the $n_s$ - $r$ plane~\cite{2019BAAS...51g.286L}.
The 7  purple lines in the figure, were derived in this paper using M-theory compactified on $G_2$ holonomy manifolds, octonions, Fano plane and error correcting codes.
}
\label{fig:LB}
\end{figure}

\noindent These benchmark points were  plotted  in the $n_s$ - $r$ plane in  Fig. 9 in~\cite{Kallosh:2019hzo}. We reproduce this figure here in Fig.~\ref{7disk2}.   \be
 n_s\approx 1-{2\over N_{e}} \qquad r\approx 3\alpha {4\over N_{e}^2}.
 \ee
\noindent Here $N_{e}\approx 55$ is the number of e-foldings of inflation. There is no significant dependence on the properties of a potential due to an attractor behavior of the theory.
The future cosmic microwave background (CMB)   missions   \cite{Hui:2018cvg,Ade:2018gkx,Abazajian:2016yjj,Shandera:2019ufi,Ade:2018sbj,Hazumi:2019lys,Hanany:2019lle}      will map polarized fluctuations in the search for the signature of gravitational waves from inflation. In particular,   LiteBIRD is designed to discover or disfavor the best motivated inflation models which were presented in  Fig.~A.2 of  \href{https://ui.adsabs.harvard.edu/abs/2019BAAS...51g.286L/abstract}{LiteBIRD} \cite{2019BAAS...51g.286L},
 which we reproduce here in our Fig. \ref{fig:LB}.
It is important here that in  single modulus  $\alpha$-attractor inflationary models~\cite{Kallosh:2013yoa}
 the prediction for observable $r$  depends only on the parameter $3\alpha$, which enters  in  the \K\, potential as $K=-3\alpha \log (T+\bar T)$.
 
  In this paper we have shown that the upper benchmark for $\alpha$-attractor inflationary E-model  with $3\alpha=7$  has a  simple relation to octonions. The fact that $G_2$ is the automorphism group of the octonions
was known since 1914~\cite{Cartan:1914}. Here we have found that    the octonions give us a powerful tool for constructing superpotentials for compactification of M-theory on manifolds with $G_2$ holonomy and enforcing  $\cN = 1$ supersymmetric minima.

In particular, we have found that it is easy to use
 Cartan-Schouten-Coxeter  conventions~\cite{Cartan:1926,Coxeter:1946},   the corresponding Fano plane for example in Fig.~\ref{fig:Fano1}, and the cyclic Hamming (7,4) error correcting code in Figs.~\ref{fig:Planat}. All these relations, starting with associative octonion triads, codewords, quadruples  and superpotential are shown in eq. \rf{Hamming_CW}. Analogous superpotentials leading to the same cosmological models can be obtained for any of the 480 octonion conventions.

\vspace{-.5cm}

In all these   cases  we found that the 7-moduli model in M-theory compactified on  a 7-tori with $G_2$ holonomy has  a
 Minkowski vacuum with 1 flat direction and the resulting single modulus \K\, potential $K=-7 \log (T+\bar T)$. In such case, the prediction  (when the flat direction is uplifted and the proper cosmological model is constructed in Sec. \ref{sec7}) for $N_{e}\approx  53$ is
 \be
 n_s\approx 1-{2\over N_{e}}\, ,  \qquad r\approx 7 {4\over N_{e}^2}\approx 10^{-2} \ .
\label{3a7} \ee

This is the top red line in Fig. \ref{7disk2} and the top purple line in Fig. \ref{fig:LB}. If gravitational waves from inflation are detected at the level close to $r\approx 10^{-2}$, one will be able to associate this fact with M-theory cosmology and $G_2$ symmetry of octonions.

The benchmarks below the top one, with $3\alpha=6,5,4,3,2,1$ were also studied in this paper.  The corresponding superpotentials were obtained by dropping some terms corresponding to certain codewords in the cyclic Hamming (7,4) code. We have found that the 7-moduli model in M-theory compactified on a 7-tori with $G_2$ holonomy has  a
 Minkowski vacuum with  two flat directions and with one of the following  \K\, potentials:    $K=-6 \log (T_{(1)}+\bar T_{(1)}) -  \log (T_2+\bar T_2)$, $K=-5 \log (T_1+\bar T_1) -  2\log (T_2+\bar T_2)$, and
$K=-4 \log (T_1+\bar T_1) - 3 \log (T_2+\bar T_2)$. It is explained in Sec. \ref{sec6} how the models with $3\alpha=6,5,4,3,2,1$ can be obtained starting with these \K\, potentials. These models have $G_2$ symmetry broken in a specific way, but they still originate from octonionic superpotentials.

To summarize, in in  Sec. \ref{sec7} of this paper we have  built  cosmological models with $3\alpha=7,6,5,4,3,2,1$, which are now directly  related to the 7-moduli M-theory model and 7 octonions with their $G_2$ automorphism symmetry, and (7,4)  Hamming error correcting codes.

We also stress here that some of  these benchmark points with $3\alpha<7$ can have a different origin. For example, the case $3\alpha=3$ is a feature of the Starobinsky model \cite{Starobinsky:1980te}, the Higgs inflation model \cite{Salopek:1988qh, Bezrukov:2007ep}, and  the conformal inflation model  \cite{Kallosh:2013hoa}. The case  $3\alpha=6$ is the feature of the string theory fibre inflation model \cite{Cicoli:2008gp,Kallosh:2017wku}.
The case $3\alpha=1$  is the lowest benchmark point of these 7 targets, the black line in Fig. \ref{7disk2} and the bottom purple line in Fig. \ref{fig:LB}.  This model  is known to represent the maximal superconformal theory in 4 dimensions~\cite{Kallosh:2015zsa,Achucarro:2017ing} with a single complex scalar and $K=- \log (T+\bar T)$. In this case the prediction for $N\approx  60$ is
\be
 n_s\approx 1-{2\over N}\, , \qquad r\approx  {4\over N^2}\approx 10^{-3} \ .
\label{a7} \ee
Meanwhile, in this paper  we  found that the  benchmark cosmological models with $3\alpha=7,6,5,4,3,2,1$ all have a natural origin in M-theory compactified on a manifold with $G_2$ holonomy. Therefore they are associated with octonions, whose  automorphism group is $G_2$. These benchmarks  are  targets for future detection of primordial gravitational waves.

According to the latest
 Planck results \cite{Planck:2018jri},  $\alpha$-attractor models ~\cite{Kallosh:2013yoa} are in good agreement with data available now, in particular for discrete set of values for $3\alpha = 7,6,5,4,3,2,1$ motivated by maximal supersymmetry. More recently we have learned that the cosmological $\alpha$-attractor models can be used in the context of dark energy models, especially interesting in case that future observational data will show the deviation from $w=-1$ equation of state. See for example \cite{Akrami:2018ylq}, where a dark energy model with $3\alpha=7$ was constructed which predicts a deviation from cosmological constant with the asymptotic equation of state  $w=-0.9$. These models will be tested by future dark energy probes including satellite missions like Euclid.

 Quite recently in \cite{Braglia:2020bym} a proposal was made how to construct early dark energy (EDE) models based on $\alpha$-attractors. These models appear to ease the Hubble tension raised by the discrepancy between low-redshifts distance-ladder measurements and a Hubble constant $H_0$ determined from cosmic microwave background (CMB) data. It has been noticed in  \cite{Braglia:2020bym} that their EDE models allow a range for $\alpha$ which also includes the discrete values  $3\alpha = 7,6,5,4,3,2,1$ motivated by maximal supersymmetry.

It is interesting that in all  $\alpha$-attractor models for inflation, for dark energy and for early dark energy, kinetic terms are always the same, defined by  $K=-3\alpha \log (T+\bar T)$. We discussed in this paper how this feature is derived from M-theory compactified on $G_2$ manifold with maximal supersymmetry spontaneously broken to the minimal $\cN=1$ supergravity  with octonion superpotentials $\mathbb{WO}$.

In the fundamental part of the model with a flat direction, the parameters are $M_{Pl}$ in 4d and the scale of the compactified manifold.
We added to the fundamental models a phenomenological $\cN=1$ supergravity potential $V$ so that these models agree with   the observational data.  The phenomenological plateau potentials slightly deforming a flat direction  have the following energy scale: $V_{\rm infl} \approx 10^{-10}M^4_{Pl}$, $V_{\rm EDE} \approx 10^{-110}M^4_{Pl}$, $V_{\rm DE} \approx 10^{-120}M^4_{Pl}$,
  for inflation, for the  EDE and for the current acceleration caused by dark energy or cosmological constant, respectively. They show the deviation from the core fundamental M-theory at some very small scales. 
  It would be very important to get  the future cosmological data for all these models to see  if octonions might be relevant to cosmology.

\section { Summary of the main  results}\label{sec9}
 We proposed an octonionic  superpotential for the effective 7 moduli $\cN=1$ supergravity associated with M-theory, compactified on a $G_2$ holonomy manifold with K\"ahler potential and superpotential of the form:
 \be
K_{7{\rm mod}}= - \sum_{i=1}^7 \log\left(  T^i + \overline{T}^i\right)\, , \quad   \mathbb{WO}= {1\over 2}  \mathbb{M}_{ij} T^i T^j  =\sum_{i=1}^{7}   T^{i}(T^{i+2}- T^{i+1}) \ .
\label{KW}\ee
The above superpotential is one of the  two linearly independent 14-term superpotentials  in Cartan-Shouten-Coxeter convention given in equation  \rf{cwgen1}.
A significant formal part of the paper explains the derivation of the superpotential for any of the  octonion conventions,  relation between  Fano planes, error correcting codes and superpotentials.

An important fact here  is the existence  of  480  different conventions for octonion multiplication. We have found that 
the  octonion conventions other than Cartan-Shouten-Coxeter convention lead to  superpotentials different from  the ones in eq. \rf{KW}. Therefore we have carefully examined these different conventions, described the relations between them and studied how  these different choices affect the  physical results.
We have found that the eigenvalues of the matrix $ \mathbb{M}_{ij} $ defining the superpotential are the same  for all  conventions, reflecting simply the fact that they differ from   \rf{KW} by a change of variables consistent with the symmetries of the supergravity Lagrangian.   Accordingly, the vacuum structure resulting from our octonion superpotentials is the same for all 480 octonion conventions. This universality of the physical results, despite apparently different  superpotentials for various octonion conventions, is one of the nontrivial results of our investigation.

The model in eq. \rf{KW} as well as the ones for all other 480 octonion conventions have a Minkowski vacuum with one flat direction. It is the basis for the cosmological models constructed in Sec. \ref{sec7}. When we use  error correcting codes to drop some terms from the superpotential, we find Minkowski vacua with two flat directions. All these models after the phenomenological potential is added lead to  observable predictions, associated with the set of B-mode targets for $\alpha$-attractors.

The new feature of our  cosmological models is the fact that the total potential with octonionic superpotentials along the inflaton direction with $\mathcal{W}^{\rm oct} = \partial_i  \mathcal{W}^{\rm oct} = 0$,  $T = \bar T$ is given by
\begin{equation}
V=F (T, \bar T) =  \Lambda + V_{\rm infl}(T, \bar T) 
\label{generalinfl1}
\end{equation}
This is a remarkable simplification. The inflaton dynamics  is completely independent on the detailed structure of the full M-theory potential. The only information about this potential that is important for cosmological applications is the identification of its supersymmetric Minkowski flat directions.

Observational predictions of $\alpha$-attractors are very stable with respect to the choice of the phenomenological potential $V_{\rm infl}(T, \bar T)$, they are mostly determined by the M-theory related kinetic terms for the inflaton fields corresponding to these flat directions,
\be
K= - m \log\left(  T_{(1)} + \overline{T}_{(1)}\right) -  n \log\left(  T_{(2)} + \overline{T}_{(2)}\right)
\ee
with $m=0, n=7; m=1, n=6; m=2, n=5; m=3, n=4$. This clarifies the relation between  B-mode targets and M-theory  compactified on a $G_2$ holonomy manifold. Inflation along various flat directions with these kinetic terms leads to  $3\alpha = 7,6,5,4,3,2,1$ and therefore to 7 possible values of the tensor to scalar ratio $r = 12\alpha/N_{e}^{2}$  in the range  $10^{-2}\gtrsim r \gtrsim  10^{-3}$, which should be accessible to  future cosmological observations.


\section*{Acknowledgement}
We are grateful to N. Bobev, A. Braun, G. Dall'Agata, S. Ferrara, F. Finelli, T. Fischbacher, R. Flauger, M. Freedman,  K. Pilch, E. Plauschinn,  A. Van Proeyen,  N. Warner and  T. Wrase  for stimulating discussions. MG, RK, AL  and YY are supported by SITP and by the US National Science Foundation Grant  PHY-1720397, and by the  Simons Foundation Origins of the Universe program (Modern Inflationary Cosmology collaboration). RK and AL  are also supported  by the Simons Fellowship in Theoretical Physics. YY is also supported by JSPS KAKENHI, Grant-in-Aid for JSPS Fellows JP19J00494. MG  and YY would like to thank the hospitality of  SITP where  this work was  initiated.

\appendix

\section{Octonionic  Superpotentials in Common Conventions  }\label{secA}

 In this appendix using the general construction outlined in section 5.1 we will give  superpotentials in various  octonion conventions  involving the structure constants of octonions
such that they can be written as a sum of seven terms  of the form
$$\sum  (T^i -T^j) (T^k -T^l)$$
with each term corresponding to a line  in the Fano plane, and  that lead to a Minkowski vacuum with one modulus. This Fano plane in turn will be related to the codewords in the Hamming code (7,4) as was done for the CSC labelling in  Sec.  \ref{sec4}. 

 The general formula for the superpotential given in equation \ref{superpotentialG} involves the cyclic permutation operator $P$ that generates the cyclic group ${\cal Z}_7$.  For the  Gunaydin-Gursey (GG) \cite{Gunaydin:1973rs} labelling of the real octonions the operator $P$ is
$
P_{GG} = (1243657)
$.
For Cartan-Schouten-Coxeter  (CSC) labelling  \cite{Cartan:1926,Coxeter:1946} the operator $P$ is simply
$
P_{C} = (1234567)
$.  
For the Cayley-Graves labelling of octonions the cyclic permutation operator is
$
P_{CG}=(1245736)
$.

\subsection{ Cartan-Schoutens-Coxeter convention}

Let us now apply this general formalism to the octonions in CSC conventions with cyclic permutation operator $P_{CSC}=(1234567)$.
Let us choose the associative  triad $(124)$ and its permutations to label the superpotentials.
 Applying the rules explained in section 5.1  we get the superpotential
    \bea
   W_{CSC}(124)&=&  \sum_{r=0}^{6} (P_{CSC})^r f_{124} (T^5-T^6)(T^3-T^7) \\ &=& \sum_{n=0}^{6}  f_{n+1,n+2,n+4} f_{n+1,n+3,n +7} f_{n+1,n+5,n+6} (T^{n+3}-T^{n+7})(T^{n+5}-T^{n+6}) \nonumber \label{Cox1}
    \eea
\bea
W_{CSC}(241)&= & \sum_{r=0}^{6} (P_{CSC})^r f_{241} (T^3-T^5)(T^6-T^7) \\ \label{Cox2}
    & =&  \sum_{n=0}^{6}  f_{n+1,n+3,n+7} f_{n+1,n+2,n +4} f_{n+1,n+5,n+6} (T^{n+2}-T^{n+4})(T^{n+5}-T^{n+6}) \nonumber
    \eea
    \bea
     W_{CSC}(412) &=& \sum_{r=0}^{6} (P_{CSC})^r f_{412} (T^5-T^7)(T^6-T^3) \\
    & = &\sum_{n=0}^{6}  f_{n+1,n+5,n+6} f_{n+1,n+2,n +4} f_{n+1,n+3,n+7} (T^{n+2}-T^{n+4})(T^{n+3}-T^{n+7}). \nonumber
    \eea
    Explicitly we have
    \bea
 W_{CSC}(124)& =&  (T^4 - T^1) (T^6 -T^7) + (T^5 - T^2) (T^7 -
      T^1) + (T^6 - T^3) (T^1 - T^2)
      \cr
      && + (T^7 - T^4) (T^2 - T^3) + (T^1 -
     T^5) (T^3 - T^4) \cr
     &&+ (T^2 - T^6) (T^4 - T^5) + (T^3 - T^7) (T^5 - T^6),
      \eea

    \bea
  W_{CSC}(241)&=& (T^2 - T^4) (T^5 -T^6) + (T^3 - T^5) (T^6 - T^7) + (T^4 - T^6) (T^7 -
      T^1)\cr
      && + (T^5 - T^7) (T^1 - T^2) + (T^6 - T^1) (T^2 - T^3) \cr
     && + (T^7 -
      T^2) (T^3 -T^4) + (T^1 - T^3) (T^4 - T^5),
      \eea

\bea
W_{CSC}(412) &=& (T^4-T^1) (T^3 - T^5) + (T^1-T^3)) (T^2-T^6) + (T^5-T^2) (T^4 -
    T^6)\cr && + (T^2-T^4) (T^3-T^7) + (T^6-T^3) (T^5-T^7) \cr
    &&+ (T^1-T^5) (T^7-T^2) + (T^6-T^1) (T^7-T^4).
    \eea
These three superpotentials have one modulus and their spectra are identical
    and  satisfy
   \be
W_{CSC}(124) + W_{CSC}(241) +W_{CSC}(412) =0.
\ee
So far we have preserved the set of associative triads. Some additional superpotentials can be easily obtained when
 the odd permutations of the triad ($124$), namely $ W_{CSC}(142) , W_{CSC}(214) $ and $ W_{CSC}(421) $ is performed.  This operation leads to a specific change of octonion conventions, as we explained in Sec. \ref{sec5}.

 The superpotentials from odd permutations  are however not independent of the ones given above and are related as follows:
 \bea &&W_{CSC}(124) = -W_{CSC}(142), \cr
&&  W_{CSC}(241) = -W_{CSC}(214),\cr
&&  W_{CSC}(412) = -W_{CSC}(421).
    \eea
 If one writes the superpotential $W(kij)$ in terms of a matrix $\mathbb{W}(kij)$ defined by
\be
W(kij) = V^T \mathbb{W}(kij) V
\ee
where $V^T= (T^1,T^2,T^3,T^4,T^5,T^6,T^7)$ one finds that the matrices $\mathbb{W}(kij), \mathbb{W}(ijk) $ and $\mathbb{W}(jki)$  all have the same eigenvalues, namely  one zero eigenvalue and three doubly degenerate non-zero eigenvalues.

\subsection{  Gunaydin-Gursey convention  }

\begin{figure}[H]
\centering
\includegraphics[scale=0.6]{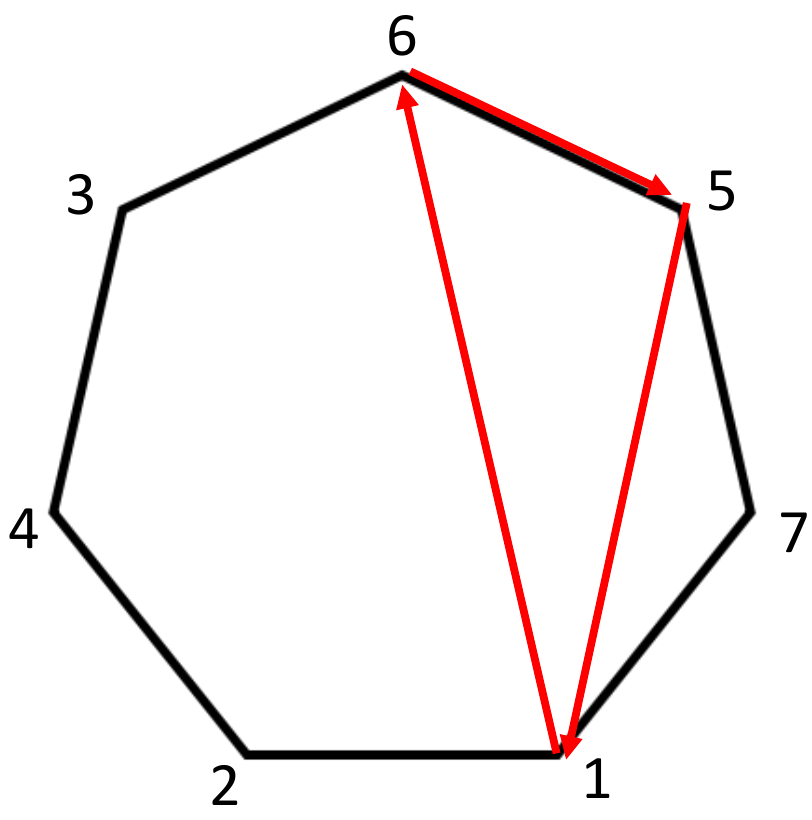}
\caption{\footnotesize The clockwise oriented heptagon for Gunaydin-Gursey octonions. The order is  in agreement with $P_{GG} = (1243657)$. The first triad 123 is shown by red arrows. The second will be 257 as we can see from the figure, etc.}
\label{fig:GG}
\end{figure}

  Consider the octonion multiplication in G\"unaydin-G\"ursey convention \cite{Gunaydin:1973rs} with structure constants in eq. \rf{multGG}.
The cyclic permutation operator in this case is
\be
P_{GG} = (1243657).
\ee
 It defines the order of vertices on the heptagon in Fig. \ref{fig:GG}. Note that given an associative triad , say (165), one can label the superpotential $W_{GG}(165)$ by any of the triads in the orbit of $(165)$ under the action of the Abelian group generated by the cyclic permutation operator $P_{GG}$:
\bea
T1 &=& \{ (165), \, P_{GG}(165)= (257) , \, P_{GG}^2(165)=(471), \, P_{GG}^3(165)= (312), \, P_{GG}^4(165)= (624), \nonumber \\ && \, P_{GG}^5(165)=(543) , \, P_{GG}^6(165)= (736) \}   .\label{triadT1}
\eea
We show this in Fig. ~\ref{fig:GG}.
Given the  associative triad $(165)$ in GG convention  its  associated quadruple is  $(4723)$  and we obtain the superpotential
\bea
WGG1&\equiv & W_{GG}(165)= \sum_{n=0}^6 (P_{GG})^n \{ f_{165} (T^4-T^7)(T^2-T^3) \} \nonumber \\ &=&f_{165} f_{147}f_{123}  (T^4 - T^7) (T^2 - T^3) + f_{257} f_{231}f_{246}  (T^3 - T^1) (T^4 - T^6) \nonumber \\ && +  f_{471} f_{462}f_{435} (T^6 -T^2) (T^3 - T^5) + f_{312} f_{354}f_{367} (T^5 - T^4) (T^6 - T^7)   \nonumber  \\ &&+  f_{624} f_{673}f_{651}(T^7 - T^3) (T^5 - T^1) +  f_{543} f_{516}f_{572} (T^1 - T^6) (T^7 - T^2)   \nonumber \\ && + f_{736} f_{725}f_{714} (T^2 - T^5) (T^1 - T^4).   \label{WGG1}
\eea
Now since all $f_{ijk}$ in cyclic order $(ijk)$  are equal to 1, we have
\bea
 && WGG1\equiv W_{GG}(165)     =   (T^4 - T^7) (T^2 - T^3) +   (T^3 - T^1) (T^4 - T^6) \cr
 \cr
 &&  +  (T^6 -
      T^2) (T^3 - T^5)     +  (T^5 - T^4) (T^6 - T^7)   +  (T^7 - T^3) (T^5 - T^1) \cr
      \cr
      && +  (T^1 - T^6) (T^7 - T^2)   + (T^2 - T^5) (T^1 - T^4).
\eea
One can write the superpotential $WGG1$ as
\be
WGG1 = V^T (\mathbb{MGG}1 ) V
\ee
where $V^T = (T^1, T^2, T^3, T^4, T^5,T^6,T^7)$. The matrix $\mathbb{MGG}1$ is symmetric with zeros along the diagonal and has a sum of terms in each horizontal line vanishing, as necessary for Minkowski vacua.
\be \mathbb{MGG}1 =  \left(\begin{array}{ccccccc}
0&0&1&-1&-1&1&0\\
0&0&-1&0&1&1&-1\\
1&-1&0&0&-1&0&1\\
-1&0&0&0&1&-1&1\\
-1&1&-1&1&0&0&0 \\
1&1&0&-1&0&0&-1\\
0&-1&1&1&0&-1&0
\end{array}\right).
\label{matrixMGG}\ee
The matrix $\mathbb{MGG}1$  has one zero eigenvalue corresponding to a Minkowski ground state  and three non-zero degenerate eigenvalues given by the roots of the cubic equation as discussed in other cases of octonion superpotentials.

By  permuting indices of the triads in  \eqref{triadT1} such that $(kij) \rightarrow (ijk)$  and reordering them  so that the first triad is $(123)$  we obtain another superpotential.
The following set of triads are in  the orbit of $(123)$ under the Abelian group generated by $P_{GG}$
\be
T2 = \{(123) , \, (246), \, (435), \, (367),\, (651), \, (572), \, (714) \}.  \label{triadT2}
\ee
Summing  over the set of triads \rf{triadT2} in the formula \rf{superpotentialG} we get another superpotential, different from $ WGG1 $,  which we label as $WGG2$
\bea
&&WGG2 \equiv W_{GG}(123)=  (T^6 - T^5) (T^4 - T^7) + (T^3 - T^1) (T^5 - T^7) \cr
\cr
&& + (T^6 - T^2) (T^7 - T^1) +
 (T^1 - T^2) (T^5 -
      T^4)+ (T^2 -
      T^4) (T^7 - T^3) \cr
      \cr
      && + (T^1 - T^6) (T^4 - T^3)  + (T^3 - T^6) (T^2 - T^5).
 \label{WGG2}     \eea
One finds that the eigenvalues of  the matrix $\mathbb{MGG}2$ of the superpotential $WGG2$  coincide with those of $\mathbb{MGG}1$.

Similarly by permuting the indices $(kij) \rightarrow (jki)$, in the set \rf{triadT1} and reordering the resulting triads we get another ordered set of triads
\be
T3 = \{(147), \, (231), \, (354), \, (462), \, (516), \, (673) , \, (725) \}  .\label{triads3}
\ee
By summing over the triads \rf{triads3} in \rf{superpotentialG} we get the third superpotential
  \bea
     && WGG3 \equiv  W_{GG}(231) =  (T^4 - T^6) (T^5 - T^7) +(T^6 - T^5) (T^2 - T^3)\cr
      \cr
      && + (T^1 - T^2) (T^6 -
      T^7)    + (T^3 - T^5) (T^7 - T^1)  + (T^7 - T^2) (T^4 - T^3) \cr
      \cr
      &&+ (T^2 -
      T^4) (T^5 - T^1) + (T^3 - T^6) (T^1 - T^4).
 \label{WGG3}     \eea
 Again  one finds that the eigenvalues of the associated matrix  $\mathbb{MGG}3$  are  identical to those of $\mathbb{MGG}1$.

By odd permutation of the associative triads $(kij)\to(ikj)$, $(kij)\to(jik) $ and $ (kij)\to (kji)$  we obtain the following sets of triads which correspond to a different set of octonion notation:
   \bea
  T4 =\{  (132), \, (264),\, (453), \, (376),  \, (615), \, (527), \, (741) \}  \\
 T5 = \{(213), \, (426), \,  (345), \, (637), \, (561), \, (752), \,  (174) \} \\
  T6 = \{ (321), \, (642), \, (534),\, (763), \,  (156), \, (275), \, (417)  \}.
   \eea
Inserting the resulting  reordered triads in \rf{superpotentialG}   one obtains three superpotentials  which we label as $WGG4, WGG5$ and $WGG6$.
   \bea
  && -WGG4 = (T^4 - T^2) (T^3 - T^7) + (T^3 - T^4) (T^6 -
      T^1) + (T^5 -T^6) (T^7 - T^4)\\
     &&  +(T^6 -
     T^3) (T^5 - T^2) + (T^1 -T^7) (T^2 - T^6) + (T^7 - T^5) (T^1 -T^3) +  (T^2 - T^1) (T^4 - T^5) \nonumber \\
     \cr
    &&-WGG5 = (T^6 - T^4) (T^7 - T^5) +   (T^5 - T^3) (T^1 -T^7) + (T^1 - T^5) (T^4 -
       T^2)\\ &&  + (T^7 - T^6) (T^2 - T^1) +  (T^4 -T^1) (T^6 - T^3) + (T^2 -
      T^7) (T^3 -T^4) + (T^3 -T^2) (T^5 - T^6) \nonumber \\
      \cr
   && -  WGG6 = (T^7 - T^4) (T^3 - T^2) + (T^1 - T^3) (T^6 -T^4)+ (T^4 - T^5) (T^7 - T^6) \\ && + (T^2 - T^6) (T^5 -T^3)  + (T^6 - T^1) (T^2 -
      T^7)  + (T^3 - T^7) (T^1 - T^5)  + (T^5 -
      T^2) (T^4 - T^1)  \nonumber
      \eea

      The above superpotentials are however not all independent. It is easy to check that
      \be
      WGG1 +WGG2 +WGG3=0 \quad , \quad
      WGG4+WGG5 +WGG6 =0\ .
      \ee
      Furthermore
      \be
      WGG3=-WGG5\,   , \quad
      WGG2=-WGG4\,  , \quad
      WGG1=-WGG6 \ .
    \label{456}  \ee
Therefore out of six superpotentials defined by a triad and its permutations as given by equation \rf{superpotentialG} only two of them are linearly independent.

\subsection{  Cayley-Graves octonions }

\begin{figure}[H]
\centering
\includegraphics[scale=0.6]{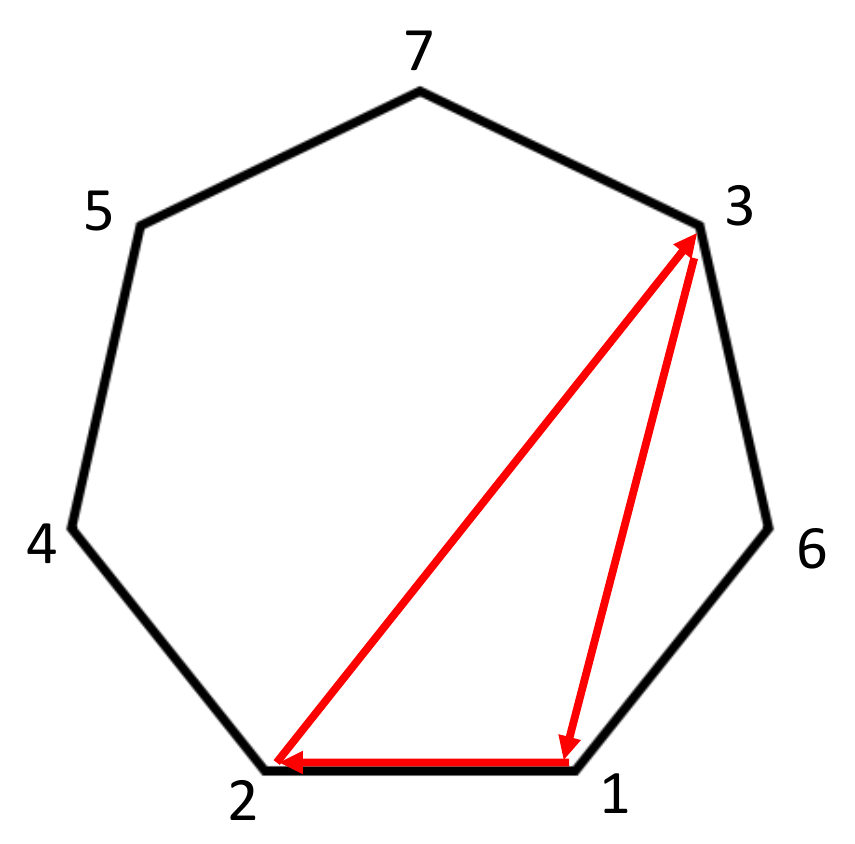}
\caption{\footnotesize The clockwise oriented heptagon for Cayley-Graves conventions. The order is  in agreement with $P_{CG} = (1245736)$. The first triad 123 is shown by red arrows. The second will be 246 as we can see from the figure, etc.}
\label{fig:CG}
\end{figure}

The cyclic permutation operator is
$
P_{CG}=(1245736)
$
Acting on the associative triad $(123)$ one generates the following triads:
$
(123), ( 246), (451), (572), (734), (365), (617)$. We show this in Fig.~\ref{fig:CG}.
With this information at hand we can produce an example of the general formula \rf{superpotentialG} for the  Cayley-Graves octonions:
\bea
&& W_{CG}(123)= \sum_{n=0}^6 (P_{GG})^n \{ f_{123} (T^4-T^5)(T^7-T^6) \} \nonumber \\ &=&f_{123} f_{145}f_{176}  (T^4 - T^5) (T^7 - T^6) + f_{246} f_{231}f_{257}  (T^3 - T^1) (T^5 - T^7) \nonumber \\ && +  f_{451} f_{462}f_{473} (T^6 -T^2) (T^7 - T^3) + f_{572} f_{514}f_{536} (T^1 - T^4) (T^3 - T^6)   \nonumber  \\ &&+  f_{734} f_{761}f_{725}(T^6 - T^1) (T^2 - T^5) +  f_{365} f_{312}f_{347} (T^1 - T^2) (T^4 - T^7)   \nonumber \\ && + f_{617} f_{624}f_{653} (T^2 - T^4) (T^5 - T^3) .  \label{WCG1}
\eea

Explicitly we have 
\bea
&&W_{CG}(123)=   (T^4 - T^5) (T^7 - T^6) +  (T^3 - T^1) (T^5 - T^7)
 +   (T^6 -T^2) (T^7 - T^3) \\
 &&+  (T^1 - T^4) (T^3 - T^6) +  (T^6 - T^1) (T^2 - T^5) +  (T^1 - T^2) (T^4 - T^7)  +  (T^2 - T^4) (T^5 - T^3).  \nonumber \label{WCG2}
\eea
By cyclic permutation of the defining triad one can define another superpotential
labelled by the triad  $(176)$: 
\bea
&&W_{CG}(176)=   (T^2 - T^3) (T^4 - T^5) +  (T^4 - T^6) (T^5 - T^7)
 +   (T^1 -T^2) (T^6 - T^5) \\
 &&+  (T^5 - T^1) (T^7 - T^3) +  (T^7 - T^2) (T^3 - T^6) +  (T^2 - T^4) (T^1 - T^7)  +  (T^3 - T^4) (T^6 - T^1).  \nonumber \label{WCG3}
\eea


 \section{Relations between most commonly used octonion conventions}\label{secB}
We present here  relations between most commonly used octonion conventions.
These can be transferred to relation between moduli in case that there is no sign flip between octonions, since the real part of moduli is positive.

 \bea\label{perm}
\left(\begin{array}{cc}{\rm CSC} & {\rm GG} \\1 & 1 \\2 & 2 \\ 3 &  4 \\4 & 3 \\5 & 6 \\ 6 & 5 \\7 & 7\end{array}\right)   \qquad
\left(\begin{array}{ccc} {\rm RCSC}  & {\rm  CG}   \\1 & 4  \\2 & 6  \\3 & 5  \\4 & 1  \\5 & 7  \\6 & 2  \\ 7 & 3 \end{array}\right)
\eea

In the first group in eq. \rf{perm} we compare  Cartan-Schouten-Coxeter conventions  \cite{Cartan:1926,Coxeter:1946}   with  Gunaydin-Gursey conventions \cite{Gunaydin:1973rs}. In the second  group in eq. \rf{perm1}  we compare  Reverse-Cartan-Schouten-Coxeter notations  with  Cayley-Graves   notations.   Both involve no flip of the sign of any octonions. Therefore we can use these relations to  change  variables in the Cartan-Schouten-Coxeter type superpotential to get the superpotential in Gunaydin-Gursey conventions, and in
Reverse-Cartan-Schouten-Coxeter notations  to get the  Cayley-Graves superpotential.

 \bea
\left(\begin{array}{cc}{\rm OK} & {\rm GG} \\1 & 1 \\2 & 2 \\ 3 &  4 \\4 & 3 \\5 & 6 \\ 6 & 5 \\{\color{red}7 } & {\color{red}-7 }\end{array}\right)   \quad
\left(\begin{array}{cc}{\rm CSC}  &{\rm  CG} \\1 & 1 \\2 & 2 \\3 & 5 \\4 & 3 \\5 & 7 \\6 & 6 \\ {\color{red}7 }& {\color{red} -4}\end{array}\right)
\label{perm1}\eea
In the second group in eq. \rf{perm1} the relabeling of moduli requires the sign change and we do not use if for generating new superpotentials. We can avoid the need to use them since, as we see in examples in eq.  \rf{perm}, we can start with  superpotentials in CSC conventions and generate 240 additional ones, including the Gunaydin-Gursey case, and with superpotential in RCSC conventions and  generate the other group of 240 superpotentials, including Cayley-Graves case, without flipping signs of the moduli.

 \subsection{From Cartan-Schouten-Coxeter to  Gunaydin-Gursey   superpotential} \label{sec3.2}

Here we explain how eqs. \rf{perm},  \rf{perm1} were derived.

The choice of triads in  \cite{Gunaydin:1973rs}
  is
\be
 f_{ijk}^{\rm GG} =+1 \qquad {\rm for}  \qquad \{ijk\} = \{ (123), (435), (516), (624), (572), (471), (673)     \}.
\label{multGG} \ee

\begin{figure}[H]
\centering
\includegraphics[scale=0.6]{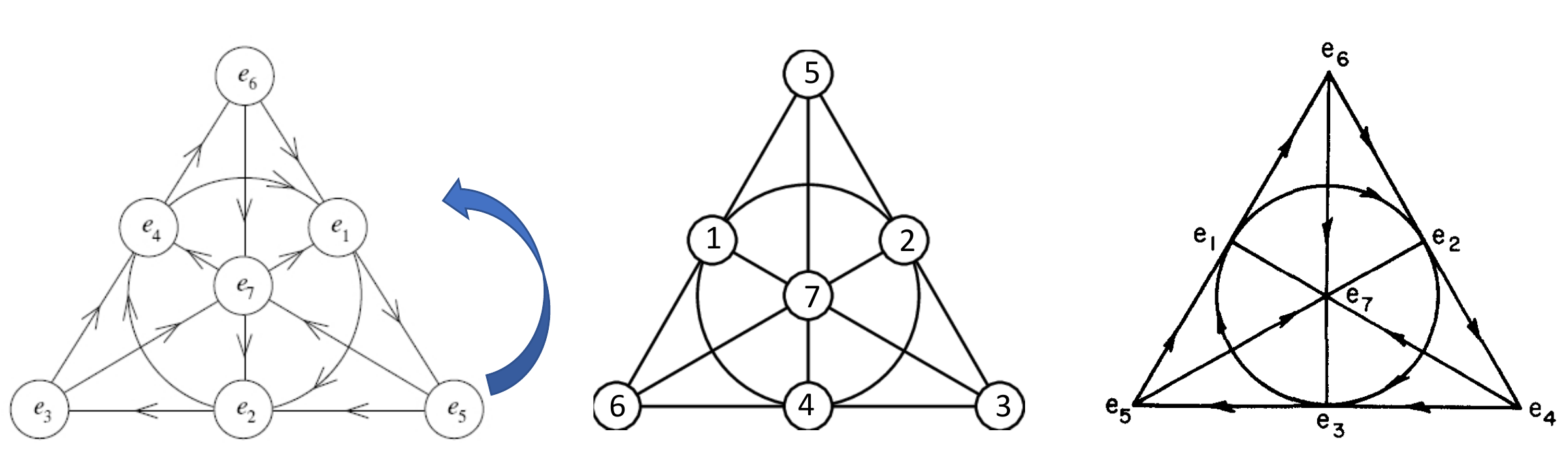}
\caption{\footnotesize  An oriented Fano plane in  \cite{Baez:2001dm} for octonions in \cite{Cartan:1926,Coxeter:1946} is the starting point at the very left of the Figure. We rotate is so that the lower left corner becomes the upper vertex of the triangle, this gives the Fano plane in the middle. The orientation of all lines remains the same. The Fano plane at the  right is the one taken from \cite{Gunaydin:1973rs}. It is clear that  relabeling  the indices as $3\leftrightarrow 4,5\leftrightarrow 6$, we can bring the Fano plane in the middle to the one on the right.
}
\label{fig:Cox_Murat}
\end{figure}
These octonions are related to the ones in \cite{Cartan:1926,Coxeter:1946}, and shown here in eq. \rf{cycl}, by   relabeling
\be
3\leftrightarrow 4, \quad 5\leftrightarrow 6.
\label{GG}\ee
To explain this it is useful to start with the Cartan-Schouten-Coxeter Fano plane in the form given by   \cite{Baez:2001dm} and first rotate it counter-clockwise, as we show in Fig. \ref{fig:Cox_Murat}. The rotated Fano plane, in the middle of the figure,  preserves all orientation of the 7 lines, one can see that  Fano planes at the left and at the right have the same orientation.
Therefore, one  can now  easily compare the Fano plane in the middle  with the Gunaydin-Gursey Fano plane  \cite{Gunaydin:1973rs}, which is at the right of the Fig. \ref{fig:Cox_Murat}.
The superpotential for GG conventions is now derived from $\mathbb{WO}_{\rm cw}$ using the change of variables in eq. \rf{GG} and we find
    \bea\label{WGG}
&&  \mathbb{WO}_{\rm GG}= (T^2 - T^3) (T^5 -T^6) +(T^3 - T^5) (T^6 - T^7) + (T^3 - T^5) (T^7 -
      T^1)\\
 \cr
      && + (T^6 - T^7) (T^1 - T^2) + (T^5 - T^1) (T^2 - T^4) + (T^7 -
      T^2) (T^4 -T^3) + (T^1 - T^4) (T^3 - T^6).  \nonumber
      \eea
      It is clear how to get also $ \mathbb{WO}_{\rm GG}'$ and $\mathbb{WO}_{\rm GG}^{''}$.
\noindent One could have instead used the general formula \rf{superpotentialG} for any set of octonion conventions. In such case, starting from
      \rf{multGG} one can get \rf{WGG} directly. We present this derivation in Appendix \ref{secA}.

          \subsection{From Reverse Cartan-Schouten-Coxeter to Cayley-Graves superpotential }

\begin{figure}[H]
\centering
\includegraphics[scale=0.55]{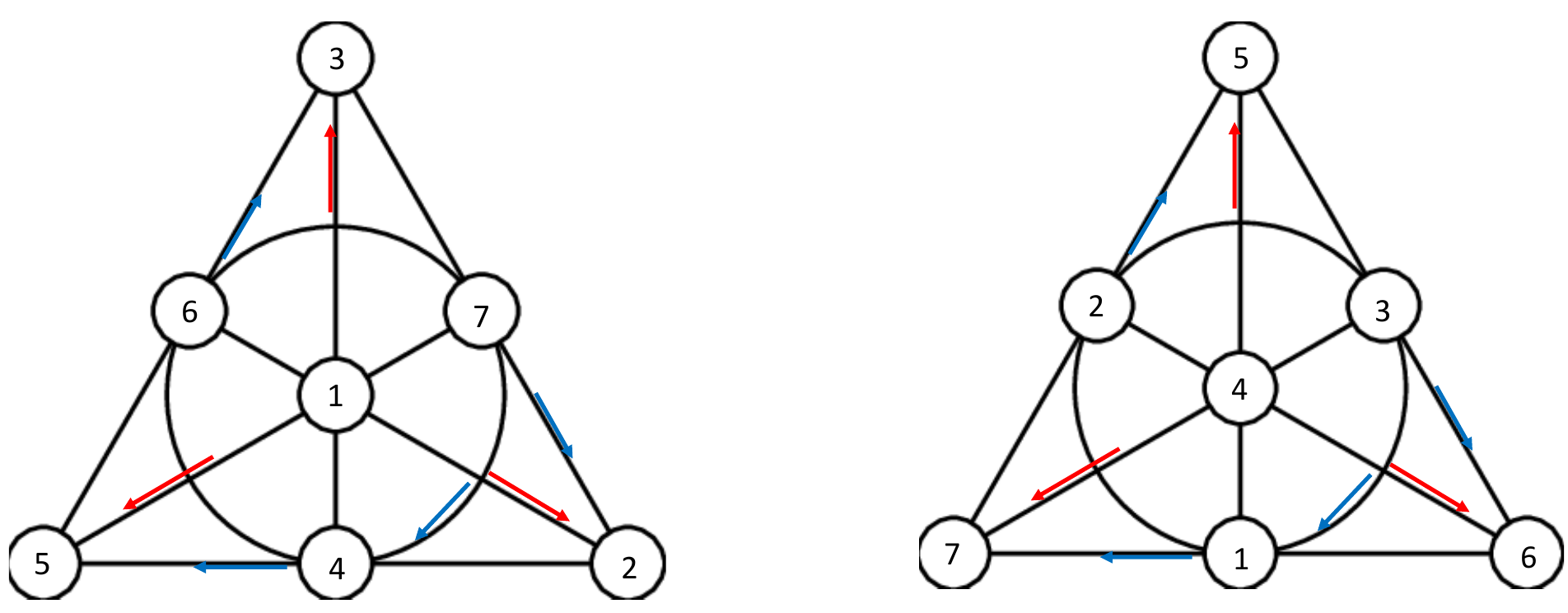}
\caption{\footnotesize  The  Fano plane at the left  is for RCSC triads  126, 237, 341, 452, 563, 674, 715.
At the right we show the oriented  Fano plane for Cayley-Graves octonions, with
123, 145, 624,  653, 725, 734,  176 triads. It has the same orientation for all 7 lines as the one for RCSC.
They are related to each other by the following relabeling from Cayley-Graves to RCSC: 5 to 3, 2 to 6, 7 to 5, 4 to 1, 3 to 7,  6 to 2, 1 to 4.}
\label{fig:New}
\end{figure}

   In Fig. \ref{fig:New} we show the Fano planes for Reverse Cartan-Schouten-Coxeter conventions and for  Cayley-Graves octonion conventions.
These two Fano planes have the same orientation on
all 7 lines. Therefore a relabeling of indices without any sign flips relates these sets of octonions. This relation is
\bea
&&e_7 ^{RCSC}= e_3^{CG},  \qquad e_6 ^{RCSC}= e_2^{CG}, \qquad e_5 ^{RCSC}= e_7^{CG}, \qquad e_4
^{RCSC}= e_1^{CG},  \qquad \cr
\cr
&&  e_3 ^{RCSC}= e_5^{CG} ,  \qquad  e_2 ^{RCSC}= e_6^{CG} ,  \qquad  e_1 ^{RCSC}= e_4^{CG}
\eea

We take our counterclockwise $ \mathbb{WO}_{\rm ccw} $ in eq. \rf{cwgen1}
and replace $1\leftrightarrow 4,  2\leftrightarrow 6,  3\rightarrow 5,  5\rightarrow 7, 7\rightarrow 3$ according to Fig. \ref{fig:New} and we find the Cayley-Graves superpotential
 \bea
&&   \mathbb{WO}_{\rm ccw}|_{RCSC\Rightarrow CG}= T^4( T^6- T^1)+ T^6(T^5-T^7) +T^5(T^1-T^2) +T^1(T^7-T^3) \cr
\cr
&&   +T^7(T^2-T^4) +T^2(T^3-T^6) +T^3(T^4-T^5) \Rightarrow   \mathbb{WO}_{\rm CG}(176)
 \eea
 This is to be compared with the formula in eq. \rf{WCG3} where the corresponding superpotential was obtained directly using CG conventions.

We can also take our counterclockwise $ \mathbb{WO}^{''}_{\rm ccw} $ in eq. \rf{cwgen3}, use the same change of variables and get another Cayley-Graves superpotential
     \bea
    \label{ccwRepl}
&&   \mathbb{WO}^{''}_{\rm ccw}|_{RCSC\Rightarrow CG}=  (T^5 - T^1) T^7  + (T^1 - T^7) T^2
      + (T^7 - T^2) T^3 + (T^2 - T^3) T^4
 \cr
 \cr
      &&    + (T^3 -
      T^4) T^6  + (T^4 - T^6) T^5  +(T^6 - T^5) T^1 \Rightarrow  \mathbb{WO}_{\rm CG}(123).
      \eea
 This is to be compared with the formula in eq. \rf{WCG2} where the corresponding superpotential was obtained directly using CG conventions.

Thus we see here that one can use any of the two ways of identification of the superpotential for Cayley-Graves type octonions and they produce exactly the same results.

\section { More on   octonion multiplication tables and Fano planes}\label{secC}

\noindent A  useful tutorial on  octonions http://www.7stones.com/Homepage/octotut0.html\\
 is presented by Geoffrey Dixon. Some of these properties help us to explain the properties of octonion superpotentials  in our 7-moduli models.

 \begin{table}
\centering
\begin{tabular}{|c||c|c|c|c|c|c|c|}\hline
$ $ & $e_{1}$& $e_{2}$& $e_{3}$ & $e_{4}$& $e_{5}$& $e_{6}$& $e_{7}$\\\hline \hline
$e_{1}$ & $ $& $ $& $e_{7}$ & $ $& $ $& $ $& $ $\\\hline
$e_{2}$  & $ $& $ $& $ $ & $e_{1} $& $ $& $ $& $ $\\\hline
$e_{3}$  & $ $& $ $& $ $ & $ $& $e_{2} $& $ $& $ $\\\hline
$e_{4}$  & $ $& $ $& $ $ & $ $& $ $& $e_{3} $& $ $\\\hline
$e_{5}$  & $ $& $ $& $ $ & $ $& $ $& $ $& $ e_{4}$\\\hline
$e_{6}$  & $e_{5} $& $ $& $ $ & $ $& $ $& $ $& $ $\\\hline
$e_{7}$  & $ $& $e_{6} $& $ $ & $ $& $ $& $ $& $ $\\\hline
\end{tabular} \qquad \begin{tabular}{|c||c|c|c|c|c|c|c|}\hline
$ $ & $e_{1}$& $e_{2}$& $e_{3}$ & $e_{4}$& $e_{5}$& $e_{6}$& $e_{7}$\\\hline \hline
$e_{1}$  & $ $& $e_{6} $& $ $ & $ $& $ $& $ $& $ $\\\hline
$e_{2}$ & $ $& $ $& $e_{7}$ & $ $& $ $& $ $& $ $\\\hline
$e_{3}$  & $ $& $ $& $ $ & $e_{1} $& $ $& $ $& $ $\\\hline
$e_{4}$  & $ $& $ $& $ $ & $ $& $e_{2} $& $ $& $ $\\\hline
$e_{5}$  & $ $& $ $& $ $ & $ $& $ $& $e_{3} $& $ $\\\hline
$e_{6}$  & $ $& $ $& $ $ & $ $& $ $& $ $& $ e_{4}$\\\hline
$e_{7}$  & $e_{5} $& $ $& $ $ & $ $& $ $& $ $& $ $\\\hline
\end{tabular}
\caption{ Table on the left is a $C^{(+)}$ table  associated with the clockwise heptagon to be used later  in  Sec. \ref{sec4.1}. It shows a part of the multiplication table in Fig. \ref{Table:Mult}, where we keep only the information on associative triads, which makes the pattern of multiplication clear. For example, we see that $e_1 e_3=e_7$ etc. On the right there is a table for $D^{(+)}$ associated with the counterclockwise heptagon to be used later  in  Sec. \ref{sec4.2}.}
\label{tab:CD}
\end{table}

From all 480  tables shown in http://tamivox.org/eugene/octonion480/index.html there are 2 favorite ones, which are called $C^{(+)}$ and $D^{(+)}$. Since all 480 tables were computer generated, these favorite ones show up as  $C^{(+)}$, the one which came as a number   406 out of 480 and $D^{(+)}$, as a number 145.
These are dual to each other, each sharing the same elegant properties. If one has
\be
e_a e_b=\pm e_c\, ,  \qquad a,b,c = 1,\dots, 7\, ,  \qquad e_a=e_{a+7}
\ee
it implies the index doubling property
\be
e_{2a} e_{2b}=\pm e_{2c}\, ,   \qquad e_{4a} e_{4b}=\pm e_{4c}
\ee
and the index cycling property
\be
e_{a+k} e_{b+k}=\pm e_{c+k}\, .
\ee
We show in Table \ref{tab:CD} excerpts from $C^{(+)} $ and $D^{(+)}$  that make the pattern of multiplication clear.

$C^{(+)}$ set corresponds to octonion multiplication table in Fig. \ref{Table:Mult}, it is a Cartan-Schouten-Coxeter set \cite{Cartan:1926,Coxeter:1946}, associated with the clockwise heptagon in Sec. \ref{sec4.1}.
After some permutations of octonions without  sign flips it becomes a Gunaydin-Gursey set \cite{Gunaydin:1973rs}.

$D^{(+)}$ set corresponds to a set of octonion conventions which we call Reverse-Cartan-Schouten-Coxeter set, associated with the counterclockwise heptagon in Sec. \ref{sec4.2}. After some  permutations without  sign flips it becomes a Cayley-Graves set
\cite{Hamilton:1848,Cayley:1845} and with other permutations without  sign flips also an Okubo set \cite{Okubo:1990nv}.
The corresponding Cayley-Graves  octonion multiplication table is shown in Fig. \ref{Table:StationQ}.

For our work it is important that starting from $C^{(+)}$ one can get 240 (including  the original $C^{(+)}$) different tables by permutations alone. It is based on the fact that  {\it an even number of sign changes is accessible via permutations alone}.
For example, one can change the signs of some 4 octonions, the remaining 3 belong to associate triads. There are 7 possible changes like that. For octonions in eq. \rf{cycl} we take the 1st triads 124, and flip the signs of 3,5,6,7. But since the triads are not changed, it means that these 4 sign flips can be compensated by some permutations without sign flips. This explains why, given a multiplication table of octonions, the corresponding Fano plane is not unique.

\begin{figure}[H]
\centering
\includegraphics[scale=0.45]{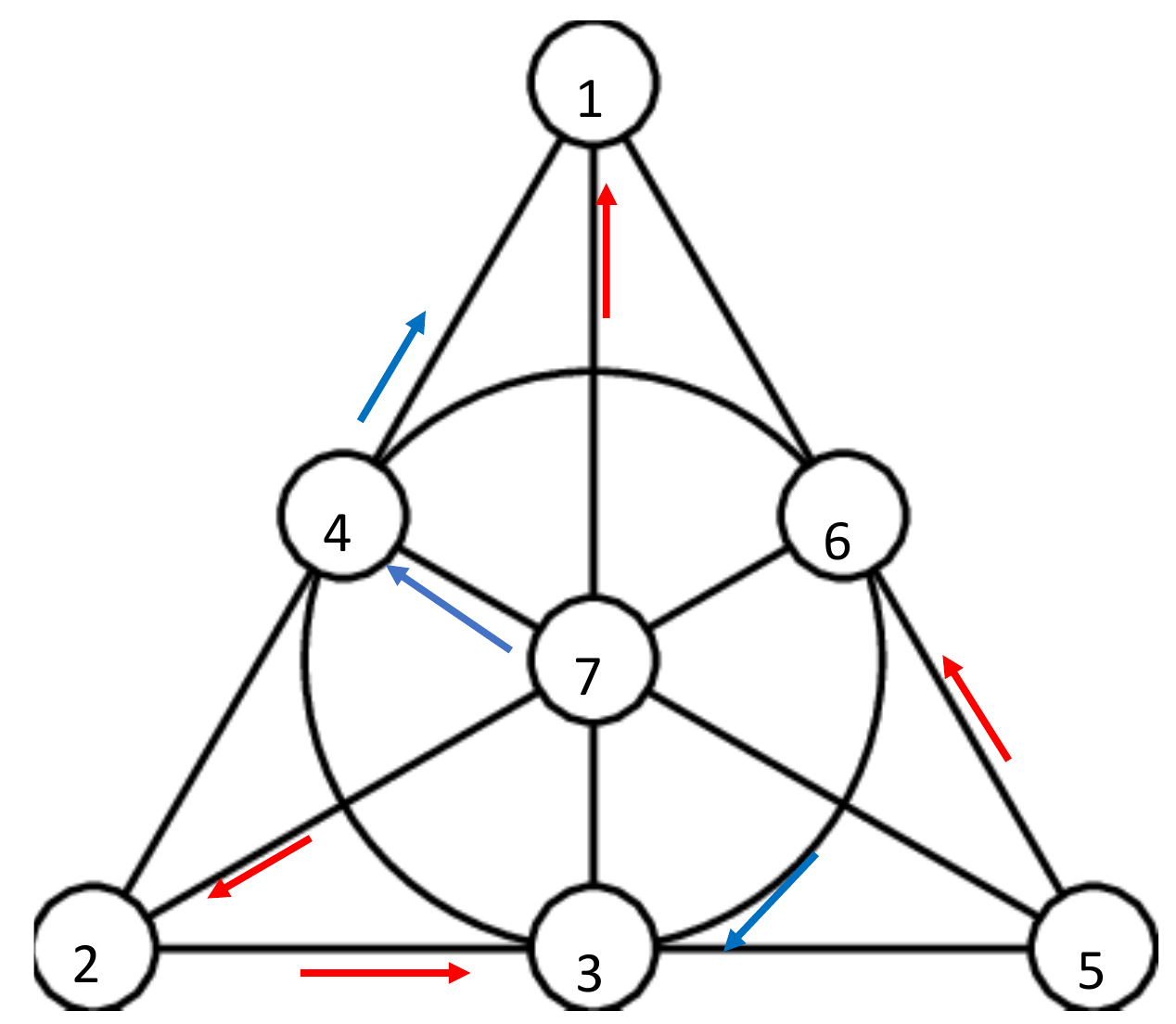}
\caption{\footnotesize  An oriented Fano plane corresponding to   Fig. 1 in  \cite{Coxeter:1946}. The difference with a Fano plane in  \cite{Baez:2001dm}, shown in Fig.  \ref{fig:Fano1}  here, includes a relabeling of some points as well as  opposite direction of  4 arrows. But these two Fano planes are for the same octonion multiplication table in Fig. \ref{Table:Mult}.}
\label{fig:Cox}
\end{figure}
For example, the original Fano plane in  \cite{Coxeter:1946} shown in Fig. \ref{fig:Cox} for the same set of triads is different from the one in  \cite{Baez:2001dm} shown in Fig.  \ref{fig:Fano1}. The difference includes the opposite orientation of 4 lines, as well as relabeling.

\begin{figure}[H]
\centering
\includegraphics[scale=0.45]{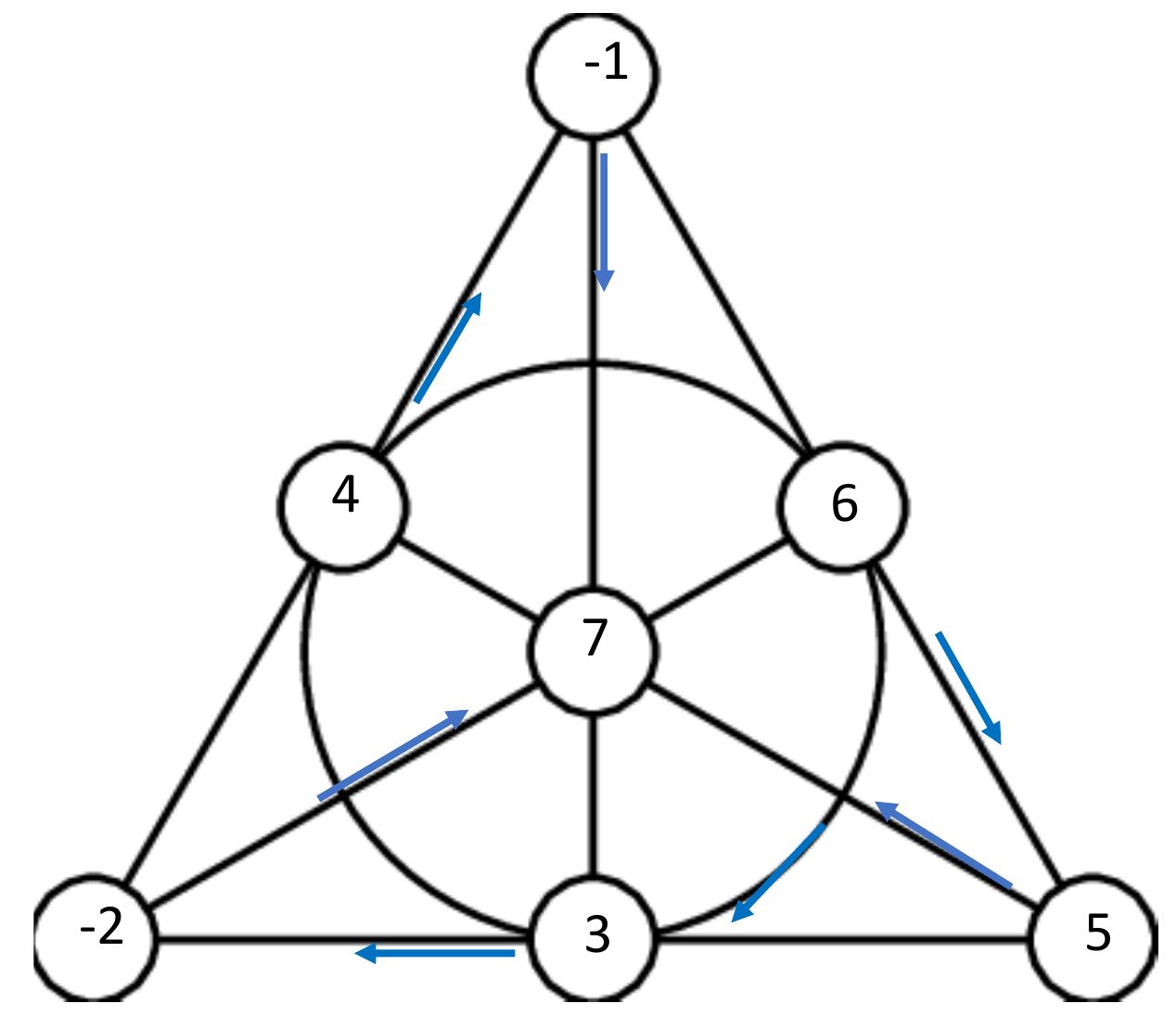}
\caption{\footnotesize  An oriented Fano plane corresponding to   Fig. \ref{fig:Cox}, where we have flipped the signs on $1$ and on $2$. Simultaneously we have flipped the direction of the red arrows in Fig. \ref{fig:Cox} when the line has an odd number of sign changes.  This  Fano plane is for the same octonion multiplication table in Fig. \ref{Table:Mult}.}
\label{fig:CoxBaez}
\end{figure}

 In general, one finds that oriented Fano planes with reversed orientation of even number of lines, can be presented in  the form with the standard orientation. An example is given in  Figs. \ref{fig:Fano1}, \ref{fig:Cox}, \ref{fig:CoxBaez}.

 We can read the associative triads from all 3 Fano planes  in Fig.  \ref{fig:Fano1}, in Fig. \ref{fig:Cox} and in Fig. \ref{fig:CoxBaez}. They are always the same, up to recycling in each triad.
\be
(124), (235), (346), (457), (561), (672), (7
13) \ee
The relation between these Fano planes is realized as follows. From Fig.  \ref{fig:Fano1} to   Fig. \ref{fig:Cox}
we have a  relabeling $
6\leftrightarrow 1, 3\leftrightarrow 2
$ and the change of 4 directions.
 From Fig.  \ref{fig:Fano1} to Fig. \ref{fig:CoxBaez} we have a  relabeling as well as 2 sign flips
\be
6\rightarrow -1\qquad 3\rightarrow -2 \qquad 1\rightarrow 6 \qquad 2\rightarrow 3
\ee
Starting from $C^{(+)}$ one can get 240 (including  $C^{(+)}$) different tables by permutations alone, without sign flips.

\section{  From cyclic to non-cyclic Hamming code}\label{secD}
Here we will show  an explicit  transformation from Cayley-Graves  octonions to Cartan-Schouten-Coxeter notation and that it also relates cyclic to non-cyclic Hamming codes. \be
e_i = \sum _{k=1}^7 T_{ik} \, j_k
\ee
or \be
e =  T \, j
\ee
\be
T = \left(\begin{array}{ccccccc}1 & 0 & 0 & 0 & 0 & 0 & 0 \\0 & 1 & 0 & 0 & 0 & 0 & 0 \\0 & 0 & 0 & 0 & 1 & 0 & 0 \\0 & 0 & 1 & 0 & 0 & 0 & 0 \\0 & 0 & 0 & 0 & 0 & 0 & 1 \\0 & 0 & 0 & 0 & 0 & 1 & 0 \\0 & 0 & 0 & -1 & 0 & 0 & 0\end{array}\right).
\label{MurT}\ee
The two bases $j_i$ and $e_i$ are related by a rotation operator  $T$, which involves a permutation of octonions, including a sign flip. Here $T$ is an orthogonal matrix $TT^T=1$. This is an example of a permutation of octonions with a sign flip.

We will now show that the the original, non-cyclic Hamming code in Fig. \ref{fig:HammG_16_H}, or in eq. \rf{NC}  which shows just a 7 codewords, and the cyclic Hamming code in Fig. \ref{fig:Planat}   are related by the same rotation $T$. Let us denote the entries of the original, non-cyclic Hamming code  as $h_i$
\bea
&& [ h_1 \dots h_7] \Rightarrow  1110000, \, 1001100, \, 0101010, \cr
\cr
&& 0010110, \, 0100101, \, 0011001, \, 1000011
\label{GrCayh}\eea
and the entries of the cyclic Hamming code  as $c_i$.
\bea
&&[ c_1 \dots c_7] \Rightarrow  1101000, \, 0110100, \, 0011010, \cr
\cr
&& 0001101, \, 1000110, \, 0100011, \, 1010001.
\label{Coxc}\eea
Then one finds
\be
h_i = \sum _{k=1}^7 T_{ik} \, c_k\, ,  \qquad c_i = \sum _{k=1}^7 T_{ki} \, h_k
\ee
or \be
h =  T \, c\, ,  \qquad c =  T^T \, h.
\ee
This means that the rotation operator $T$ in eq. \rf{MurT} acts on a code as follows:
\be
 [ h_1 \dots h_7] \qquad \Longrightarrow \qquad [ c_1 \dots c_7].
\ee
For each codeword it means
\bea
&& 1110000 \qquad \longrightarrow \qquad 1101000 \cr
\cr
&& 1001100 \qquad \longrightarrow \qquad  0110100 \cr
\cr
&& 0101010 \qquad \longrightarrow \qquad 0011010 \cr
\cr
&& 0010110 \qquad \longrightarrow \qquad 000110-1 \cr
\cr
&& 0100101 \qquad \longrightarrow \qquad 1000110 \cr
\cr
&& 0011001 \qquad \longrightarrow \qquad 010001-1 \cr
\cr
&& 1000011 \qquad \longrightarrow \qquad 101000-1 .
\label{cToc}\eea
Note that since Fano plane is a projective plane\footnote{ See also \cite{Planat} where the relevant issues of quantum information, Galois fields, finite geometry, Fano plane and cyclic binary Hamming (7,4) code are discussed in detail.} over a field of characteristic two, we have $-1= +1$ (mod 2). The right hand side of eq. \rf{cToc} becomes a set of 7 codewords of a cyclic code
\be
1101000, \, 0110100, \, 0011010, \, 0001101, \, 1000110, \, 0100011, \, 1010001.
\ee
Modulo this subtlety we see that the cyclic codewords and original Hamming codewords are related by a rotation which involves a permutation of octonions, including a sign flip.

One can also relate the cyclic Hamming codewords $(c_1....c_7)_C$ to the non-cyclic Hamming codewords $(h_1....h_7)_H$ by the action of the permutation operators $T_P=(1)(2)(6)(3574)$ :
\bea
T_P(1101000)_C=(1110000)_H \cr
T_P(0110100)_C=(0100101)_H \cr
T_P(0011010)_C=(0010110)_H \cr
T_P(0001101)_C=(0011001)_H \cr
T_P(1000110)_C=(1000011)_H \cr
T_P(0100011)_C=(0101010)_H \cr
T_P(1010001)_C=(1001100)_H.
\eea

\section{Octonions and dS vacua stabilization in M-theory}\label{secE}
In this paper we mainly concentrated on finding supersymmetric flat directions in M-theory, and the  benchmark cosmological models with $3\alpha=7,6,5,4,3,2,1$. However, in addition to it, our results, in combination with the nonperturbative corrections to the superpotential studied in \cite{Kallosh:2019zgd,Cribiori:2019drf,Cribiori:2019hrb}, provide us with  a simple analytical method to find local supersymmetric Minkowski vacua, which can be uplifted to metastable dS vacua.

Consider, for example, the M-theory superpotential $W^{\rm oct}=\sum _{\{ijkl \}} (T^i -T^j) (T^k -T^l)$ \rf{ourW} in combination with the nonperturbative racetrack superpotential $W_{KL}(T^{1})=W_{0} +Ae^{-aT^{1}}- Be^{-bT^{1}}$ used in the KL mechanism of vacuum stabilization \cite{Kallosh:2004yh}:
\be
W=W^{\rm oct}+W_{KL}(T^{1})= \sum _{\{ijkl \}} (T^i -T^j) (T^k -T^l) + W_{0} +Ae^{-aT^{1}}- Be^{-bT^{1}}.
\label{ourW2} \ee
where
\be\label{w0stab}
W_0=  -A \left({a\,A\over
b\,B}\right)^{a\over b-a} +B \left ({a\,A\over b\,B}\right) ^{b\over b-a}  \ .
\ee
The octonion superpotential $W^{\rm oct}$ and all of its derivatives vanish along the flat direction $T^i=T^j$, ${\rm Im} \, T^i=0$. Meanwhile the
superpotential $W_{KL}(T^{1})$ and all of its derivatives vanish for all $T^i=T^j$, ${\rm Im} \, T^i=0$, but only at the point
\be
\label{fullcompl}
T^{1} = {1\over a-b}\left(\ln  {a\,A\over b\,B}\right)\ .
\ee
Therefore the point where  all $T^i$ are equal to each other,
\be
 T^i  = {1\over a-b}\left(\ln  {a\,A\over b\,B}\right)\ ,
\ee
is the only point where the full superpotential $W$ and  its derivatives vanish. This means that this point corresponds to the local supersymmetric Minkowski vacuum, which can be uplifted to a metastable dS vacuum following \cite{Kallosh:2004yh,Kallosh:2019zgd,Cribiori:2019drf,Cribiori:2019hrb}. The same result can be obtained if one uses the KL superpotential  $W_{KL}(T^{i})$ with respect to any other superfield $T^i$.

\bibliographystyle{JHEP}
\bibliography{lindekalloshrefs}
\end{document}